\documentclass[iop,tighten]{emulateapj}

\usepackage{graphicx}
\usepackage{amssymb}
\usepackage{amsmath}
\usepackage{gensymb}
\usepackage{apjfonts}
\usepackage{natbib}
\usepackage{times}
\usepackage{microtype}
\usepackage{footmisc}

\newcommand{\chandra}{\textit{Chandra }}
\newcommand{\hst}{\textit{HST }}
\newcommand{\fuse}{\textit{FUSE }}
\newcommand{\xmm}{\textit{XMM-Newton }}
\newcommand{\suzaku}{\textit{Suzaku }}

\newcommand{\rxte}{\textit{RXTE }}
\newcommand{\asca}{\textit{ASCA }}
\newcommand{\beppo}{\textit{BeppoSAX }}

\newcommand{\gmm}{$\Gamma$ }
\newcommand{\chisq}{$\chi\sp{2}$ } 
\newcommand{\xspec}{\textsc{xspec} }
\newcommand{\nh}{$N\sb{\rm H}$ }
\newcommand{\nhgal}{$N\sb{\rm H}\sp{\it Gal}$ }
\newcommand{\msol}{M_\odot}

\newcommand{\flux}{{\rm erg\,s^{-1}cm^{-2}}}
\newcommand{\kms}{{\rm km\,s^{-1}}}
\newcommand{\kmss}{{\rm km\,s^{-2}}}
\newcommand{\ang}{\text{\normalfont\AA}}

\newcommand{\ka}{K$\alpha$ }
\newcommand{\kb}{K$\beta$ }

\newcommand{\nsig}{$N\sb{\sigma}$ }
\newcommand{\mnsig}{$|N\sb{\sigma}|$ }
\newcommand{\nv}{N\,\textsc{v}}
\newcommand{\civ}{C\,\textsc{iv}}
\newcommand{\siiv}{Si\,\textsc{iv}}

\begin{document}

\submitted{Accepted 2014 October 15}

\shorttitle{Absorption in NGC 3783}
\shortauthors{Scott et al.}

\title{Long-term X-ray stability and UV variability of the ionized absorption in NGC 3783}

\author{A.~E.~Scott\altaffilmark{1,2}, 
  W.~N.~Brandt\altaffilmark{1,2}, 
  E.~Behar\altaffilmark{3},
  D.~M.~Crenshaw\altaffilmark{4},
  J.~R.~Gabel\altaffilmark{5},
  R.~R.~Gibson\altaffilmark{6},\\
  S.~Kaspi\altaffilmark{3},
  S.~B.~Kraemer\altaffilmark{7},  
  T.~J.~Turner\altaffilmark{8}}

\affil{
$^{1}$Department of Astronomy \& Astrophysics, 525 Davey Laboratory, Pennsylvania State University, University Park, PA 16802, USA; amyscott@psu.edu\\
$^{2}$Institute for Gravitation and the Cosmos, Pennsylvania State University, University Park, PA 16802, USA\\
$^{3}$Department of Physics, Technion, Haifa 32000, Israel\\
$^{4}$Department of Physics and Astronomy, Georgia State University, 25 Park Place, Suite 605, Atlanta, GA 30303, USA\\
$^{5}$Physics Department, Creighton University, Omaha, NE 68178, USA\\
$^{6}$Department of Astronomy, University of Washington, Box 351580, Seattle, WA 98195, USA\\
$^{7}$Institute for Astrophysics and Computational Sciences, Department of Physics, The Catholic University of America, Washington, DC 20064, USA\\
$^{8}$Department of Physics, University of Maryland Baltimore County, 1000 Hilltop Circle, Baltimore, MD 21250, USA
}


\begin{abstract}
We present the results of recent \chandra High-Energy Transmission
Grating Spectrometer and \textit{Hubble Space Telescope} Cosmic
Origins Spectrograph observations of the nearby Seyfert 1 galaxy
NGC~3783 which shows a strong, non-varying X-ray warm absorber and
physically related and kinematically varying UV absorption.  We
compare our new observations to high-resolution, high signal-to-noise
archival data from 2001, allowing a unique investigation into the
long-term variations of the absorption over a 12~yr period.  We find
no statistically significant changes in the physical properties of the
X-ray absorber, but there is a significant drop of $\sim 40\%$ in the
UV and X-ray flux, and a significant flattening of the underlying
X-ray power-law slope.  Large kinematic changes are seen in the UV
absorbers, possibly due to radial deceleration of the material.
Similar behavior is not observed in the X-ray data, likely due to its
lower velocity resolution, which shows an outflow velocity of
$v\sb{\rm out}\sim-655\;\kms$ in both epochs.  The narrow iron~\ka
emission line at 6.4~keV shows no variation between epochs, and its
measured width places the material producing the line at a radial
distance of $\sim0.03$~pc from the central black hole.\\
\end{abstract}


\keywords{galaxies: active --- galaxies: Seyfert --- galaxies: individual (NGC~3783) --- X-rays: galaxies}


\section{Introduction}
\label{section:intro}
There is increasing evidence that supermassive black holes (SMBHs)
co-evolve with the bulges of their host galaxies.  The tight
correlation between the velocity dispersion of stars in the bulge and
the mass of the black hole at the center (M$-\sigma$; e.g.,
\citealt{ferrarese00,gebhardt00,gultekin09,mcconnell13,kormendy13}),
suggests regulatory feedback
(e.g.,~\citealt{silk98,king03,king05,mcquillin13}) from the Active
Galactic Nucleus (AGN), likely in the form of jets or winds.  Such
winds give rise to blueshifted absorption lines seen in the UV spectra
of approximately half of Seyfert galaxies
(e.g.,~\citealt{crenshaw99}), commonly the same AGN which also show a
`warm absorber' (WA; \citealt{halpern84}) in their soft X-ray spectra
(e.g.,~\citealt{reynolds97}).  This, along with the similarity of the
outflow velocities inferred, suggests that the UV and X-ray absorbers
are at least partly related physically and may be different
manifestations of the same outflow
(e.g.,~\citealt{mathur94,shields97,crenshaw99}).

Despite their importance and prevalence, the properties of these AGN
winds are still not fully understood and their origin within the AGN
structure is unclear.  Two main scenarios suggest that the winds
responsible for the warm absorber features are produced at very
different locations.  \citet{krolik01} propose that a
multi-temperature wind is formed from photoionized evaporation of
material off the inner edge of the torus located at parsec-scales from
the central SMBH.  However, \citet{elvis00} suggests that a wind
emerges vertically from a narrow region on the accretion disk at
$\sim1000\,R\sb{g}$ (corresponding to $5\times10\sp{-4}$~pc for a
$10\sp{7}\msol$ black hole) and is radially accelerated by radiation
pressure and bent to an angle of $\sim 60\degree$.  Such a wind may
produce the signature of a warm absorber at pc-scales when it collides
with the shell of gas that has been swept up by radiation pressure and
surrounds the black hole at 1--100~pc \citep{king14}.

Determining the radial distance of the absorbing material from the
central SMBH is therefore crucial for understanding its origin within
the AGN structure.  However, this quantity is not directly measurable
(see e.g.,~\citealt{crenshaw12}).  Since the thickness of the absorber
cannot exceed its distance from the SMBH an absolute maximum distance
can be determined ($r\sb{\rm max}\le L\sb{\rm ion}/\xi N\sb{\rm H}$;
\citealt{turner93}).  A minimum can be estimated from the size of the
\civ\, broad-line region as this is always at least partially covered
(e.g.,~\citealt{crenshaw12}) or by assuming the outflow is travelling
at the local escape velocity (e.g.,~\citealt{tombesi13}).  More
stringent limits can be obtained from absorption of excited states
(e.g.,~\citealt{kaastra04}) or from variability of the absorption.  In
particular, if the absorbers do not respond to changes in the ionizing
continuum, an upper limit on the density of the absorber is obtained,
giving a lower limit on its radial distance.

Variability has been observed previously in WA; indeed, the first WA
was identified after an apparent increase in the absorption column
density ($N\sb{\rm H}$) occurred over $\sim 1$~yr in MR~2251-178
\citep{halpern84,pan90}.  Further observations with \asca showed
variations in \nh and the ionization parameter ($\xi$) on timescales
of months to yrs, which were explained by material moving in and out
of the line-of-sight \citep{kaspi04}.  However, absorption in this AGN
has also been observed to remain stable over a period of $\sim 9$~yrs
since 2002 \citep{reeves13}.  There are many other examples of AGN
whose WA show variability, covering a range of timescales.  For
example, short timescale absorption variability was observed in
MCG-6-30-15, with a factor of 2 increase in \nh seen over a period of
3 weeks \citep{fabian94}, changes in individual absorption lines have
been observed over days \citep{gibson07}, and occultation events have
been observed to occur over $\lesssim 8$ hrs
\citep{mckernan98,marinucci14}.  Longer timescale variability is
observed in Mrk~348, with a decrease in \nh and $\xi$ in some of its
absorption components occurring over 6 yrs \citep{marchese14}, and
Mrk~704 shows very large changes (up to factors of $\sim 8$ depending
upon which spectral model is adopted) in the covering factor, $f$,
$N\sb{\rm H}$, and $\xi$ over 3~yrs \citep{matt11}.  Similarly,
\citet{longinotti13} report the re-emergence of a WA in Mrk~335 after
$\gtrsim 10$~yrs of observations in which no absorption was detected,
indicating both large and long-term variations.  NGC~3516 shows
absorption which is variable over a range of timescales with an
occultation occurring over hours \citep{turner08}, changes in
ionization state occurring over months \citep{turner05} and the
appearance of fast kinematic components happening after years
\citep{holczer12}.  However, similar to MR~2251-178, other AGN show
periods in which their WA do not appear to vary; e.g.,~NGC~5548 has
shown periods of 3~yrs (1999--2002) with no apparent change in its
properties \citep{steenbrugge05}, but others (2002--2005) do show a
decrease in the ionization \citep{detmers08}.  Later monitoring of
this AGN in 2007 with \suzaku over month-timescales again shows the WA
properties to be stable, despite large variations in the continuum
flux \citep{krongold10}.  Recent monitoring with \xmm during 2013
shows the nucleus of NGC~5548 to be obscured by a long-lasting, clumpy
stream of ionized gas, thus leading to a further change in the X-ray
absorption \citep{kaastra14}.  The absorption components in Mrk~509 do
not show short-term variation over a monitoring period of 36 days,
despite an increase in soft X-ray flux of 60\% during the campaign.
No long-term (3 to 9 yrs) changes are detected either, suggesting the
absorbing material lies at large ($>$pc) distances from the black hole
\citep{kaastra12}.  Intrinsic UV absorption has also been observed to
vary in many sources, due to changes in the ionization state of the
material or motion of material across the line of sight giving a
variable column density
(e.g.,~\citealt{maran96,shields97,kraemer01,kraemer02}).

NGC~3783 is a nearby ($z=0.00976\pm0.00009$; \citealt{RC3}), optically
bright ($V\simeq13$~mag) Seyfert~1 with a black hole mass of $M\sb{\rm
  BH}=(2.98\pm0.54)\times10\sp{7}\msol$ \citep{peterson04}.  It has
been studied extensively in the \mbox{X-ray} band and has a strong
X-ray warm absorber
(e.g.,~\citealt{turner93,george98,derosa02,kaspi00,kaspi01,kaspi02,blustin02,behar03,brenneman11}).
The first high-resolution grating observations from \chandra showed
many blueshifted, narrow absorption lines from hydrogen- and
helium-like ions of Ar, S, Si, Mg, Ne, and O, and L-shell transitions
of Fe~\textsc{xvii} to Fe~\textsc{xxi} as well as weak emission lines
from O and Ne \citep{kaspi00}.  Further photoionization calculations
indicated these were due to two phases of gas with different covering
factors and different ionization levels \citep{kaspi01}.
\citeauthor{kaspi02}~(\citeyear{kaspi02}; hereafter K02) presented a
high signal-to-noise (S/N), high-resolution X-ray spectrum from 900~ks
of \chandra data obtained in 2001.  Narrow absorption lines from Fe,
Ca, Ar, S, Si, Al, Mg, Ne, O, N, and C were again identified, with an
average outflow velocity of $v\sb{\rm out}=-590\pm150\;\kms$.  The
lines showed an asymmetry in their profiles, thought to be due to the
two separate absorbing systems.  Absorption lines from lower
ionization state ions were also observed, along with M shell
transitions of Fe.  Global photoionization modelling of this data set
by \citet{netzer03} found it to be best described by three distinct
absorption components with a large range in ionization state (${\rm
  log}\,U\sb{\rm OX}=-2.4,-1.2,-0.6$)\footnote{Although the
  conversions between different definitions of the ionization
  parameter depend upon the spectral energy distribution assumed for
  the source, approximate relations are: ${\rm log}\,U={\rm
    log}\,\xi-1.5$ \citep{crenshaw12} and ${\rm log}\,U={\rm
    log}\,U\sb{\rm OX}+1.99$ \citep{krongold03}.\label{ionnote}}, but
a narrower range in column density (${\rm log\,}N\sb{\rm
  H}=21.9,22.0,22.3$), with each split into two kinematic components.
However, the same data set was modelled by \citet{krongold03} using
only a two-phase absorber (${\rm log}\,U\sb{\rm OX}=-2.8,-1.2$, ${\rm
  log\,}N\sb{\rm H}=21.6,22.2$) and the same outflow velocity for each
component ($v\sb{\rm out}\sim750\;\kms$).  Similar models were found
for \xmm Reflection Grating Spectrometer (RGS; \citealt{RGS}) data in
which absorption lines from Fe, Ar, S, Si, Mg, Ne, O, N, and C were
all identified.  Two phases of gas, with ${\rm
  log}\,\xi=2.4,0.3$\footref{ionnote}, ${\rm log\,}N\sb{\rm
  H}=22.45,20.73$, and an outflow velocity of $\sim800\;\kms$ were
used \citep{blustin02}.  \citet{holczer07} improved upon the
traditional multi-component modelling by reconstructing a continuous
distribution of column density with ionization parameter, also found
in other Seyfert outflows \citep{behar09}.  No short-term variability
of the absorption was found in either the \chandra or \xmm data.  The
lack of response by the absorbers to changes in the continuum allowed
lower distance limits of $r\ge3.2,0.6,0.2$~pc to be placed on the
components found in the \chandra data \citep{netzer03}, and
$r\ge0.5$~pc and $r\ge2.8$~pc for the high and low ionization
components identified in the \xmm data \citep{behar03}.  Longer
timescale variations were observed with early \asca data when the
opacity of the WA decreased by a factor of $\approx 2$ over an 18
month period \citep{george98}.  Changes in the ionization and column
density may have also occurred in the 8 yrs between 2001 and a \suzaku
observation in 2009 \citep{brenneman11}. The WA shows no signs of
variability within this 210~ks observation \citep{reis12}.

NGC~3783 was the subject of an intensive monitoring campaign in the UV
from $2000-2002$.  Combining 18 epochs from the Space Telescope
Imaging Spectrograph (STIS; \citealt{STIS}) on board the
\textit{Hubble Space Telescope} (\textit{HST}) and 5 epochs from the
\textit{Far Ultraviolet Spectroscopic Explorer} (\textit{FUSE}),
\citeauthor{gabel03a}~(\citeyear{gabel03a}; hereafter G03a) presented
a high S/N spectrum covering $905-1730$\,\ang.  This showed
blueshifted absorption from O\,\textsc{vi}, \nv, \civ,
N\,\textsc{iii}, C\,\textsc{iii}* and Ly$\alpha$ to Ly$\epsilon$ and
the previously detected kinematic components with outflow velocities
of $-1320\;\kms$ and $-548\;\kms$ (components~1 \& 2;
\citealt{maran96,crenshaw99}), $-724\;\kms$ (3; \citealt{kraemer01}),
and $-1027\;\kms$ (4; \citealt{kaiser02,gabel02}).
\citeauthor{gabel03b}~(\citeyear{gabel03b}; hereafter G03b) also found
component 1 to be decelerating, and that the deceleration was
increasing.  This may be due to the directional shift of an absorber
with respect to our line of sight or different regions of a continuous
flow crossing our sightline.  No decelerations were detected in the
other kinematic components.  Component 1 has also been shown to be
composed of two distinct, but co-located regions (1a and 1b;
\citealt{kraemer01}).  Modelling by \citet{gabel05} found an
ionization parameter of ${\rm log}\,U\sim -0.5$\footref{ionnote} for
the UV components 1b, 2, and 3, which is consistent with that of the
lowest-ionization X-ray absorber and distances of the components were
estimated to be $r\sb{1}\sim 25$\,pc, $r\sb{2}\le25$\,pc and
$r\sb{3}\le50$\,pc.

The wealth of high S/N, high-resolution spectroscopy available for
NGC~3783 has allowed a detailed characterization of its extremely rich
X-ray absorption spectrum and UV absorbing material.  Our aim is to
extend the work of these studies and investigate the variability of
the absorption to determine how the material evolves over long
timescales.  We therefore obtained a new 170~ks \chandra High-Energy
Transmission Grating Spectrometer (HETGS; \citealt{hetgs}) observation
(taken in 2013) to compare with the 900~ks archival spectrum from 2001
in order to study the evolution of the absorbers over more than a
decade at high-resolution.  Previous high-resolution studies of
NGC~3783 have only probed timescales of up to $\sim 1$~yr.  We also
obtained new \hst Cosmic Origins Spectrograph (COS; \citealt{COS})
data taken within a few days of our X-ray observation to investigate
similar long-term variation in the UV absorption, and further
investigate the relationship between the X-ray and UV absorbers.  A
description of our X-ray and UV observations and data reduction is
given in \S\ref{section:obs}.  We present our main analysis including
an investigation of the broadband spectral shape, individual
absorption lines and the kinematics of the absorbers in
\S\ref{section:results}.  We summarize our conclusions in
\S\ref{section:sum}.\\


\section{Observations}
\label{section:obs}

\subsection{\chandra data}
NGC~3783 has been observed by \chandra \citep{chandra} once in 2000
and extensively in 2001, resulting in a combined exposure time of
888.1~ks.  These data were originally reported in K02 and are
re-analyzed here.  In this work we also report results from two new
\chandra observations taken in March 2013 with a combined exposure
time of 160.7~ks.  Details of all these observations are given in
Table~\ref{table:obslog}.

\begin{table}[h]
\vspace*{-0.4cm}
\centering
\caption{\chandra observation log}
\label{table:obslog}
    \begin{tabular}{lllcc}
      \hline \hline
      \multicolumn{1}{l}{Obsid} &
      \multicolumn{1}{l}{Seq No.} &
      \multicolumn{1}{l}{Start Time} &
      \multicolumn{1}{c}{Exposure time} \\
      \multicolumn{1}{l}{ } &
      \multicolumn{1}{l}{ } &
      \multicolumn{1}{l}{(UTD)} &
      \multicolumn{1}{c}{(ks)} \\
      \hline
      00373   & 700045   & 2000 Jan 20 23:33:16   &  56.4 \\
      02090   & 700280   & 2001 Feb 24 18:44:57   & 165.1 \\
      02091   & 700281   & 2001 Feb 27 09:17:51   & 168.9 \\
      02092   & 700282   & 2001 Mar 10 00:31:15   & 165.5 \\
      02093   & 700283   & 2001 Mar 31 03:36:27   & 166.1 \\
      02094   & 700284   & 2001 Jun 26 09:57:42   & 166.2 \\
      \hline
      14991   & 702799   & 2013 Mar 25 16:49:21   &  59.0 \\
      15626   & 702799   & 2013 Mar 27 14:08:06   & 101.7 \\
      \hline
    \end{tabular}
\end{table}

\chandra performed the observations with the HETGS \citep{hetgs},
which consists of two separate gratings, the High Energy Grating (HEG)
and the Medium Energy Grating (MEG), in the focal plane of the
Advanced CCD Imaging Spectrometer (ACIS; \citealt{acis}) which images
the resulting spectra.  All 8 observations were reduced in a uniform
and standard way using \textsc{ciao} version 4.5 \citep{ciao} and
version 4.5.6 of the calibration database.  The X-ray spectra were
extracted for each grating arm, HEG and MEG, and for both the positive
and negative first orders.  Higher order spectra, which include $\le
6\%$ of the counts in the first order, were not considered.  Auxiliary
response files and redistribution matrices were generated using
\textsc{fullgarf} and \textsc{mkgrmf} respectively.  The $\pm1^{\rm
  st}$ order spectra were combined using a weighted mean with ${\rm
  weights}=1/\sigma\sp{2}$ where $\sigma=1+\sqrt{N+0.75}$
\citep{gehrels86}\footnote{In figures, the upper error is determined
  with $\sigma\sb{+}=1+\sqrt{N+0.75}$ and the lower error with
  $\sigma\sb{-}=N-N(1-\frac{1}{9N}-\frac{1}{3\sqrt{N}})^3$.} and $N$
is the number of counts.  The spectra from each of the 6 observations
in 2000/2001 were combined (and hereafter will be referred to as the
2001 data).  The two observations in 2013 were also combined and they
did not show any significant differences before addition.  HEG data in
the wavelength range $1.25-13$\,\ang\, were combined with MEG data in
the range $1.7-26$\,\ang, again using a weighted mean.  The total
number of counts in the resulting 0.5--10.0~keV combined spectra were
$909,903$ for the 2001 data and $73,762$ for 2013.  No background
counts were subtracted in this analysis as their contribution is known
to be small ($<0.5$\% of the signal in HETGS/ACIS).  The spectra are
corrected for a Galactic \nh contribution estimated to be
\nhgal$=1.4\times10\sp{21}\,\textrm{cm}\sp{-2}$ using equation (7) in
\citet{willingale13} and the values $E(B-V)=0.12$ \citep{alloin95} and
N$\sb{\rm HI}=9.91\times10\sp{20}\textrm{cm}\sp{-2}$ from the
Leiden/Argentine/Bonn survey (LAB; \citealt{kalberla05}).

\begin{figure*}
  \includegraphics[width=1.0\textwidth]{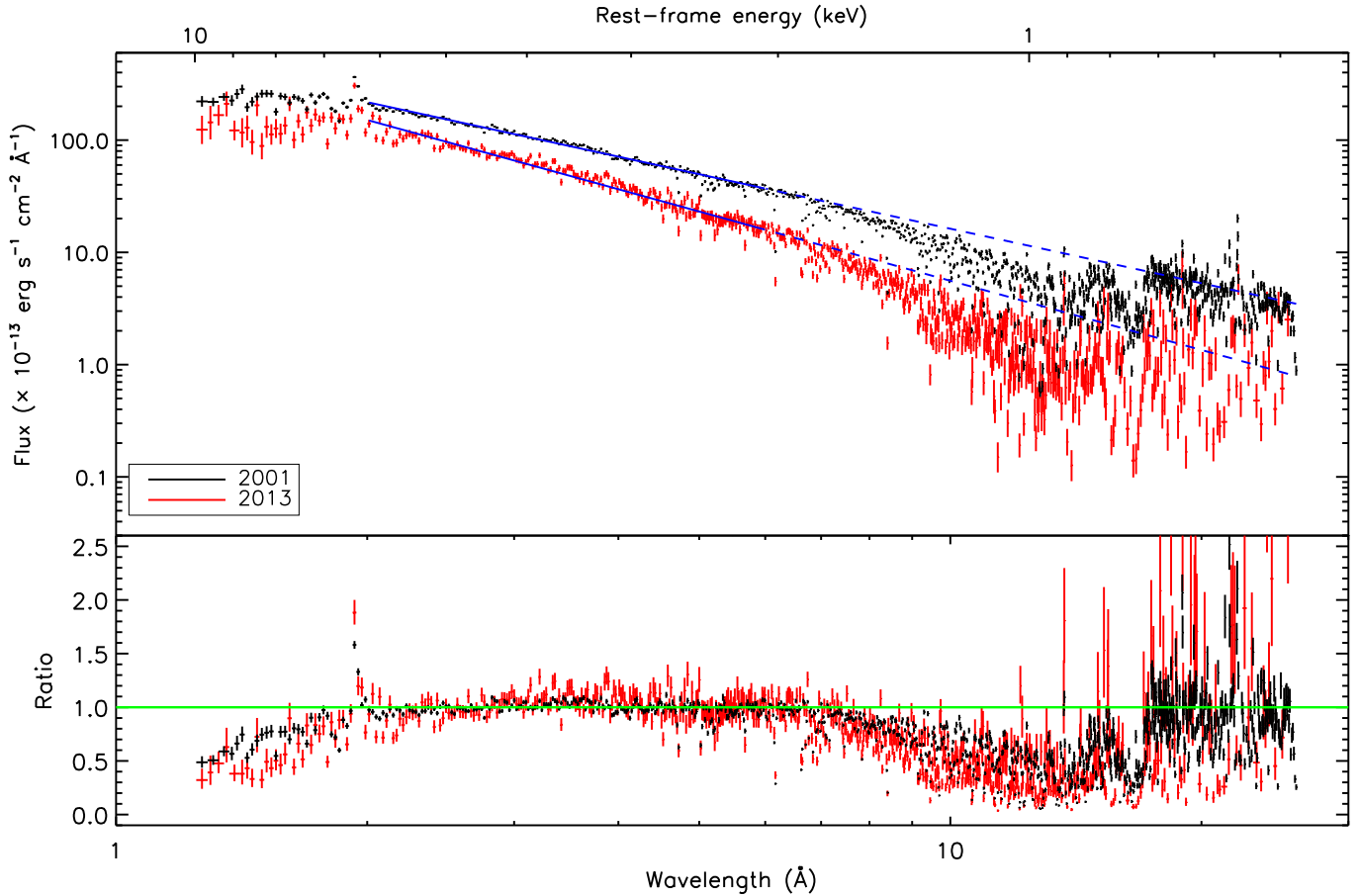}
  \caption{Broadband X-ray spectra of NGC~3783.  Top - the combined
    HEG \& MEG, $\pm1^{\rm st}$ order, fluxed spectra from the 2001
    (black) and 2013 (red) observations.  They are plotted with a
    minimum bin size of 0.02\,\ang, which is increased as necessary so
    as to ensure at least 10 counts per bin in the 2013 spectrum, and
    80 counts per bin for the 2001 spectrum.  Shown in blue (solid) is
    the power-law fit to the data included in a Line Free Zone and
    within the wavelength range $2-6$\,\ang\,($\sim 2-6$~keV).  This
    fit is also shown extrapolated to longer wavelengths (blue,
    dashed).  A more detailed view of each spectrum from
    $2-11$\,\ang\, is given in Figure~\ref{fig:lines_all}.  Bottom -
    The ratio of the data to the power-law model shown in the top
    panel.  The warm absorber at longer wavelengths/lower energies is
    clearly visible.  The lower ratios at $\lambda<2\,\ang$ are not
    caused by significant continuum absorption, but are a result of
    not modeling the reflection component in this figure.  The spectra
    at $\lambda\gtrsim11$\,\ang\, show many emission lines, likely
    from the narrow-line region (NLR).\\}
  \label{fig:full_x_spec}
\end{figure*}

The top panel of Figure~\ref{fig:full_x_spec} shows the `fluxed',
$\pm1^{\rm st}$ order, combined HEG \& MEG spectra for the 2001 data
(black), and the 2013 data (red) plotted in the rest-frame of
NGC~3783.  The spectra are shown with a minimum bin size of
0.02\,\ang, which is twice the bin size used in subsequent analyses.
The binning is also increased further where required, to ensure at
least 10 counts per bin in the 2013 spectrum and 80 counts per bin in
the 2001 spectrum.  This improves the clarity of the figure, allowing
the absorption and emission lines to be seen clearly (a more detailed
view of individual absorption lines is given in
Figure~\ref{fig:lines_all}).  The spectra do not show features
resulting from the ACIS chip gaps, which would be wider than any of
the spectral features we investigate.

\subsection{\hst data} 
NGC~3783 has been observed multiple times in the UV.  In this work we
present the most recent observation taken in March 2013 with COS
\citep{COS} on board \textit{HST}.  The observation was made within
days of the recent \chandra observations described in the previous
section.  The observations were performed using the G130M and G160M
gratings each with a resolving power of $\lambda/\Delta\lambda \simeq
16,000-21,000$ and covering a wavelength range of 900--1450\,\ang\,
and 1405--1775\,\ang\, respectively.  The exposure time of the
observation was 3.884\,ks, as listed in Table~\ref{table:hst_obslog}.

\begin{table}[h]
\vspace*{-0.4cm}
\centering
\caption{UV observations}
\label{table:hst_obslog}
    \begin{tabular}{cccc}
      \hline \hline
      \multicolumn{1}{c}{Epoch} &
      \multicolumn{1}{c}{Date} &
      \multicolumn{1}{c}{Exposure} &
      \multicolumn{1}{c}{Instrument/} \\
      \multicolumn{1}{c}{ } &
      \multicolumn{1}{c}{ } &
      \multicolumn{1}{c}{Time (ks)} &
      \multicolumn{1}{c}{ Grating} \\
      \hline
      1$^a$  & 2000 Feb 27     &   5.400  & STIS, E140M       \\ 
      2$^a$  & 2001 Feb -- Apr &  63.784  & STIS, E140M       \\
      3$^a$  & 2002 Jan 06     &   4.942  & STIS, E140M       \\
      4$^b$  & 2004 May 5      &  24.338  & \fuse             \\
      5$^b$  & 2011 Mar 23     &   2.088  & STIS, G430M       \\
             & 2011 May 26     &   4.553  & COS, G130M/G160M  \\
      6      & 2013 Mar 30     &   3.884  & COS, G130M/G160M  \\
      \hline
    \end{tabular}
    \begin{center}
    \textbf{Notes.} $^a$Presented in G03b.  $^b$Epochs to be presented
    in J. Gabel et al. (in preparation).
    \end{center}
\vspace*{-0.2cm}
\end{table}

\begin{figure*}[h]
  \centering
  \includegraphics[width=0.78\textwidth]{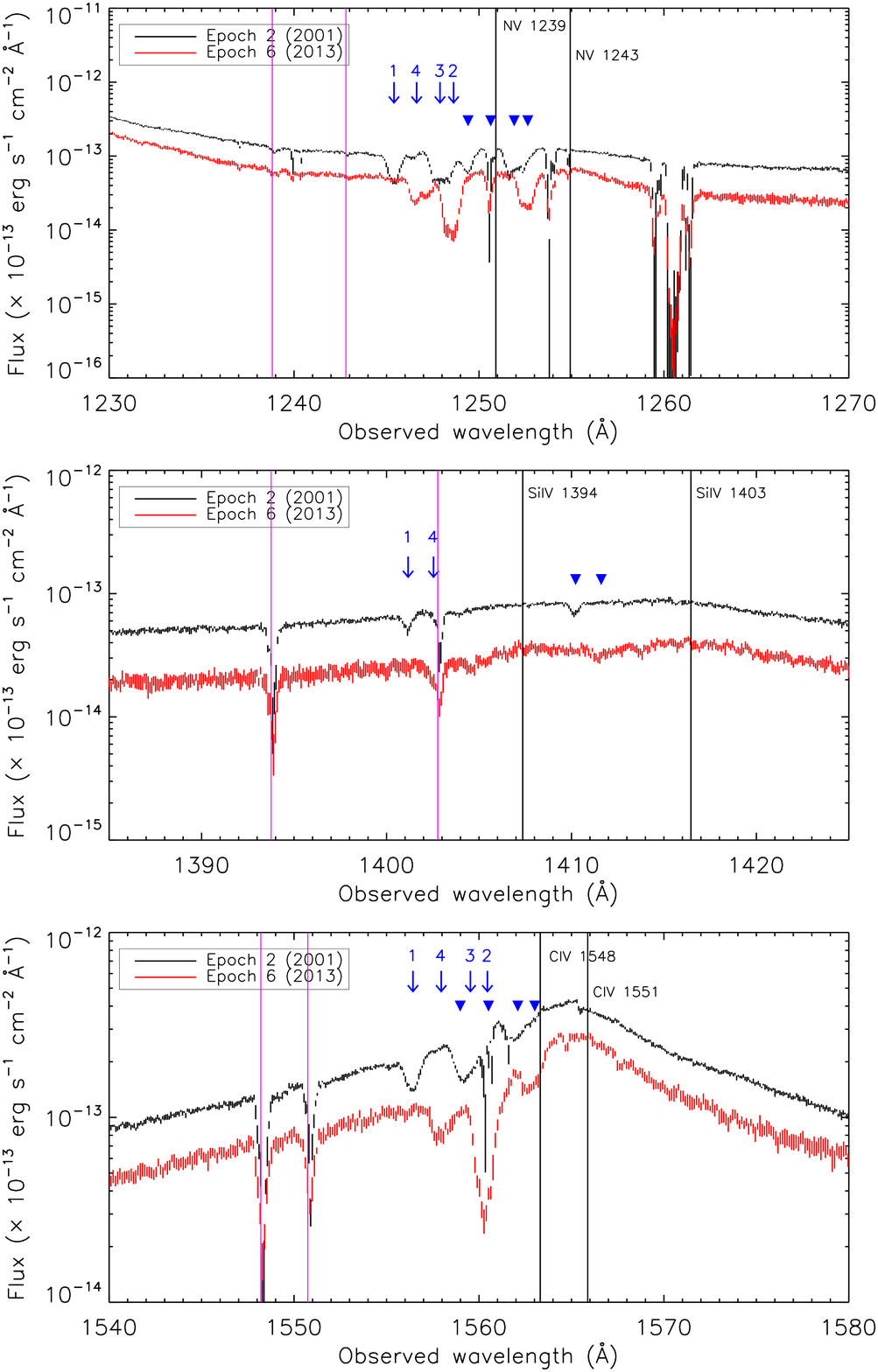}
  \caption{Sections of the UV spectra, clearly showing the drop in
    continuum flux between the two epochs and absorption troughs from
    the doublets of \nv, \siiv\, and \civ.  The data from Epoch~2
    (2001) are shown in black, while the new 2013 observations of
    Epoch~6 are shown in red.  Black vertical lines indicate the
    expected wavelength of the absorption lines in the rest-frame of
    NGC~3783 ($z=0.00976$).  The observed absorption troughs are
    noticeably blueshifted from these values and are highlighted with
    the blue arrows and triangles.  Arrows indicate blueshifted
    absorption lines due to the shorter wavelength line in the
    doublet, while triangles indicate those due to the longer
    wavelength line.  Numbers above the arrows refer to each of the UV
    kinematic components which are discussed in detail in
    Section~\ref{section:uv_kin}.  Note that not every component is
    present in both epochs.  The vertical magenta lines identify
    Galactic interstellar absorption lines due to the ions shown in each
    respective panel.  ISM lines due to other ions are also visible, but
    are not highlighted for clarity.}
  \label{fig:uv_spec}
\end{figure*}

We also re-consider earlier \hst data taken with STIS during the
extensive monitoring campaign in 2001.  These data were taken with the
E140M medium-resolution echelle grating covering a wavelength range of
1144--1710\,\ang\, with a resolving power of $R\simeq 45,800$ and were
originally reported in G03a (see that paper for more details on the
data reduction).  For consistency with G03b we refer to the UV
observations by an `epoch number' as listed in
Table~\ref{table:hst_obslog}.  In this work we consider observations
from Epoch 2, as they most closely match the observation dates of the
2001 \chandra data.  They have a combined exposure time of 63.784\,ks.
A \fuse spectrum from 2004 and \hst spectra from 2011, listed as
Epochs~4 \& 5, will be discussed in J. Gabel et al.~(in preparation).
We refer to our most recent 2013 \hst observation considered in this
work as Epoch 6.  Portions of the UV spectra are shown in
Figure~\ref{fig:uv_spec}.  Epoch~2 data from 2001 are shown in black,
and Epoch~6 data from 2013 are shown in red.\\

\section{Results}
\label{section:results}

\subsection{Broadband spectra}
\label{section:broadband}
In this section we assess changes in the broadband spectra of NGC~3783
between 2001 and 2013.  Figure~\ref{fig:full_x_spec} shows the 2001
X-ray spectrum in black and the 2013 spectrum in red, both corrected
for Galactic \nh and plotted against rest-frame energy and rest-frame
wavelength.  We do not attempt a detailed modelling of the broadband
continuum which is challenging due to the complexity of the warm
absorber and possible presence of a soft X-ray excess and Compton
reflection (e.g.,~\citealt{derosa02}) for which we would require data
at higher energies to constrain fully.  Instead, we fit a power-law
model to data in the range $2-6$\,\ang\,($\sim 2-6$~keV) which is less
affected by the absorption, allowing a simple comparison of the
spectral shape in the two epochs.  We also only include in the fitting
those regions of the spectrum which do not contain strong absorption
lines i.e.,~the Line Free Zones (LFZ) defined by \citet{kaspi01},
although we note that few zones are entirely line free, and absorption
edges still contribute to curvature in the spectrum.  The best-fit
parameters are not altered significantly when the all the data in the
energy range, not just the LFZ, are used.  The best-fitting model is
shown in Figure~\ref{fig:full_x_spec} by the solid, blue line which is
extrapolated to lower energies/longer wavelengths by the dashed, blue
line.  This figure clearly shows a drop in the X-ray flux between 2001
and 2013 and we measure $F\sb{\rm 2-6,\,2001}=(3.84\pm0.01)\times
10\sp{-11}\,\flux$ and $F\sb{\rm 2-6,\,2013}=(2.25\pm0.02)\times
10\sp{-11}\,\flux$ making the decrease highly significant\footnote{We
  also estimate $F\sb{\rm 2-10\:keV,\,2001}=(5.55\pm0.03)\times
  10\sp{-11}\,\flux$ and $F\sb{\rm
    2-10\:keV,\,2013}=(3.20\pm0.05)\times 10\sp{-11}\,\flux$,
  corresponding to a drop in flux of $42\pm1\%$.}.  We also see a
significant flattening of the continuum slope, with $\Gamma\sb{\rm
  2-6}=1.37\pm0.01$ measured in the 2001 spectrum and $\Gamma\sb{\rm
  2-6}=1.07\pm0.04$ measured in 2013.  Such variations in the
power-law slope may be intrinsic, likely related to changes in the
accretion flow close to the central SMBH, or they may be only apparent
changes.  These can be due to a change in the amount of absorbing
material along the line-of-sight to the central source, or due to a
change in the relative amounts of intrinsic and reflected emission
contributing to the power-law observed.  We comment on each of these
possibilities below.

The bottom panel of Figure~\ref{fig:full_x_spec} shows the ratio of
each spectrum to its $2-6$\,\ang\, power-law model.  This allows a
simple comparison of the absorption at long wavelengths/low energies.
The general shape appears similar in both epochs, suggesting no
dramatic change in the warm absorbers.  Although the narrow wavelength
range (2--6\,\ang) over which we measure the power-law slope contains
only a few strong absorption lines (e.g.,~from S between
3.5--5.4\,\ang), the warm absorber could still impact the continuum
through bound-free absorption (see e.g.,~Figure~5 of
\citealt{netzer03}; the Ca and Ar edges lie at 3.1\,\ang\, and
3.9\,\ang, respectively).  We therefore measure the power-law slope
over a series of progressively narrower wavelength ranges in order to
estimate better the intrinsic continuum slope and determine whether
the change in \gmm between the two epochs is real, and not an artifact
of increased absorption in 2013.  For the 2001 spectrum we measure
$\Gamma\sb{\rm 2-5\,\ang}=1.39\pm0.01$, increasing to $\Gamma\sb{\rm
  2-4\,\ang}=1.44\pm0.02$ and $\Gamma\sb{\rm 2-3\,\ang}=1.47\pm0.05$.
This shows an apparent convergence on $\Gamma\sim1.47$ when more of
the data thought to be contaminated by the spectral signature of the
warm absorbers are excluded, indicating that they do affect shorter
wavelengths/higher energies.  Although this value is steeper as
expected, it is still lower than the $\Gamma\sim 1.5-1.7$ often
reported for NGC~3783 from models of the full spectrum
(e.g.,~\citealt{blustin02,netzer03,krongold03,reeves04}).  The
power-law slope measured in the 2013 spectrum also steepens when a
narrower energy range is used ($\Gamma\sb{\rm 2-5\,\ang}=1.13\pm0.05$,
$\Gamma\sb{\rm 2-4\,\ang}=1.24\pm0.07$, and $\Gamma\sb{\rm
  2-3\,\ang}=1.27\pm0.15$), but remains significantly flatter than in
2001, suggesting the observed change is real and unrelated to the warm
absorbers.  These fitting results are not affected by the wing of Fe
K$\alpha$.

The flattening of the power-law slope we see is consistent with the
commonly observed $\Gamma$-flux relation for AGN
(e.g.,~\citealt{taylor03,lamer03,pounds04,vaughan04,miller07,gibson12})
in that the slope flattens with a decrease in flux.  This relation may
be due to an increase in the fraction of the spectrum coming from
reflection of the primary emission off some distant material, which
only responds to a decrease in continuum flux after a time delay.  We
therefore model the 2001 spectra with a power-law plus neutral
reflection model using $R=0.9$ as found in \suzaku observations
\citep{brenneman11,reynolds12} and assume $E\sb{\rm cut}=200$ keV and
${\rm cos}\,i=0.6$.  We then simulate spectra with the best-fitting
parameters from this model, but with only 60\% of the power-law flux,
consistent with the decrease previously observed.  A simple power-law
fit to these spectra over $2-6$\,\ang\, gives $\Gamma=1.34$.  Although
this is flatter than $\Gamma=1.37$ which is observed in the 2001 data,
it is not as flat as the value observed in 2013, suggesting a
fractional increase in the reflection component is unable to account
for the change in power-law slope between the epochs.  We also find
that a stronger reflection component e.g.,~$R=1.5$ is unable to
explain the change in \gmm within our data, which we note is not ideal
for constraining a reflection component.

A study of SDSS quasars with multiple \chandra observations suggests
that changes in the intrinsic power-law slope of $\Delta\Gamma=0.3$
are not uncommon \citep{gibson12} and similar, although smaller,
spectral variations have been observed {\it within} previous
observations of NGC~3783.  For example, using \xmm European Photon
Imaging Camera (EPIC; \citealt{MOS,pn}) data, \citet{reeves04}
observed a change in the power-law slope from $\Gamma=1.57$ at low
fluxes to $\Gamma=1.71$ at high fluxes, consistent with the flux-\gmm
correlation, and the spectral variability observed in \beppo
observations was also explained with a small change of
$\Delta\Gamma=0.1$ in the power-law slope \citep{derosa02}.  \suzaku
data also showed spectral variability on both short (20~ks) and longer
(100~ks) timescales due to the central source, not the WA
\citep{reis12}.

\begin{figure*}[ht!]
  \centering 
    \includegraphics[width=1.0\textwidth]{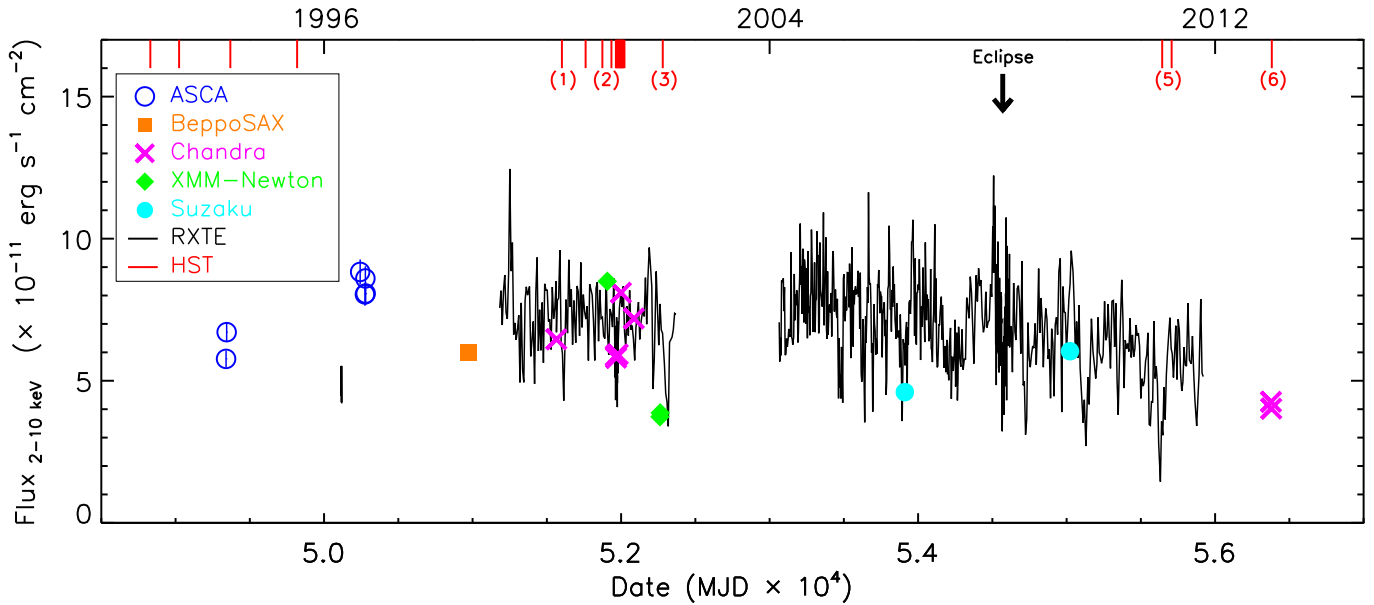}
  \caption{Long-term X-ray lightcurve for NGC~3783.  The black line
    indicates the $2-10$~keV flux determined from \rxte data
    \citep{rivers13}, which show a factor of 5 change in flux.  The
    black arrow indicates the date of the apparent eclipse event found
    in the \rxte data by \citet{markowitz14}.  Colored symbols
    indicate the $2-10$~keV flux from observations by other X-ray
    missions.  They include \asca (blue, open circles;
    \citealt{george98}), \beppo (orange square; \citealt{derosa02}),
    \xmm (green diamonds; \citealt{blustin02}, \citealt{reeves04}),
    \suzaku (cyan, filled circles; \citealt{miyazawa09},
    \citealt{brenneman11}), and \chandra (magenta crosses; this work).
    These fluxes are determined from a simple power-law fit to the
    spectra from each individual \chandra observation used in this
    work.  Note that Obsids 02090 and 02091 lie very close together on
    this figure.  The dates of \hst observations are indicated by the
    red tick marks at the top of the figure.  The numbers in brackets
    refer to the epoch numbers listed in Table~\ref{table:hst_obslog}.
    Data from \textit{HEAO-1} \citep{mushotzky80}, \textit{EXOSAT}
    \citep{turner89}, and \textit{Ginga} \citep{nandra94} are also
    consistent with the range of fluxes shown in the figure, but are
    not included for clarity.\\}
  \label{fig:lc}
\end{figure*}

Long-term \textit{Rossi X-ray Timing Explorer} (\textit{RXTE})
monitoring of NGC~3783 shows considerable X-ray continuum variations
on all timescales (e.g.,~\citealt{markowitz04}), with $2-10$~keV flux
changes as large as $\Delta F\sim5\times10\sp{-11}\,\flux$ occurring
over $\sim 20$~days.  Such changes are consistent with our
observations, and while some may be due to absorbing material, for
example the occultation event found recently by \citet{markowitz14},
others are more likely related to intrinsic variations within the
X-ray continuum emitting region than persistent, rapid changes in the
absorption.  Figure~\ref{fig:lc} shows the long-term $2-10$~keV X-ray
light-curve for NGC~3783.  The black data come from \rxte monitoring
\citep{rivers13}, red tick marks indicate dates of \hst observations,
and colored symbols show X-ray fluxes from observations by other
missions, including the magenta crosses which correspond to each of
the \chandra observations presented in this work.

Figure~\ref{fig:uv_spec} shows broad sections of the UV spectra;
Epoch~2 data (2001) are shown in black, and Epoch 6 data (2013) are
shown in red.  It shows a drop in the continuum flux similar to that
seen in the X-ray band with the average 1270--1520\,\ang\, flux in
Epoch 6 being $40\pm4\%$ of the average Epoch~2 flux over the same
wavelength range.  Similarly to the X-ray continuum, the UV continuum
is known to be highly variable (e.g.,~\citealt{reichert94}).  Also
shown are absorption features due to the doublets of
\nv\,($\lambda\lambda$ 1238.821\,\ang, 1242.804\,\ang),
\siiv\,($\lambda\lambda$ 1393.76\,\ang, 1402.77\,\ang) and
\civ\,($\lambda\lambda$ 1548.202\,\ang, 1550.774\,\ang).  Black
vertical lines indicate the expected wavelengths of these absorption
lines in the rest frame of NGC~3783, and the absorption profiles
observed are blueshifted with respect to these positions, indicating
the material is outflowing.  The absorption troughs due to the shorter
wavelength member of the doublet are highlighted by the blue arrows,
and those due to the longer wavelength line are shown with solid blue
triangles.  The numbers shown above the arrows correspond to different
kinematic components.  These will be discussed in detail in
\S\ref{section:uv_kin}, in which we present velocity profiles of the
spectra in order to investigate further the outflow kinematics.


\subsection{Individual absorption lines}
\label{section:lines}
The 900\,ks X-ray spectrum taken in 2001 shows 135 narrow absorption
lines from hydrogen-like, helium-like and lower-ionization ions of Fe,
Ca, Ar, S, Si, Al, Mg, Ne, O, N, and C, including 1s--2p, and many
higher order (e.g.,~up to 5p) transitions (K02; this work).  Many of
these absorption lines are also seen in the 2013 spectrum, although
due to its lower S/N (9 at 7\,\ang\, compared with 37 at 7\,\ang\, in
2001) only those with the highest equivalent widths (EW) are detected
with a high level of significance.  However, these still originate
from a range of ionization species and different order transitions.
We look for changes in the absorption lines between epochs, which may
be indicative of variations in the ionization state, column density,
or line-of-sight covering factor of the absorbing material.

In this section we only consider the spectra up to a wavelength of
11\,\ang\, to ensure a good number of counts per bin while being able
to maintain a bin size of $\Delta\lambda=0.01\,\ang$ required for the
identification of the absorption lines.  The 2001 spectrum contains at
least 100 counts per bin throughout this wavelength range, while the
2013 spectrum contains $>10$ counts per bin except for a few at longer
wavelengths at the locations of absorption lines (the total number of
counts included in each of the individual lines we consider in detail
is $>80$).

\begin{figure*}[h]
  \centering 
    \includegraphics[width=1.0\textwidth]{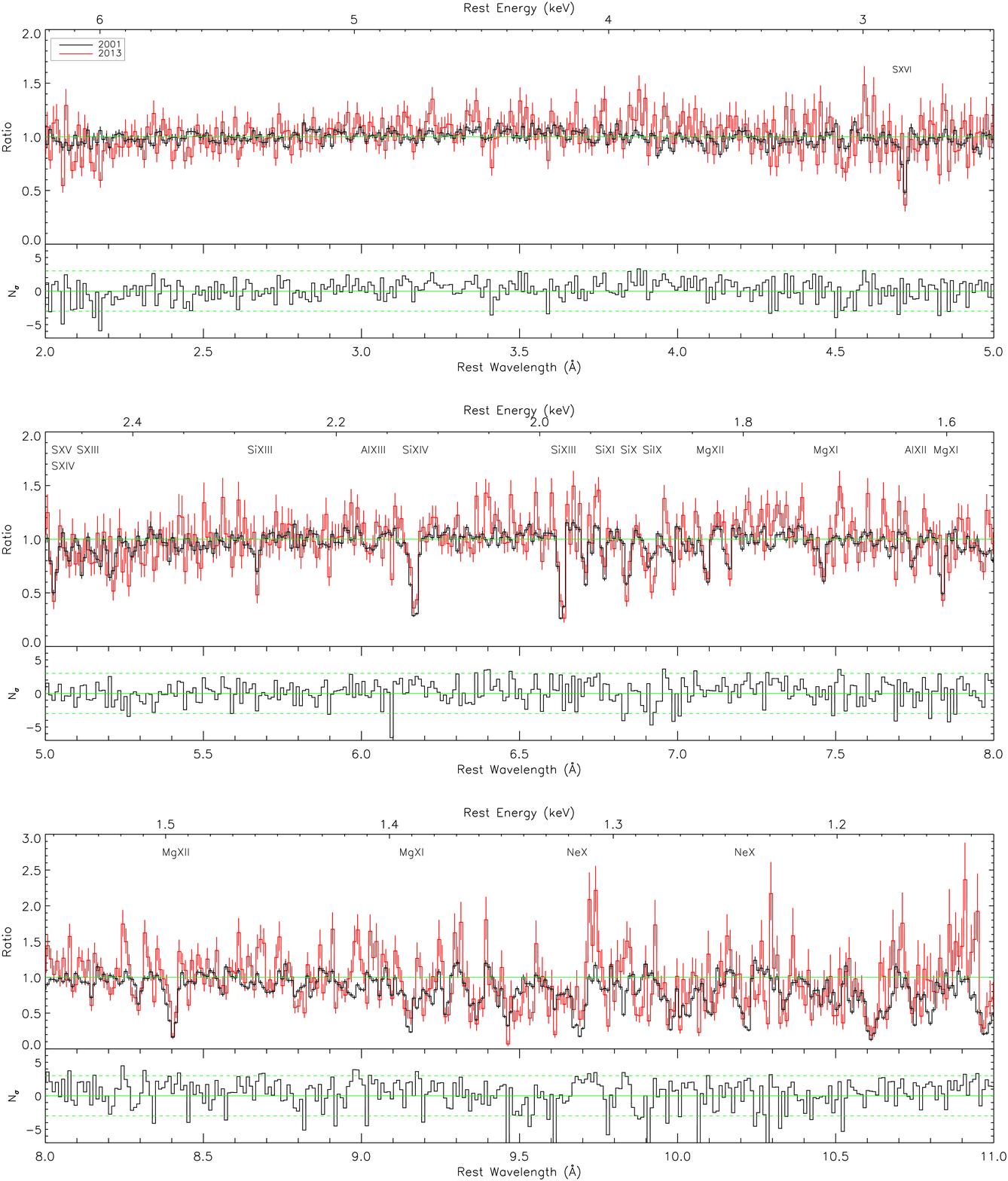}
  \caption{A comparison of the absorption lines in the spectra from
    both epochs.  The top panels plot the ratios of the spectra
    divided by the continuum, which is modelled with a power law over
    the narrow wavelength range included in each panel (except for the
    8--11\,\ang\, panel in which the continuum model is taken from a
    fit to the data in LFZ within 6--9\,\ang).  The 2001 data are
    shown in black and the 2013 data are shown in red, both with
    $1\sigma$ error bars, and are plotted against rest-frame
    wavelength and rest-frame energy.  The bottom panels show \nsig
    which is expected to lie between $\pm1$ if there is no significant
    variation between the two epochs.  The dashed green lines show
    \mnsig$=3$ outside of which the spectra are significantly
    different in that bin.  Strong, unblended absorption lines for
    which individual Gaussian fitting is carried out are labelled on
    the figure.  We caution that emission lines appear artificially
    enlarged in this ratio plot due to their additive, rather than
    multiplicative nature.}
  \label{fig:lines_all}
\end{figure*}

The spectra from both epochs are individually modelled with a power
law to give a representation of the source continuum.  We do not use a
global power law fit to the full $2-11$\,\ang\, spectrum as this does
not give a good representation of the continuum due to the spectral
curvature caused by absorption edges.  As our data are not well suited
to a detailed modelling of the continuum, in this analysis we simply
model the continuum to a level such that we can investigate variations
of the individual absorption lines.  We do this using narrow
wavelength ranges ($\Delta\lambda=3$\,\ang) corresponding to those
shown in each panel of Figure~\ref{fig:lines_all}, over which the
continuum is approximately a simple power-law.  We also only include
data in the LFZ.  At longer wavelengths not enough data remain in the
LFZ to give a reliable fit.  Therefore, in order to model the
$8-11$\,\ang\, range we extrapolate the power law fitted to data
within $6-9$\,\ang.  We therefore caution that the continuum modelling
in the $>9\,\ang$ region may not be optimal but we find no substantial
changes to our results when we use the alternative data range of
$5-9$\,\ang\, in the fit.  We estimate a $1\sigma$ error on the
continuum level by considering the distribution of residuals about the
resulting best-fit power-law.  We note that the subsequent modelling
of the individual absorption lines is dependent on the continuum used.

In order to assess any changes in the depths of the absorption lines
between epochs, we plot the ratio of the data to the continuum fit in
Figure~\ref{fig:lines_all}.  The 2001 data are shown in black, and the
2013 data are shown in red.  The bottom panel of each sub-plot shows
\nsig defined as

\begin{equation}
  N_{\sigma} = \frac{R_{\rm 2013} - R_{\rm 2001}}{\sqrt{\sigma_{\rm 2013}^{2} + \sigma_{\rm 2001}^{2}}},
  \label{eqn:nsigma}
\end{equation}
where $R_{\rm 2013}$ and $R_{\rm 2001}$ are the two ratios and
$\sigma_{\rm 2013}$ and $\sigma_{\rm 2001}$ are their errors
\citep{filizak13}.  This indicator allows a simple, visual indication
of any potential changes between epochs.  It is expected to lie
between $\pm 1\sigma$ if the two spectra are not significantly
different in that particular bin, with significant changes having
\mnsig$>3$ (these levels are indicated by the dashed green lines).
Even with no real variations in the absorption lines between the
epochs, we still expect to see \nsig values at $>1\sigma$ from purely
statistical considerations.  The \nsig values are expected to follow a
Gaussian distribution, therefore of the 300 bins within any
$3$\,\ang\, segment, at least 95 are expected to have \mnsig$>1$, 14
will have \mnsig$>2$ and 1 will have \mnsig$>3$, regardless of any
true variations of the absorption lines between epochs.  However, due
to the lower S/N of the 2013 spectrum which affects the continuum
modelling, there are many individual bins which display a significant
variation ($>3\sigma$).  These features are not due to chip gaps which
would produce features wider than the individual bin size.  The
positions of the main absorption lines we consider are labelled on
Figure~\ref{fig:lines_all}, and each typically covers $3-4$ of the
0.01\,\ang\, bins.  In some cases \nsig indicates a significant
variation in some, but not all of the bins within a line.  We
therefore instead use Gaussian fitting of the individual absorption
lines which is more informative for assessing line variations.

We attempt to model each of the S, Si, Al, Mg, and Ne absorption lines
that are listed as unblended by K02 with a Gaussian profile.  The
lines for which a Gaussian offers a good fit to the data are listed in
Table~\ref{table:lines} along with their transition (obtained from
NIST\footnote{http://physics.nist.gov/PhysRefData/ASD/lines\_form.html})
and rest-frame wavelength.  They are also labelled on
Figure~\ref{fig:lines_all}.  In general, the line parameters we obtain
for the 2001 spectrum are similar to those presented in K02.  We note
that more detailed analysis of the line profiles (see e.g.,~K02)
indicates that they are not strictly Gaussian and have an asymmetry
caused by material with different outflow velocities
(see~\S\ref{section:kinematics}).  However, assuming a Gaussian model
is adequate for our purposes here, in which we compare line flux
estimates to assess whether a particular absorption line has varied
significantly in strength between epochs.  For all lines we fit a
Gaussian profile to the 2001 data in which all parameters are free to
vary, and then fit the same Gaussian shape to the 2013 data; i.e.,\ we
fix the mean and sigma values, and allow only the line flux to vary.
This allows us to determine whether the absorption line becomes
stronger or weaker in the new data, assuming that the line profile
does not change.  These values are listed in Table~\ref{table:lines}
as the `2013 (fix)' model.  In some cases we are also able to fit a
freely varying Gaussian profile to the lines in the new data (listed
in Table~\ref{table:lines} as `2013'), with $\lambda\sb{\rm obs}$,
$\sigma$, velocity shift and line flux being determined from the
best-fitting Gaussian parameters.  The full-width at half-maximum
(FWHM) is determined from

\begin{equation}
  {\rm FWHM}\;({\rm km\,s}^{-1}) = 2\sqrt{2\,{\rm ln}\,2}\;\left(\frac{\sigma\; c}{\lambda_{\rm obs}}\right),
  \label{eqn:fwhm}
\end{equation}
where $\sigma$ and $\lambda\sb{\rm obs}$ are in units of \ang, and $c$
is in $\kms$.  In cases where the observed absorption line is thought
to be comprised of a doublet, the velocity shift quoted is calculated
with respect to the line of shortest wavelength.  In
Table~\ref{table:lines} we list the physical line fluxes of the
absorption lines in the 2013 spectra, and also quote a value which has
been scaled using a factor determined from the ratio of the continua
from both epochs at the line position ($C\sb{\rm 2001}$/$C\sb{\rm
  2013}$).  This removes the difference introduced by the lower
continuum flux level of the 2013 data and allows a direct comparison
of the line fluxes in each epoch.  The significance of each change in
$\sigma$ units, is listed.

\begin{table*}[h]
\scriptsize
\centering
\caption{Parameters of the Gaussian fits to the strongest, unblended absorption lines.}
\label{table:lines}
    \begin{tabular}{clccccclll}
      \hline \hline \\[-2.0ex]
      \multicolumn{1}{c}{Species} &
      \multicolumn{1}{l}{Data/} &
      \multicolumn{1}{c}{$\lambda_{\rm obs}$} &
      \multicolumn{1}{c}{$\sigma$} &
      \multicolumn{1}{c}{FWHM} &
      \multicolumn{1}{c}{v$_{\rm shift}$} &
      \multicolumn{2}{c}{Line flux } &
      \multicolumn{1}{c}{EW} \\
      \multicolumn{1}{c}{($\lambda_{\rm rest}$)} &
      \multicolumn{1}{l}{ Model} &
      \multicolumn{1}{c}{(\AA)} &
      \multicolumn{1}{c}{(m\AA)} &
      \multicolumn{1}{c}{(km s$^{-1}$)} &
      \multicolumn{1}{c}{(km s$^{-1}$)} &
      \multicolumn{2}{c}{($\times 10^{-7}$ ph cm$^{-2}$ s$^{-1}$)} &
      \multicolumn{1}{c}{(m\AA)} &
      \multicolumn{1}{c}{ } \\
      \multicolumn{1}{c}{ } &
      \multicolumn{1}{l}{ } &
      \multicolumn{1}{c}{ } &
      \multicolumn{1}{c}{ } &
      \multicolumn{1}{c}{ } &
      \multicolumn{1}{c}{ } &
      \multicolumn{1}{c}{Physical} &
      \multicolumn{1}{c}{Scaled} &
      \multicolumn{1}{c}{ } \\[0.5ex]
     \hline\\[-2.0ex]
S \textsc{xvi} (4.729) & 2001       & $4.7181\pm0.0006$ & $8.0\pm0.5$ & $1198\pm81$ & $-691\pm39$ & $138.8\pm8.8$ &                             & $10.8\pm1.9$ \\
  (H-like, 1s -- 2p)   & 2013 (fix) & \ldots            & \ldots      & \ldots      & \ldots      &  $76.9\pm8.6$ & $161.9\pm18.2\;(1.1\sigma)$ & $11.5\pm5.0\;(0.1\sigma)$
\\[2.8ex] 

S \textsc{xv} (5.039) & 2001       & $5.0276\pm0.0007$ & $8.5\pm0.6$ & $1187\pm89$  & $-681\pm39$  & $139.2\pm9.3$ &                             & $10.7\pm2.2$ \\
  (He-like, 1s -- 2p) & 2013 (fix) & \ldots            & \ldots      & \ldots       & \ldots       & $76.2\pm10.7$ & $154.0\pm21.5\;(0.6\sigma)$ & $7.2\pm4.8\;(0.7\sigma)$\\
                      & 2013       & $5.027\pm0.002$   & $7\pm2$     & $1014\pm217$ & $-736\pm116$ & $68.3\pm14.3$ & $83.0\pm46.0\;(1.2\sigma)$  & $8.0\pm5.7\;(0.4\sigma)$
\\[2.8ex]

S \textsc{xiv} (5.084,5.086) & 2001       & $5.074\pm0.002$ & $7\pm1$ & $1003\pm207$ & $-567\pm103$ & $48.7\pm9.5$  &                            & $3.2\pm2.6$ \\
  (Li-like, 1s -- 2p)        & 2013 (fix) & \ldots          & \ldots  & \ldots       & \ldots       & $27.8\pm13.0$ & $56.8\pm26.6\;(0.3\sigma)$ & $6.7\pm4.9\;(0.6\sigma)$ 
\\[2.8ex]

S \textsc{xiii} (5.126) & 2001       & $5.117\pm0.001$ & $6\pm1$  & $861\pm134$   & $-510\pm60$   & $71.7\pm8.7$  &                            & $5.5\pm2.0$  \\
  (Be-like, 1s -- 2p)   & 2013 (fix) & \ldots          & \ldots   & \ldots        & \ldots        & $19.3\pm12.1$ & $39.6\pm24.8\;(1.2\sigma)$ & $4.3\pm4.2\;(0.3\sigma)$\\
                        & 2013       & $5.106\pm0.008$ & $13\pm9$ & $1736\pm1186$ & $-1198\pm450$ & $40.4\pm22.4$ & $83.0\pm46.0\;(0.2\sigma)$ & $6.2\pm5.9\;(0.1\sigma)$ 
\\[2.8ex]

Si \textsc{xiv} (6.182) & 2001       & $6.1681\pm0.0002$ & $11.2\pm0.3$ & $1276\pm34$ & $-676\pm12$ & $233.1\pm5.1$ &                             & $21.4\pm2.6$ \\   
  (H-like, 1s -- 2p)    & 2013 (fix) & \ldots            & \ldots       & \ldots      & \ldots      & $79.6\pm5.6$  & $192.4\pm13.5\;(2.8\sigma)$ & $14.5\pm5.5\;(1.1\sigma)$ \\
                        & 2013       & $6.1680\pm0.0009$ & $9\pm1$      & $976\pm139$ & $-677\pm45$ & $68.4\pm7.4$  & $165.3\pm17.9\;(3.6\sigma)$ & $14.5\pm5.5\;(1.1\sigma)$
\\[2.8ex]
                         
Si \textsc{xiii} (6.648) & 2001       & $6.6340\pm0.0002$ & $7.7\pm0.3$ & $816\pm32$ & $-633\pm8$ & $161.6\pm4.3$ &                            & $13.5\pm2.5$  \\
  (He-like, 1s -- 2p)    & 2013 (fix) & \ldots            & \ldots      & \ldots     & \ldots     & $63.8\pm3.6$  & $164.2\pm9.3\;(0.3\sigma)$ & $11.6\pm4.9\;(0.3\sigma)$
\\[2.8ex]

Si \textsc{xiii} (5.681) & 2001       & $5.668\pm0.001$ & $8\pm1$ & $973\pm129$ & $-666\pm60$ & $67.0\pm8.1$ &                             & $5.9\pm2.5$  \\
  (He-like, 1s -- 3p)    & 2013 (fix) & \ldots          & \ldots  & \ldots      & \ldots      & $46.5\pm8.8$ & $104.5\pm19.7\;(1.8\sigma)$ & $5.4\pm5.1\;(0.1\sigma)$
\\[2.8ex]

Si \textsc{xi} (6.778) & 2001       & $6.7678\pm0.0009$ & $4.9\pm0.5$ & $515\pm54$ & $-451\pm39$ & $46.1\pm4.1$ &                            & $4.8\pm1.4$ \\ 
  (Be-like, 1s -- 2p)  & 2013 (fix) & \ldots            & \ldots      & \ldots     & \ldots      & $13.0\pm4.0$ & $34.0\pm10.5\;(1.1\sigma)$ & $3.1\pm2.9\;(0.5\sigma)$
\\[2.8ex]

Si \textsc{x} (6.854,6.864) & 2001       & $6.842\pm0.0003$ & $6.4\pm0.6$ & $662\pm66$  & $-508\pm14$ & $80.0\pm4.2$ &                            & $8.4\pm2.2$ \\
  (B-like, 1s -- 2p)        & 2013 (fix) & \ldots           & \ldots      & \ldots      & \ldots      & $38.4\pm3.7$ & $101.8\pm9.9\;(2.0\sigma)$ & $11.3\pm4.0\;(0.6\sigma)$ \\
                            & 2013       & $6.837\pm0.001$  & $8\pm1$     & $802\pm135$ & $-758\pm61$ & $42.3\pm6.3$ & $111.9\pm16.6\;(1.9\sigma)$ & $11.3\pm4.0\;(0.6\sigma)$
\\[2.8ex]

Si \textsc{ix} (6.923,6.939) & 2001       & $6.912\pm0.001$ & $10\pm1$ & $1058\pm104$ & $-497\pm43$  & $63.2\pm5.3$ &                             & $5.6\pm3.5$  \\
  (C-like, 1s -- 2p)         & 2013 (fix) & \ldots          & \ldots   & \ldots       & \ldots       & $44.3\pm5.6$ & $118.4\pm15.0\;(3.5\sigma)$ & $12.5\pm5.7\;(1.0\sigma)$ \\
                             & 2013       & $6.916\pm0.003$ & $15\pm3$ & $1547\pm283$ & $-302\pm125$ & $57.6\pm9.0$ & $153.9\pm24.1\;(3.7\sigma)$ & $15.2\pm8.1\;(1.1\sigma)$
\\[2.8ex]

Al \textsc{xiii} (6.053) & 2001       & $6.029\pm0.003$ & $13\pm3$ & $1574\pm354$ & $-1197\pm144$ & $43.8\pm8.9$  &                            & $3.0\pm3.6$ \\
  (H-like 1s -- 3p)      & 2013 (fix) & \ldots          & \ldots   & \ldots       & \ldots        & $12.1\pm10.6$ & $28.8\pm25.1\;(0.6\sigma)$ & $1.5\pm2.5\;(0.3\sigma)$
\\[2.8ex]

Al \textsc{xii} (7.757) & 2001       & $7.7431\pm0.0007$ & $7\pm1$ & $608\pm95$ & $-537\pm29$ & $42.3\pm4.3$ &                            & $5.2\pm1.9$  \\
  (He-like 1s -- 2p)    & 2013 (fix) & \ldots            & \ldots  & \ldots     & \ldots      & $17.7\pm4.1$ & $52.5\pm12.1\;(0.8\sigma)$ & $5.7\pm3.7\;(0.1\sigma)$
\\[2.8ex]

Mg \textsc{xii} (8.421) & 2001       & $8.4020\pm0.0004$ & $13.4\pm0.4$ & $1128\pm31$ & $-675\pm12$ & $217.1\pm5.3$ &                             & $28.6\pm2.6$  \\
  (H-like, 1s -- 2p)    & 2013 (fix) & \ldots            & \ldots       & \ldots      & \ldots      & $55.9\pm3.1$  & $194.8\pm10.9\;(1.8\sigma)$ & $10.2\pm11.3\;(1.6\sigma)$ \\
                        & 2013       & $8.404\pm0.001$   & $8\pm1$      & $662\pm83$  & $-603\pm40$ & $36.7\pm4.3$  & $128.0\pm15.0\;(5.6\sigma)$ & $14.1\pm6.8\;(2.0\sigma)$
\\[2.8ex]

Mg \textsc{xii} (7.106) & 2001       & $7.0921\pm0.0004$ & $6.0\pm0.7$ & $632\pm69$ & $-589\pm18$ & $69.5\pm4.2$ &                            & $7.9\pm1.8$ \\
  (H-like, 1s -- 3p)    & 2013 (fix) & \ldots            & \ldots      & \ldots     & \ldots      & $24.7\pm3.9$ & $67.4\pm10.6\;(0.2\sigma)$ & $6.9\pm3.5\;(0.3\sigma)$
\\[2.8ex]

Mg \textsc{xi} (9.169) & 2001       & $9.1543\pm0.0006$ & $18.9\pm0.6$ & $1454\pm44$ & $-480\pm18$ & $241.3\pm6.5$ &                             & $35.3\pm3.1$ \\
  (He-like, 1s -- 2p)  & 2013 (fix) & \ldots            & \ldots       & \ldots      & \ldots      & $41.2\pm6.3$  & $159.1\pm24.2\;(3.3\sigma)$ & $16.2\pm13.7\;(1.4\sigma)$ \\
                       & 2013       & $9.149\pm0.003$   & $12\pm3$     & $953\pm222$ & $-654\pm82$ & $34.6\pm6.4$  & $133.3\pm24.6\;(4.2\sigma)$ & $12.7\pm10.1\;(2.1\sigma)$
\\[2.8ex]

Mg \textsc{xi} (7.851) & 2001       & $7.8361\pm0.0004$ & $7.6\pm0.5$ & $685\pm41$ & $-569\pm16$ & $87.6\pm4.2$ &                           & $9.7\pm2.2$  \\
  (He-like, 1s -- 3p)  & 2013 (fix) & \ldots            & \ldots      & \ldots     & \ldots      & $29.4\pm3.7$ & $87.6\pm11.0$ (No change) & $9.0\pm4.4\;(0.1\sigma)$
\\[2.8ex]

Mg \textsc{xi} (7.473) & 2001       & $7.4595\pm0.0006$ & $6.6\pm0.6$ & $622\pm57$ & $-543\pm22$ & $62.1\pm4.2$ &                            & $7.7\pm2.3$   \\
  (He-like, 1s -- 4p)  & 2013 (fix) & \ldots            & \ldots      & \ldots     & \ldots      & $21.0\pm4.1$ & $60.0\pm11.6\;(0.2\sigma)$ & $6.3\pm4.5\;(0.3\sigma)$
\\[2.8ex]

Ne \textsc{x} (9.708) & 2001       & $9.6857\pm0.0005$ & $20.9\pm0.8$ & $1523\pm57$ & $-690\pm16$ & $290.0\pm8.5$ &                             & $41.9\pm2.1$ \\
  (H-like, 1s -- 4p)  & 2013 (fix) & \ldots            & \ldots       & \ldots      & \ldots      & $43.0\pm7.4$  & $177.6\pm30.5\;(3.5\sigma)$ & $25.7\pm10.4\;(1.5\sigma)$
\\[2.8ex]

Ne \textsc{x} (10.239) & 2001       & $10.2185\pm0.0004$ & $11.2\pm0.5$ & $770\pm37$ & $-599\pm11$ & $184.9\pm6.2$ &                             & $22.5\pm2.9$ \\
  (H-like, 1s -- 3p)   & 2013 (fix) & \ldots             & \ldots       & \ldots     & \ldots      & $29.9\pm5.6$  & $131.8\pm24.5\;(2.1\sigma)$ & $18.7\pm11.5\;(0.3\sigma)$
\\[2.0ex]
      \hline
    \end{tabular}
    \begin{center}
    \textbf{Notes.} In the `2013 (fix)' model, the Gaussian shape is
    fixed to that used for the 2001 data, leaving only the line flux
    free to vary.  For the 2013 data we quote both a physical line
    flux calculated directly from the data, and also a value that has
    been scaled using the ratio of the continuum fluxes in both
    epochs, to enable a direct comparison.  The significance of the
    flux and EW changes between epochs for each line are listed in
    parentheses.
    \end{center}
\end{table*}

Of the 19 absorption lines we consider, 5 have a line flux which
appears to vary significantly ($>3\sigma$) between 2001 and 2013.  For
3 of these, the absorption lines appear to be weaker in the 2013 data
while maintaining a similar profile shape.  These are the 1s -- 2p
transition of H-like Si~\textsc{xiv} at a rest-frame wavelength of
6.182\,\ang, the 1s -- 2p transition of He-like Mg~\textsc{xi} at
$\lambda\sb{\rm rest}=9.169\,\ang$, and the 1s -- 4p transition of
H-like Ne~\textsc{x} at $\lambda\sb{\rm rest}=9.708\,\ang$.  However,
in the original modelling of these lines, the level of the continuum
was fixed to its best-fitting value.  Therefore the errors on the line
parameters listed in Table~\ref{table:lines} are statistical errors
only, and do not include a contribution from the systematic error
resulting from the uncertainty in the continuum placement.  When the
continuum level is allowed to vary within its $1\sigma$ error bounds,
the measured differences in the line flux between the two epochs are
no longer significant for each of the 3 lines ($\le1.6\sigma$).

\begin{figure}[h]
  \centering 
    \includegraphics[width=0.48\textwidth]{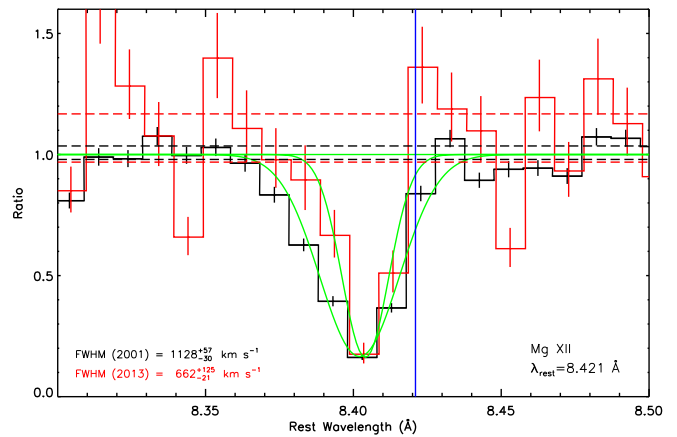}
  \caption{The Gaussian fitting of the Mg~\textsc{xii} absorption line
    at $\lambda\sb{\rm rest}=8.421$\,\ang.  The 2001 data are shown in
    black and the 2013 data are shown in red.  A small artificial $x$
    offset of +0.0005\,\ang\, has been added to the 2013 data to
    improve the clarity of the figure by reducing symbol overlap.  The
    vertical blue line indicates $\lambda\sb{\rm rest}$ from which the
    absorption line is noticeably blueshifted.  The best-fitting
    Gaussian profiles are shown in green and the parameters of the
    Gaussian fit to the 2013 data were left free to vary.  The shape
    of the line profile is significantly narrower in the 2013
    spectrum, even after accounting for the error on the continuum
    placement.  The $1\sigma$ error bounds on the continuum level are
    indicated with the dashed black and red lines.  The FWHM values
    quoted on the figure also include the systematic error associated
    with the continuum placement and therefore differ from those in
    Table~\ref{table:lines}, in which we list only the statistical
    errors.}
  \label{fig:ind_line}
\end{figure}

The remaining two of the 19 lines show a change in their profile
shape.  The C-like, Si\,\textsc{ix} doublet at $\lambda\sb{\rm
  rest}=6.923, 6.939\,\ang$ appears as a single line in the 2001
spectrum, but as two separate lines in the 2013 spectrum.  This
prevents a direct comparison between the epochs, and since we only
model the data with a single Gaussian, an apparent increase in the
line flux is obtained.  In the 2001 spectrum, the H-like,
Mg\,\textsc{xii} absorption line at $\lambda\sb{\rm rest}=8.421\,\ang$
has a broad Gaussian profile with a FWHM of $1128\pm31\,\kms$, but in
the 2013 spectrum the line is significantly narrower ($5\sigma$) with
${\rm FWHM}=662\pm83\,\kms$.  This results in a significant change in
the line flux, despite the deepest point of the line being consistent
between epochs, in both position and depth.  We allowed the continuum
level to vary in additional Gaussian fits to this line, since the
shape of the profile can also vary when a different continuum level is
used.  The variation of the FWHM is still significant between epochs
($3.6\sigma$), however, we note that this line is likely not resolved
in the 2013 spectrum.  We show an enlarged portion of the data in
Figure~\ref{fig:ind_line} including the Gaussian fitting to this line
as an example.

In addition to comparing line fluxes derived from the Gaussian
modelling, we also compare the EWs of the lines.  For consistency with
the previous work of K02 we derive this directly from the residuals,
rather than the Gaussian fit.  The EW is given by
\begin{equation}
  {\rm EW} = \sum_{i}\left(1-\frac{F_{i}}{F_{C}}\right)B_{i},
  \label{eqn:ew}
\end{equation}
where $F_{i}$ is the flux in the $i$th bin, $F_{C}$ is the continuum
level at that bin, and $B_{i}$ is the bin width (0.01\,\ang).  The
limits of summation are taken as the points at which the Gaussian
profile becomes indistinguishable from the continuum.  The error is
estimated by
\begin{equation}
  \Delta {\rm EW} = \sqrt{\left[\frac{\Delta F_{C}}{F_{C}}\sum_{i}\left(\frac{B_{i}F_{i}}{F_{C}}\right)\right]^{2}+\sum_{i}\left(\frac{B_{i}\Delta F_{i}}{F_{C}}\right)^{2}},
  \label{eqn:ew_err}
\end{equation}
where $\Delta F\sb{i}$ is the error on the flux in the $i$th bin and
$\Delta F\sb{\rm C}$ is the error on the continuum level, estimated
from the distribution of residuals about the best-fitting continuum
level.

Unlike the original Gaussian modelling used to estimate the line
fluxes, the EW determination directly includes an estimate of the
error on the continuum.  Using this criterion, none of the 19
individual absorption lines we consider show a significant change in
EW between the two epochs.  The results obtained from comparing the EW
between epochs, and looking for changes in line flux, are therefore in
good agreement.  The significances of each change in EW, all of which
are $\le 2.1\sigma$, are listed in Table~\ref{table:lines}.

As none of the 19 absorption lines we investigate shows a significant
variation between epochs considering both their line flux and EW,
dramatic changes in the physical properties of the absorbing material
are unlikely.  However, our ability to detect significantly a
variation in the line depends on the strength of the line itself
(i.e.,~we might not significantly detect a change in a line that is
weaker).  We note that the 19 lines we consider are the strongest in
the data set, despite covering a range of EW ($\sim 40\,{\rm m}\ang$
to $\sim3\,{\rm m}\ang$), and therefore these lines represent the best
chance of detecting variations.  We estimate that for the
Si~\textsc{xiv} line ($\lambda\sb{\rm rest}=6.182$\,\ang), for
example, we would be able to detect changes in the line flux of
$\gtrsim 20\%$ and $\gtrsim 50\%$ at $>1\sigma$ and $>3\sigma$
significance, respectively; continuum placement uncertainty largely
sets these lower bounds.  However, photoionization modelling reported
in \citet{netzer03} suggests that some of the strongest absorption
lines may be saturated, and are therefore less sensitive to any
changes in the column density.  They include Si~\textsc{xiv}
(6.182\,\ang), Si~\textsc{xiii} (6.648\,\ang), Si~\textsc{xiii}
(5.681\,\ang), and S~\textsc{xvi} (4.729\,\ang), although
S~\textsc{xv} (5.039\,\ang), Si~\textsc{xi} (6.778\,\ang),
Si~\textsc{x} (6.854\,\ang), and Si~\textsc{ix} (6.939\,\ang) are
thought to be unsaturated.  Given the apparent lack of variation in
the absorbing material, detailed photoionization modeling is not
required.  We find that the best-fitting model for the 2001 spectrum
determined by \citet{netzer03}, which includes the effects of three
ionized absorbers, each split into two kinematic components, provides
a good fit to the 2013 data, further indicating the constancy of the
X-ray warm absorber.

With the exception of the Fe \ka emission line which is considered in
detail in \S\ref{section:iron}, we do not investigate the properties
of emission lines, many of which have $\lambda\sb{\rm
  rest}>11\,$\ang\, where the S/N of the 2013 spectrum drops
considerably.  However, these emission lines are visible in
Figure~\ref{fig:full_x_spec}, and their contribution may be higher in
the 2013 spectrum.  This is due to the lower continuum flux level
resulting in an increase in the relative contribution from the
extended X-ray emission of the NLR.  This increased emission could
also cause apparent changes in the profiles of some absorption lines,
and may explain the narrower shape of the Mg~\textsc{xii} line we see.
Figure~\ref{fig:lines_all} also shows a Ne emission line at
$\lambda\sb{\rm rest}=9.708$\,\ang\, (Ne~\textsc{x}), which appears to
be considerably stronger in the 2013 spectrum.  However this is mostly
an artifact of taking a ratio of the data and the additive, rather
than multiplicative nature of emission lines, as the line fluxes in
the two epochs are consistent at $1.6\sigma$ when the error on the
continuum placement is also considered.


\subsection{Kinematics}
\label{section:kinematics}
\subsubsection{X-ray kinematics}
\label{section:xray_kin}
\begin{figure*}
  \centering 
    \begin{tabular}{cc}
      \includegraphics[height=0.24\textwidth]{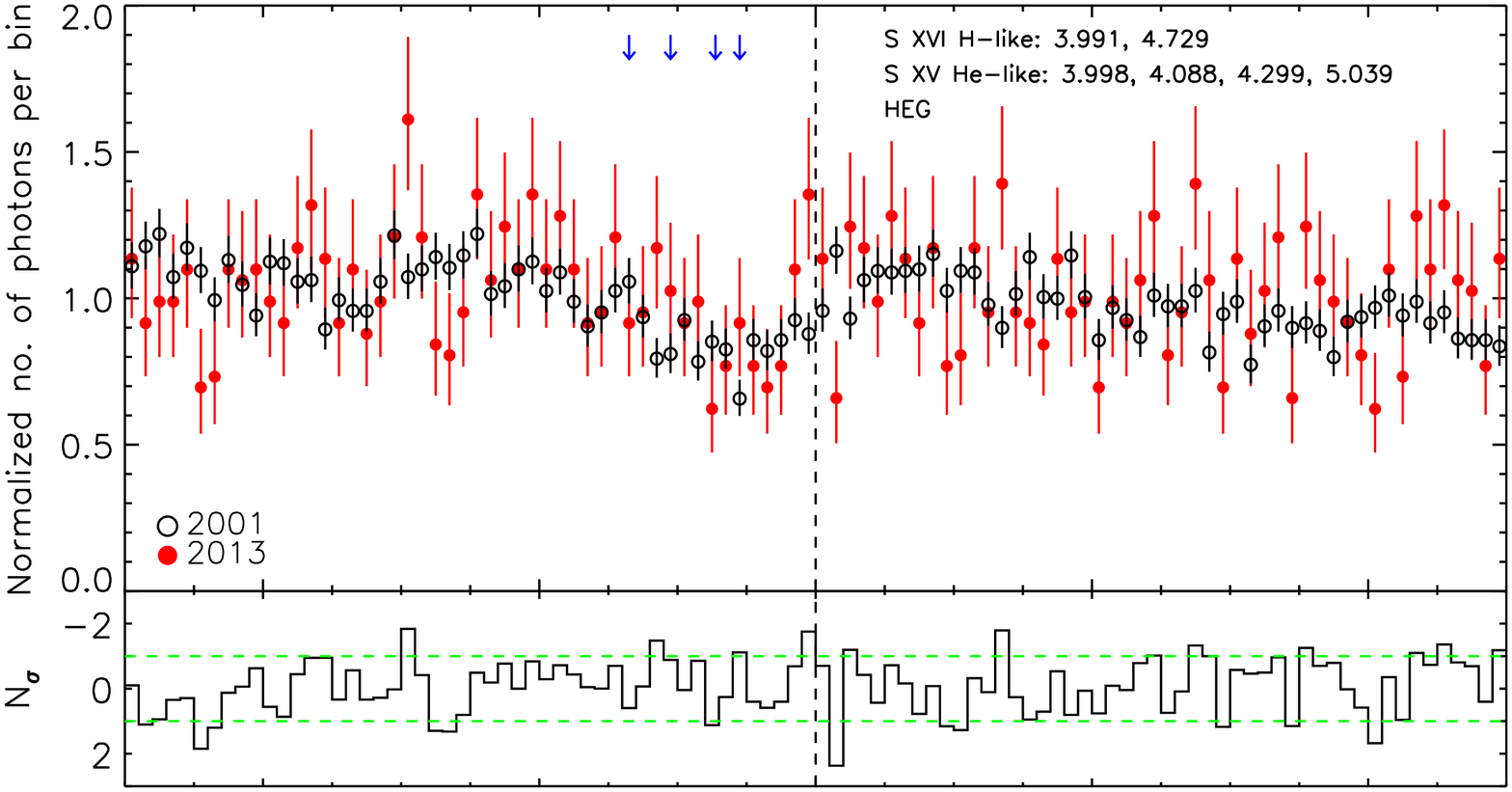} &
      \includegraphics[height=0.24\textwidth]{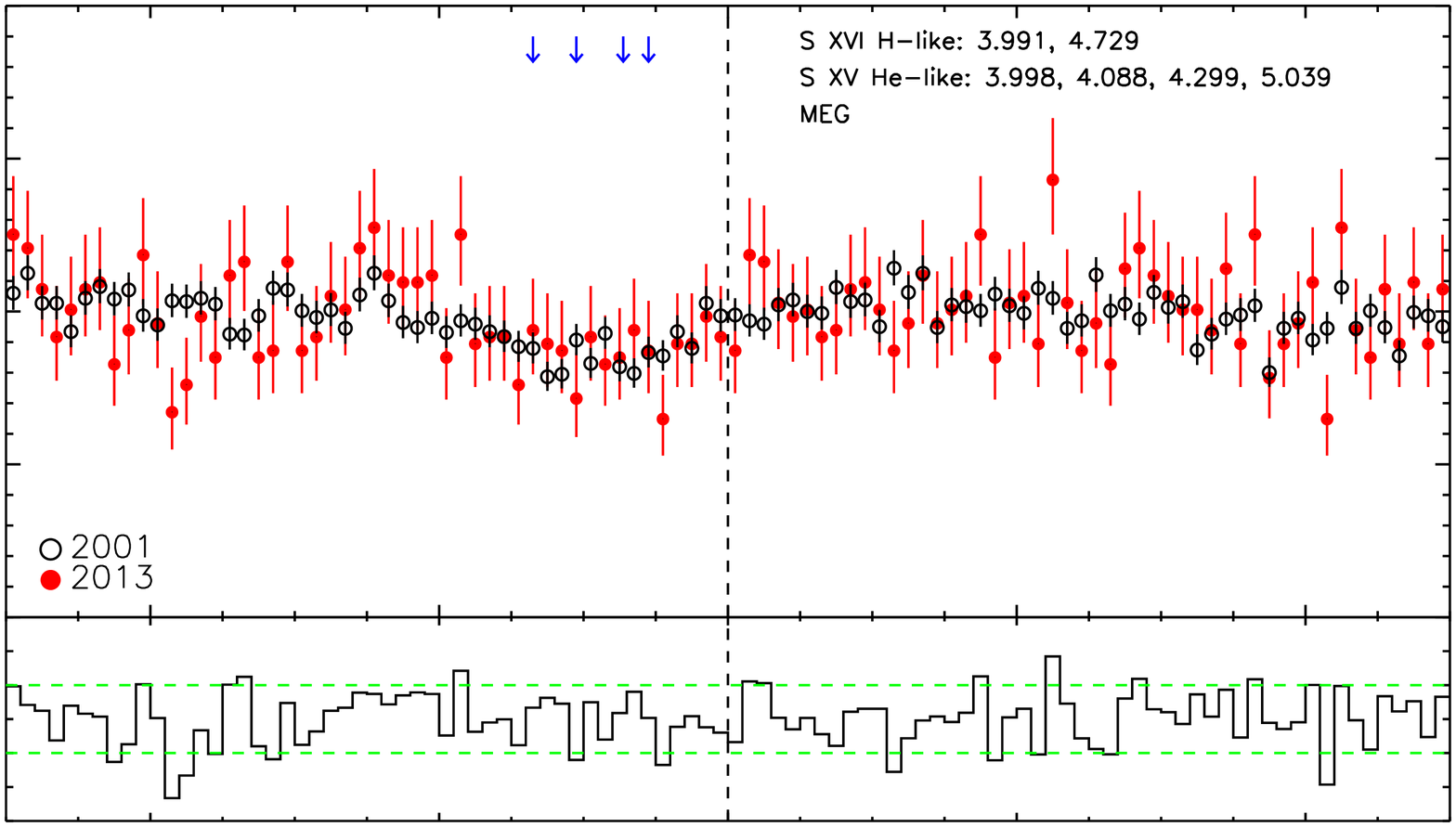} \\
      \includegraphics[height=0.24\textwidth]{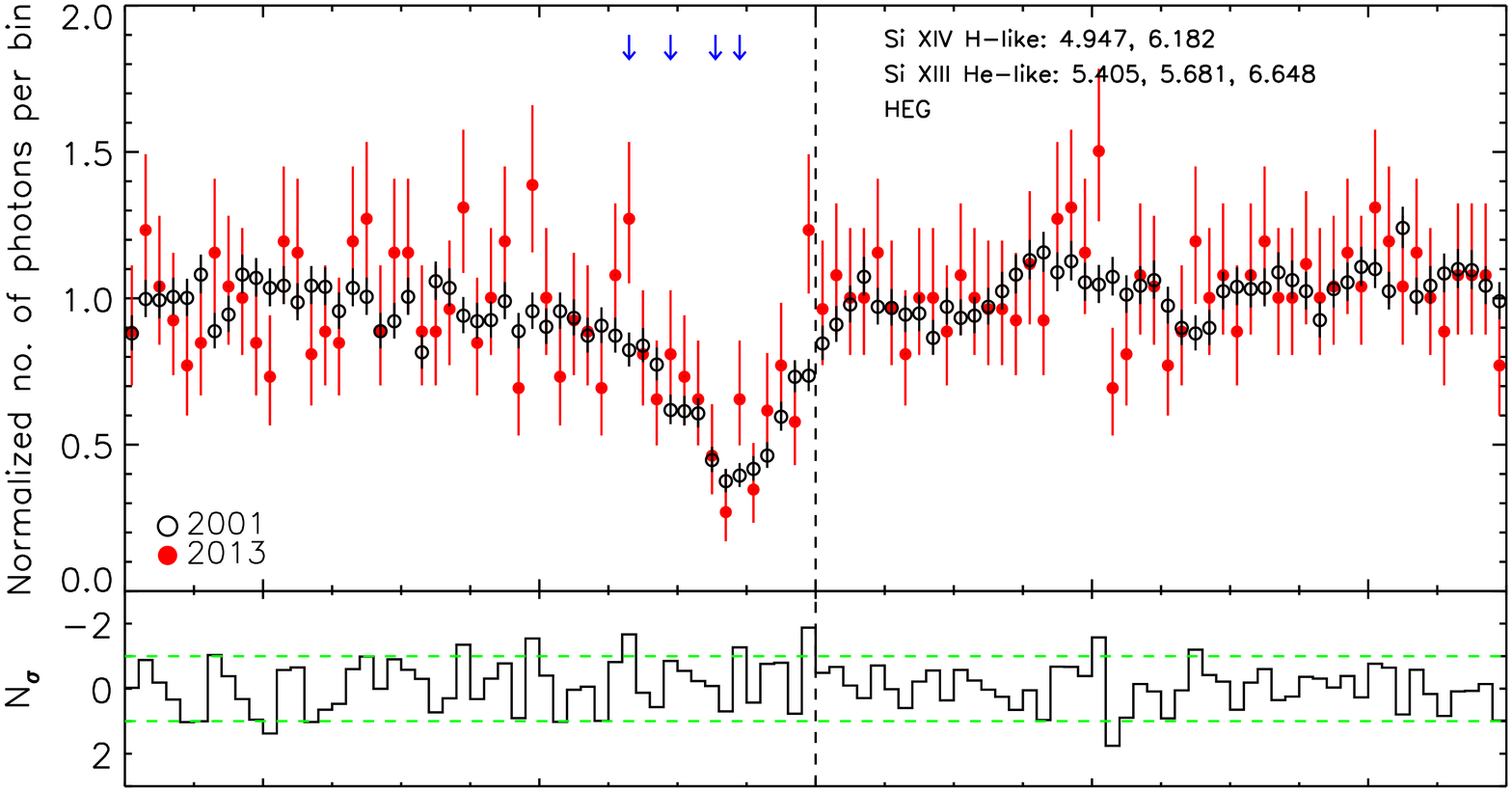} &
      \includegraphics[height=0.24\textwidth]{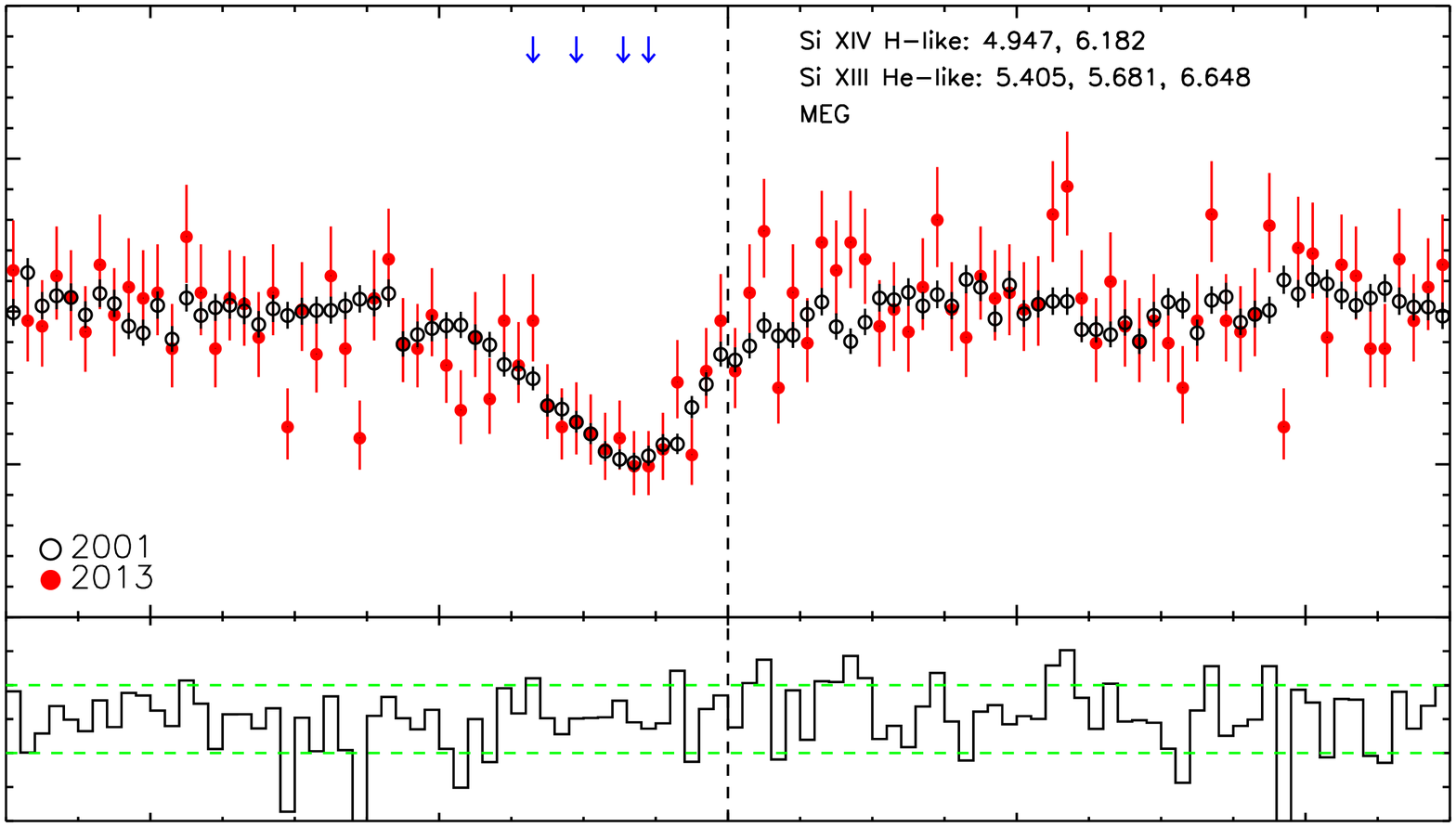} \\
      \includegraphics[height=0.24\textwidth]{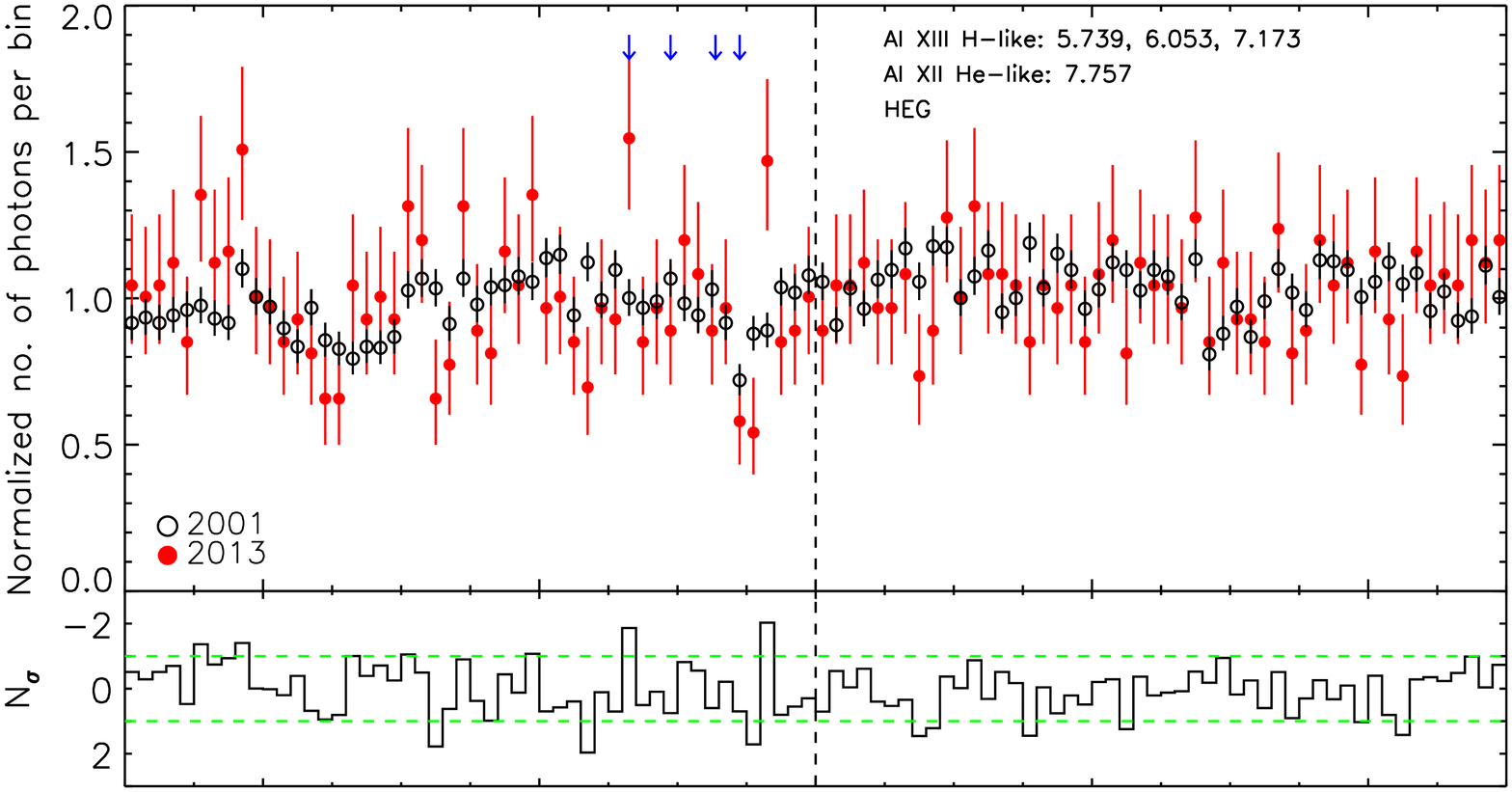} &
      \includegraphics[height=0.24\textwidth]{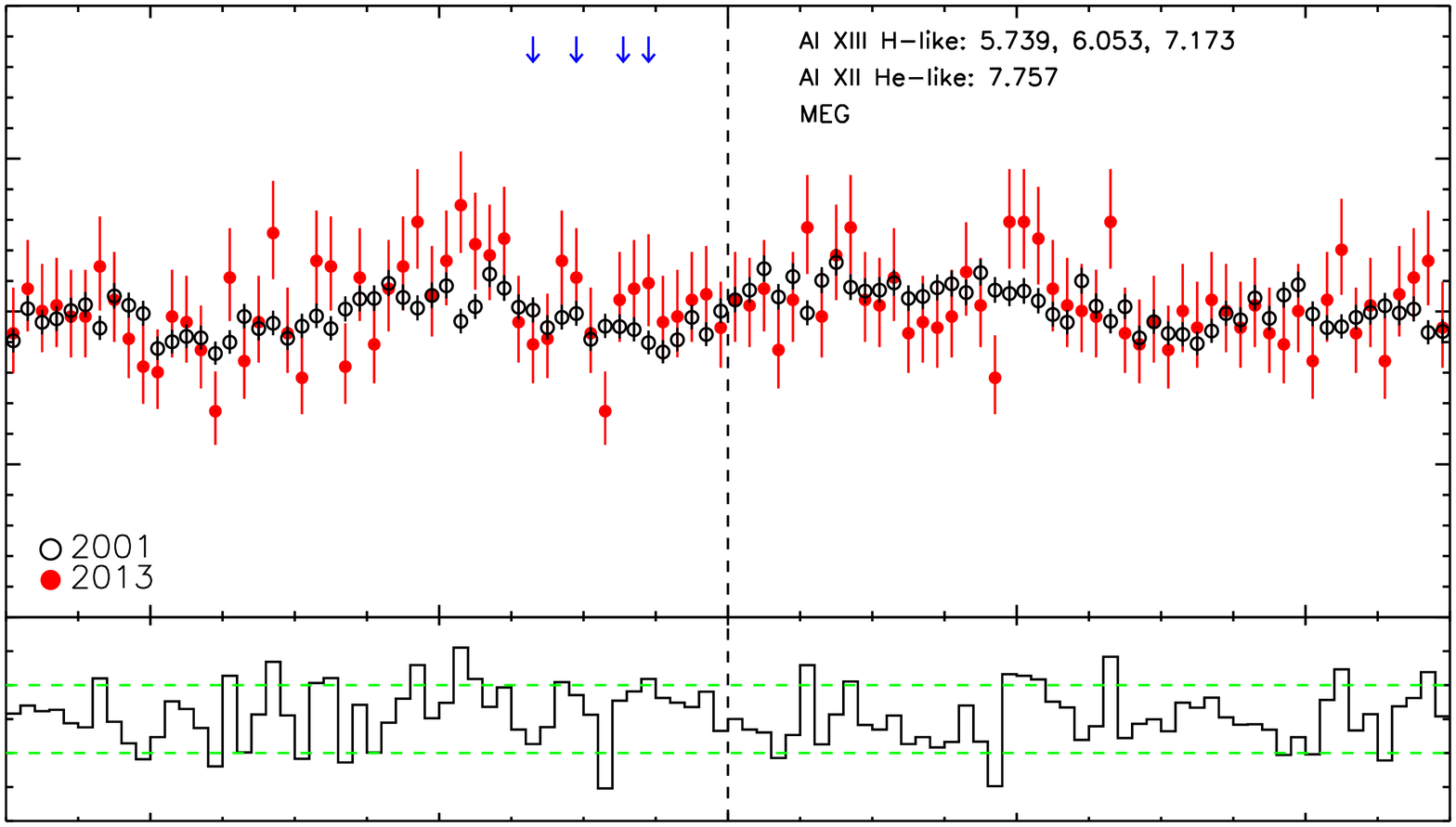} \\
      \includegraphics[height=0.24\textwidth]{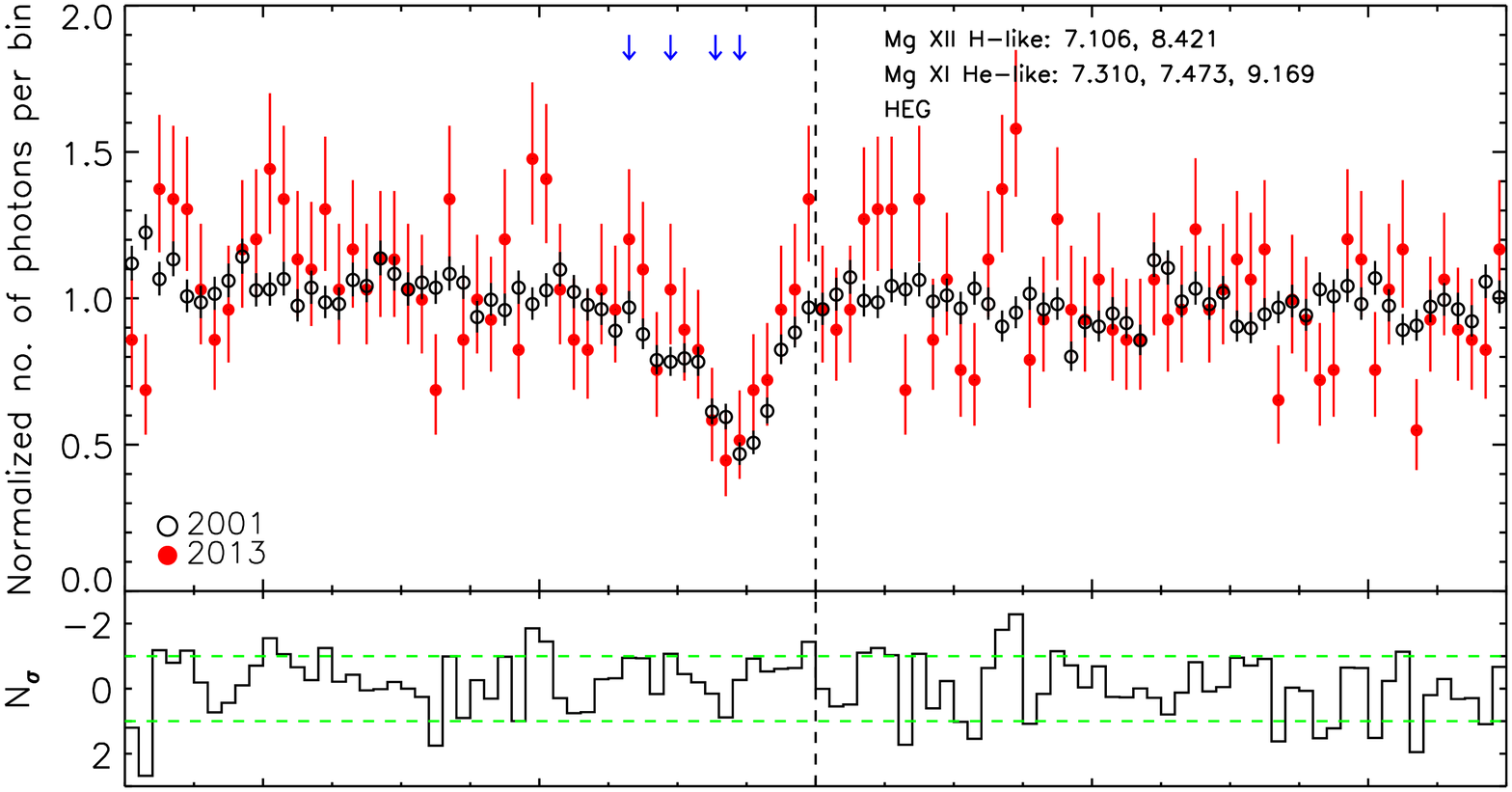} &
      \includegraphics[height=0.24\textwidth]{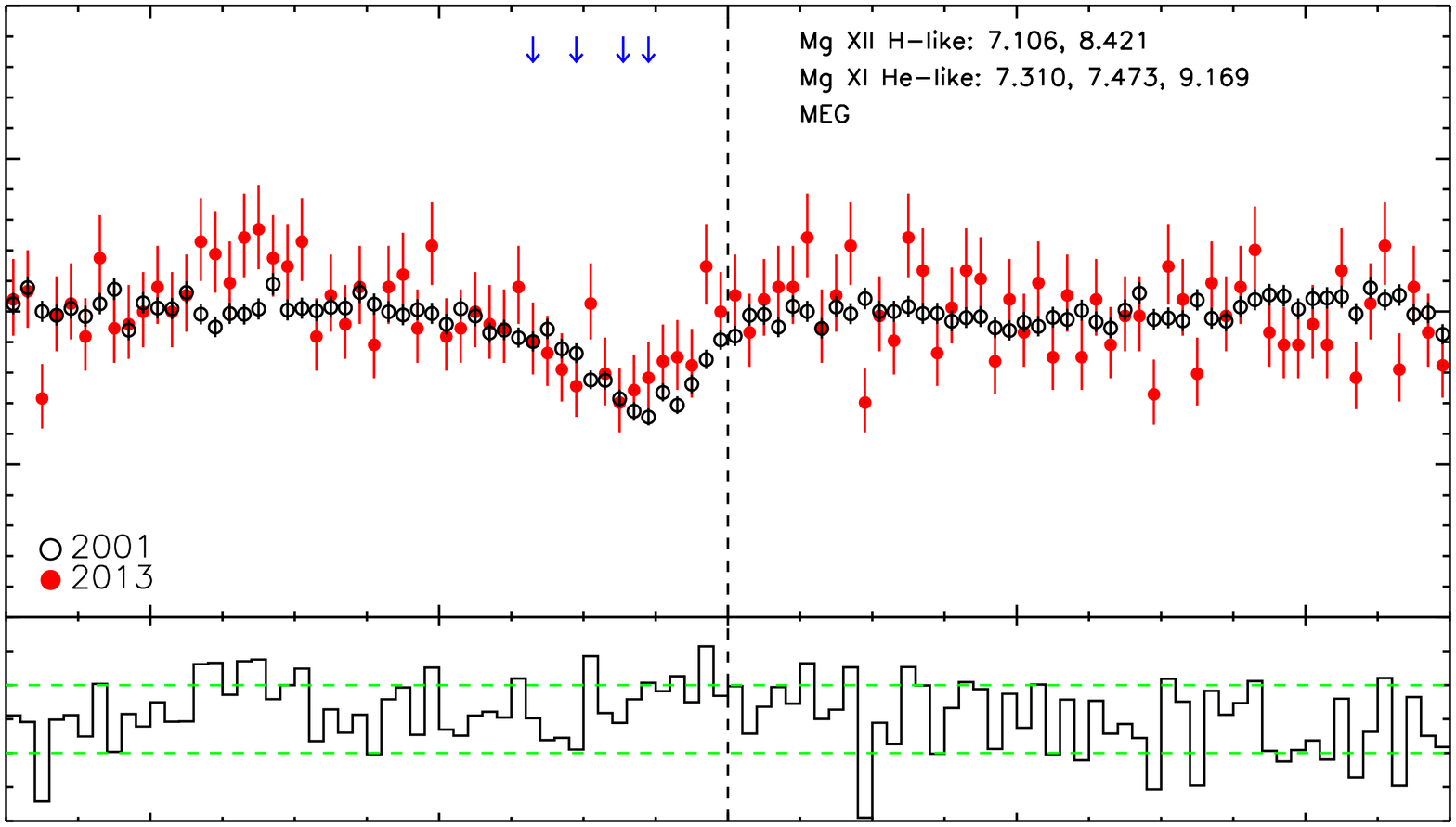} \\
      \includegraphics[height=0.272\textwidth]{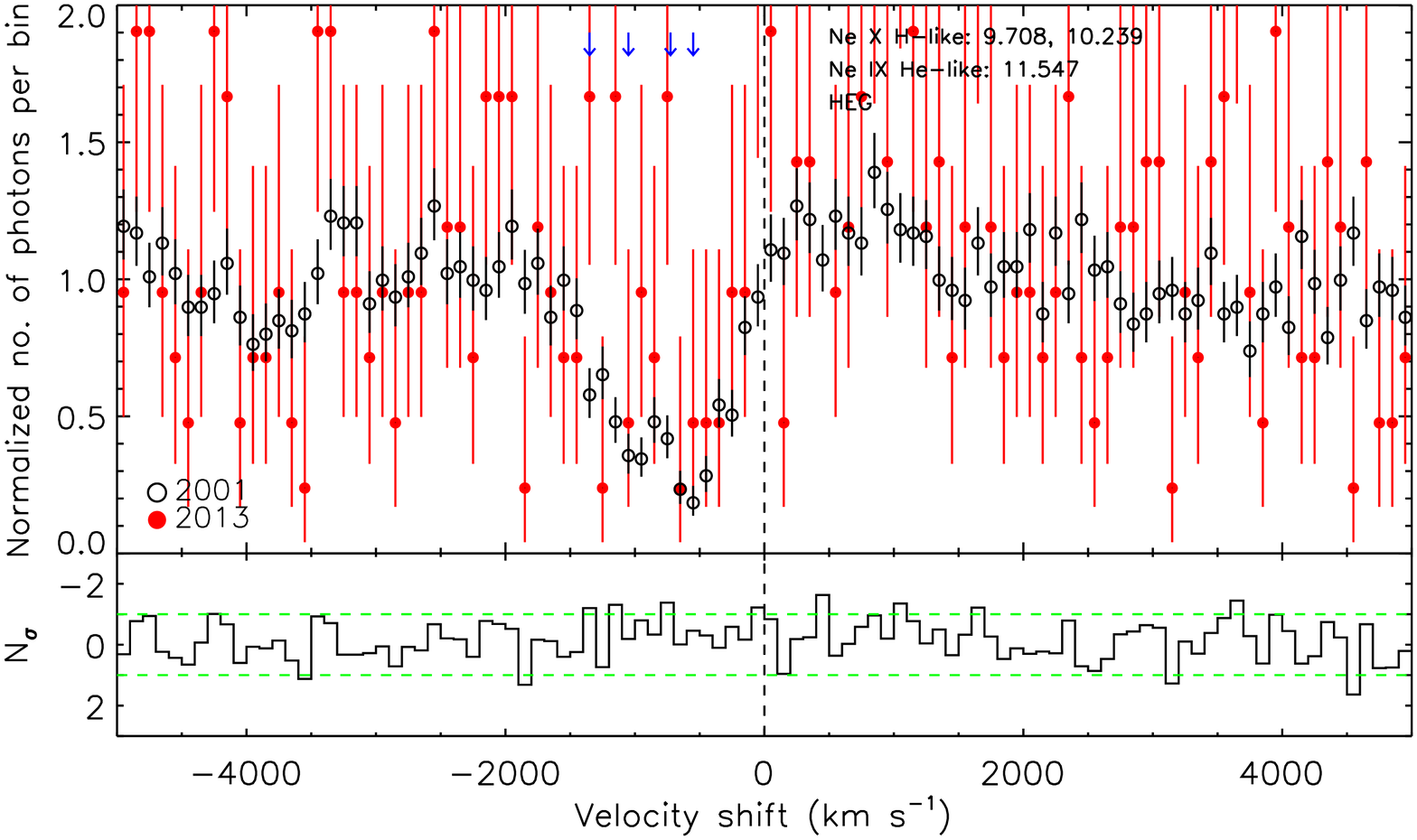} &
      \includegraphics[height=0.272\textwidth]{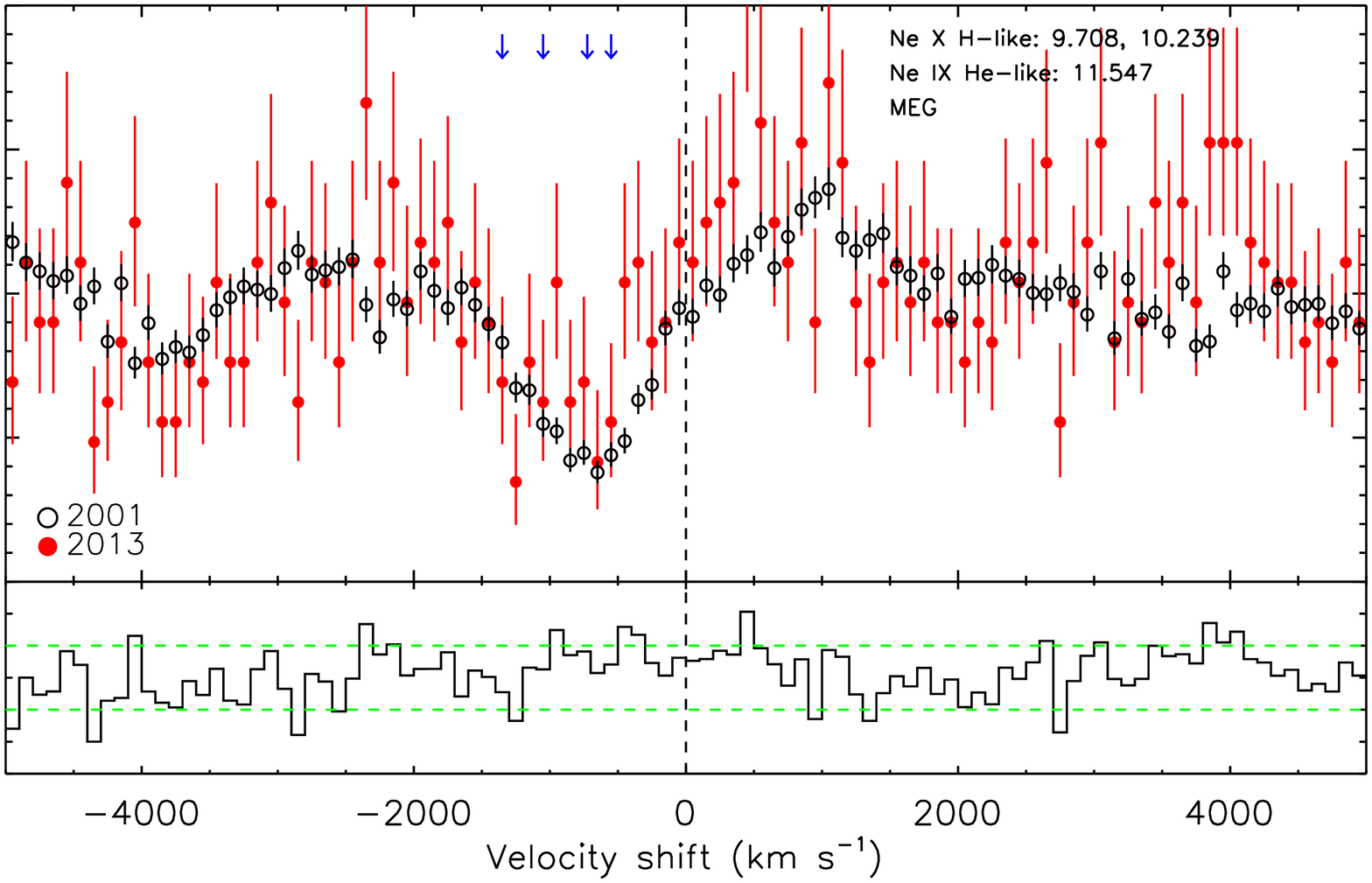} \\
    \end{tabular}
  \caption{X-ray velocity profiles.  The HEG (left) and MEG (right)
    velocity spectra are created by stacking absorption lines from H-like
    and He-like ions of the elements S, Si, Al, Mg, and Ne.  Data from
    the 2001 observations are shown by the black-open circles, and
    data from the 2013 observations are shown by the red-solid
    circles.  Error bars are computed from the number of counts
    included in each bin \citep{gehrels86}.  The bin size is
    $100\;\kms$.  The rest-frame wavelengths of the individual
    absorption lines included in the profile are listed on each figure.
    The blue arrows indicate the locations of the four kinematic
    components identified in the UV spectra, using the velocity shift
    values from G03a.}
  \label{fig:vel}
\end{figure*}

\begin{figure*}
  \centering 
    \begin{tabular}{cc}
      \includegraphics[height=0.26\textwidth]{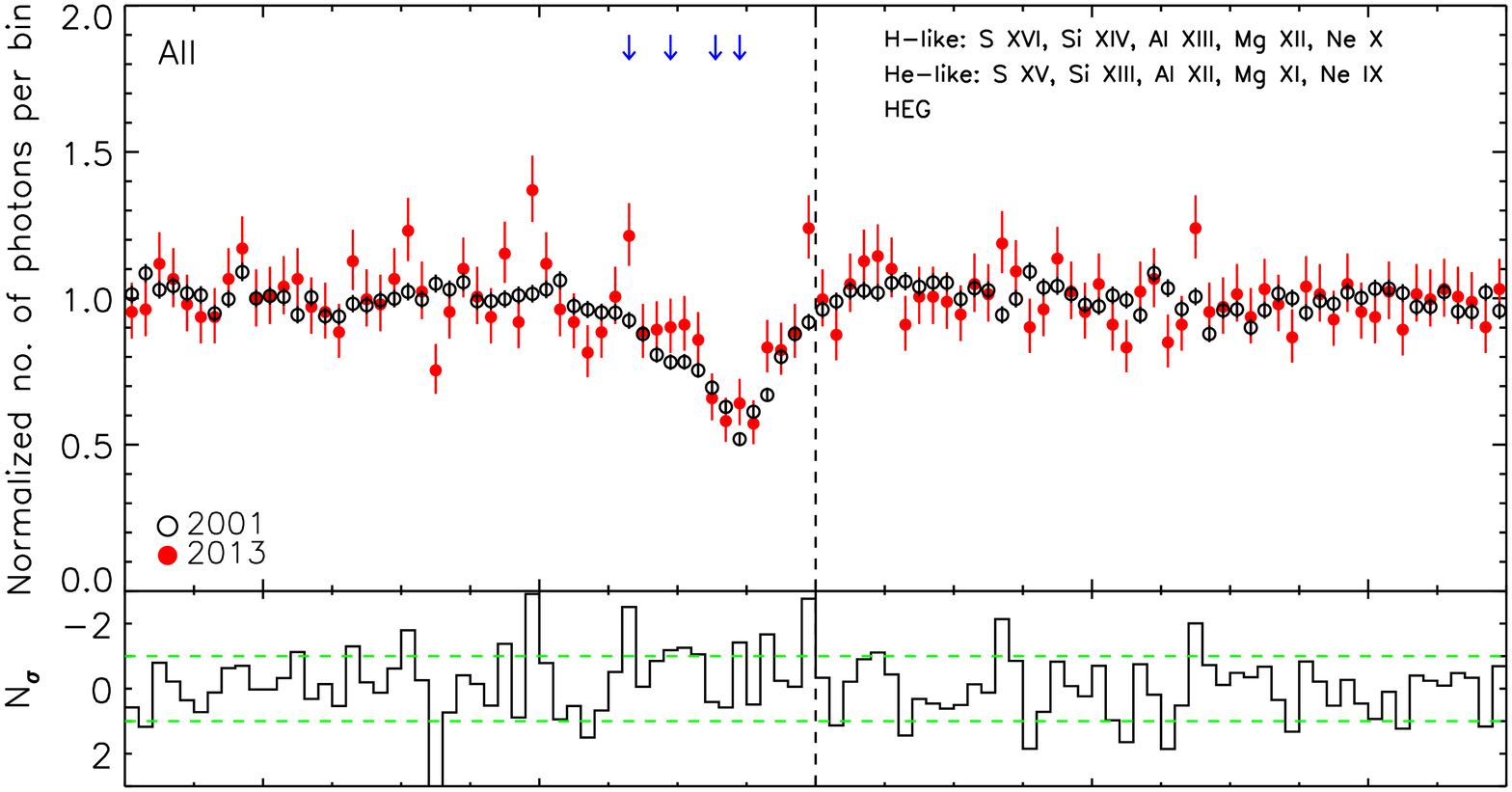} &
      \includegraphics[height=0.26\textwidth]{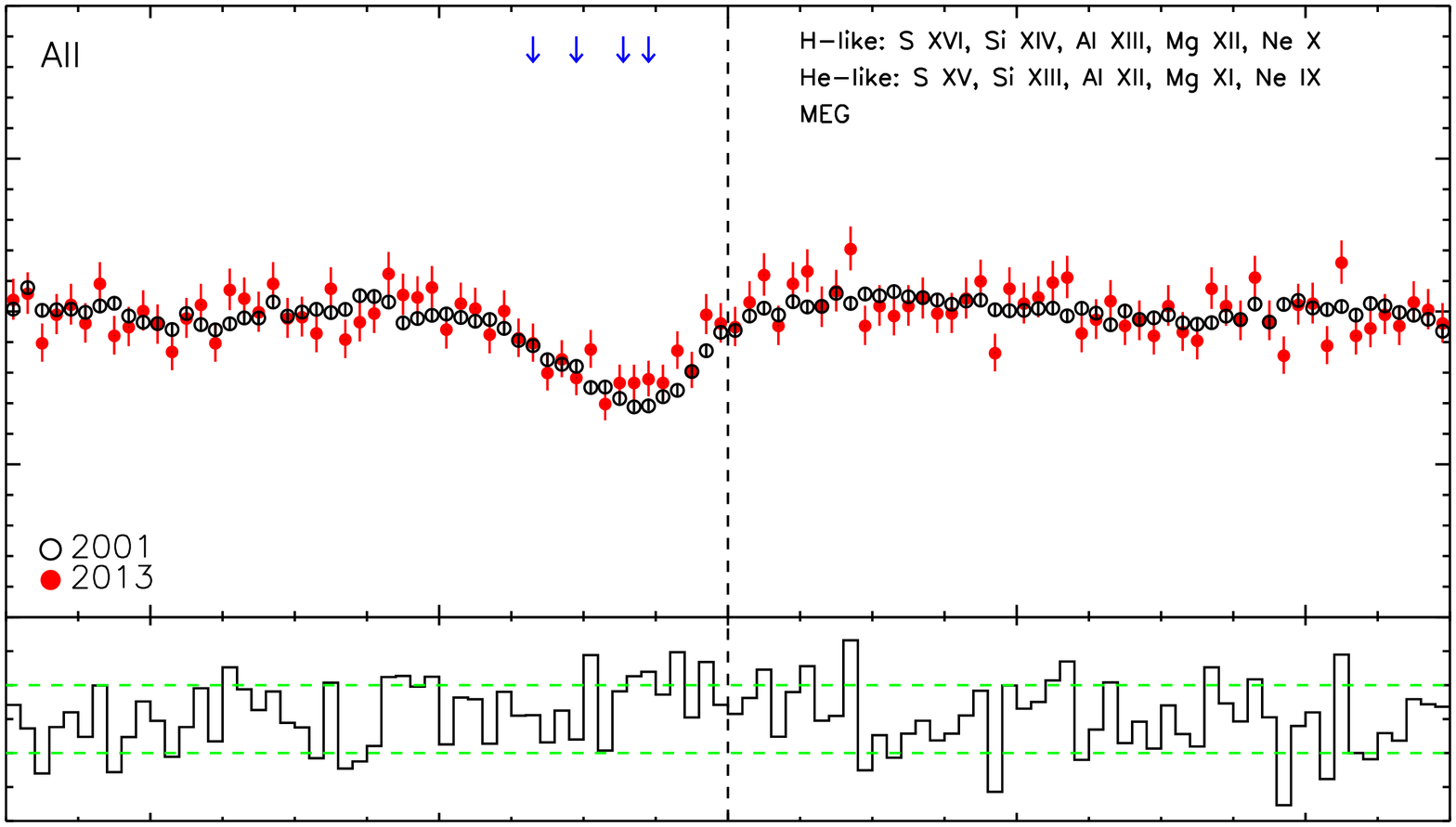} \\
      \includegraphics[height=0.294\textwidth]{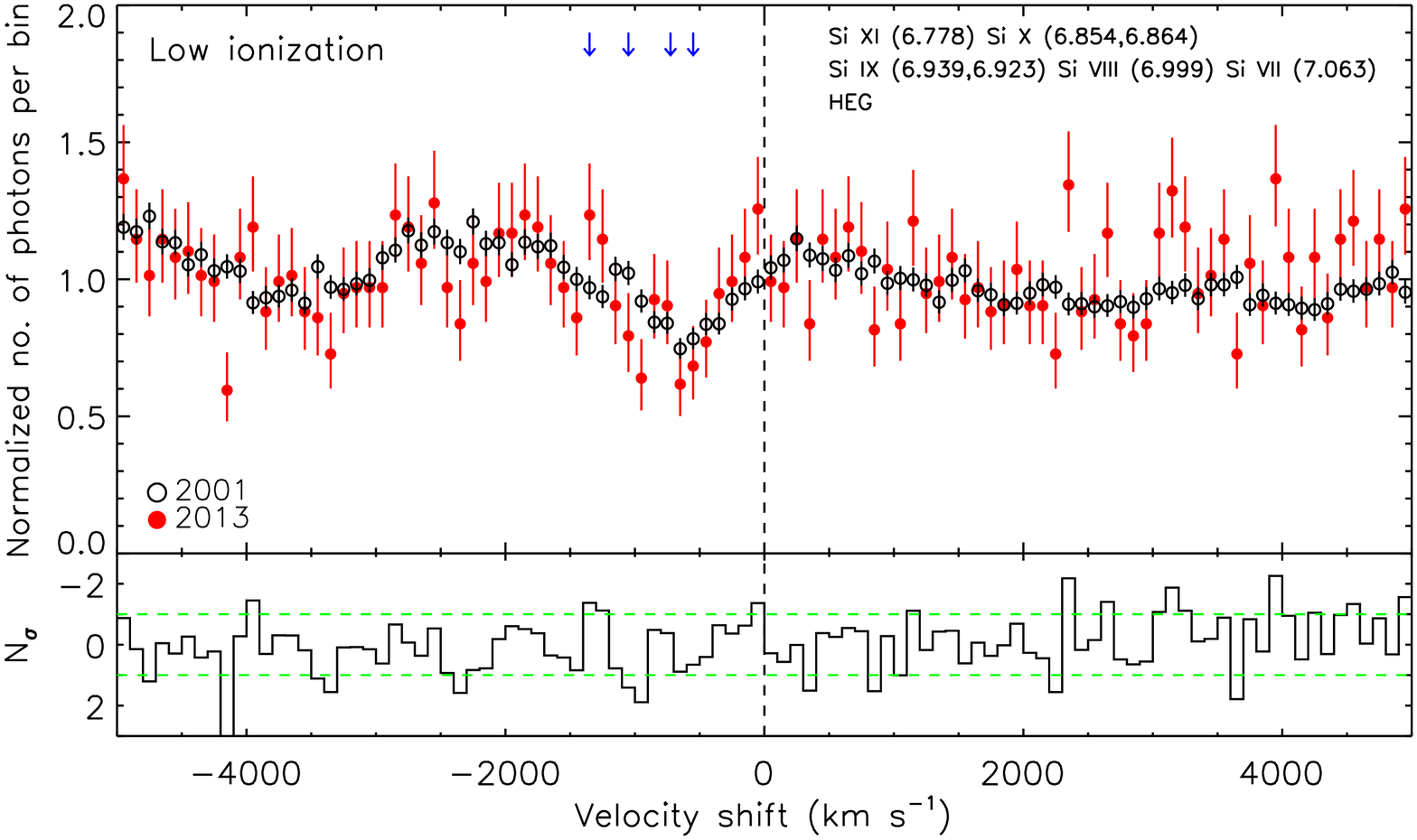} & 
      \includegraphics[height=0.294\textwidth]{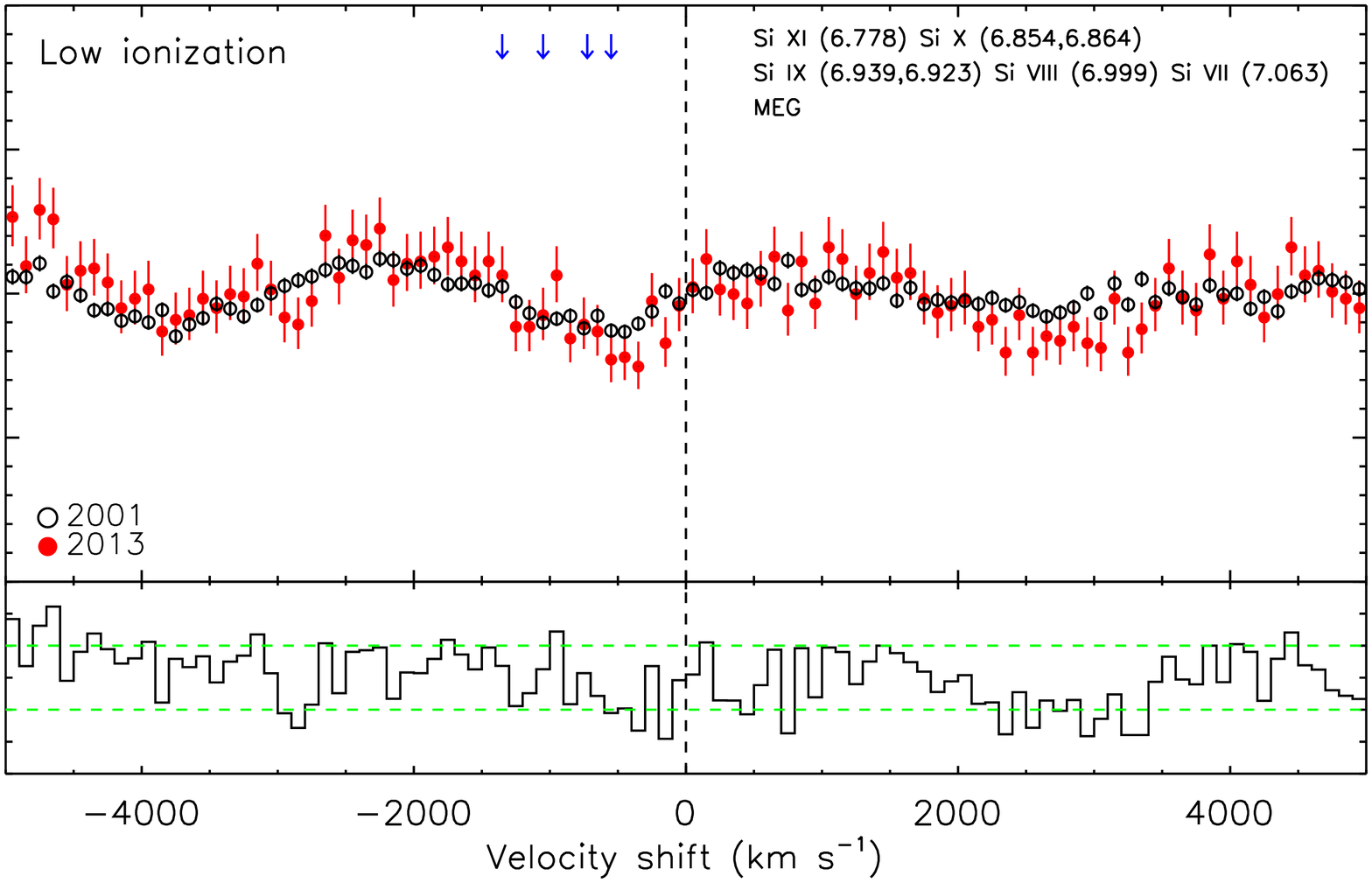} \\
    \end{tabular}
  \caption{X-ray velocity profiles.  Top - velocity spectra created by
    stacking all 23 absorption lines due to H- and He-like S, Si, Al,
    Mg, and Ne into the same profile.  Bottom - velocity spectra
    created from \textit{only} absorption lines due to lower
    ionization states of Si (Be-like, B-like, C-like, N-like, and
    O-like).  Colors and symbols are the same as in
    Figure~\ref{fig:vel}.  HEG profiles are shown on the left, MEG on
    the right.}
  \label{fig:vel2}
\end{figure*}

In this section we consider the stacking of multiple absorption lines
from the same element in order to improve the statistics, and we do this
in velocity space in order to probe the kinematics of the absorbing
material directly.  Our `velocity spectra' are built up on a
photon-by-photon basis using data directly from the event files rather
than the spectra already binned in wavelength.  Each event file is
filtered, keeping events registered in the HEG or MEG and $\pm1\sp{\rm
  st}$ orders.  Events from bad pixels are discarded and our results
are not affected by spectral lines falling into ACIS chip gaps.  The
velocity shift of each photon is determined with respect to the
expected wavelength of a number of different spectral lines.  We
consider each of the spectral lines listed by K02, excluding those
which may be blended with lines from a different element.  Lines from
H-like and He-like ions of the same element are combined to give a
larger number of photons in each bin.  While no correlation between
the ionization state and the blueshift of the lines was observed by
K02, lower blueshifts were measured for increasing transition orders
(e.g.,~1s--3p, 4p, 5p).  In our analysis we include lines from all
transition orders which may result in a wider combined absorption
profile.  The velocity profiles for S, Si, Al, Mg, and Ne are shown in
Figure~\ref{fig:vel}.  We do not show the profiles for Ca or Ar as
they have a small average number of photons per bin and do not show a
significant absorption profile.  The individual lines included in each
profile and their rest-frame wavelengths are listed on the figures.
We also create a velocity profile by stacking \textit{all} of the 23
H- and He-like absorption lines of S, Si, Al, Mg, and Ne resulting in
a 2013 HEG profile with $\sim 115$ counts per bin and a 2013 MEG profile 
with $\sim 250$ counts per bin.  These are shown in the top two panels of
Figure~\ref{fig:vel2}.  The left-hand panels in Figures~\ref{fig:vel}
and~\ref{fig:vel2} show profiles created from the HEG data, while the
MEG profiles are shown on the right.  The two grating arms are
considered separately due to their different instrumental resolutions
($\Delta\lambda\sb{\rm HEG}=0.012$\,\ang, $\Delta\lambda\sb{\rm
  MEG}=0.023$\,\ang).  The expected width of the profile created
simply by the instrumental resolution is calculated from\\

\begin{equation}
  \sigma\sb{\rm inst}\,({\rm km\,s^{-1}}) = \left(\frac{\Delta\lambda}{\lambda}\right) \left(\frac{1}{2\sqrt{2 {\rm ln} 2}}\right) \; c,
  \label{eqn:inst}
\end{equation}
where $\lambda$ is the shortest wavelength of any line included in the
stacking to give an indication of the worst resolution expected.  For
each profile, the open-black circles show data from 2001 and the
solid-red circles show data from 2013.  Error bars are determined from
the number of photons included in each velocity bin \citep{gehrels86},
which have widths of $100\;\kms$.  Each profile is divided by its own
`continuum level' (i.e., the typical number of photons in bins with
$v<-2000\;\kms$ and $v>0\;\kms$ which excludes the majority of the
absorption profile), such that the profiles from each epoch are
normalized to the same {\it y-}axis value of 1, allowing a direct
comparison.  The blue arrows indicate the velocity shifts of the four
UV absorption components using the values listed in G03a. We also plot
\nsig calculated from both the number of counts and the errors on each
bin.

\begin{table*}
\centering
\scriptsize
\caption{Gaussian fitting results for the X-ray velocity profiles}
    \begin{tabular}{lllccccccccccc}
      \hline \hline
      \multicolumn{1}{l}{ } &
      \multicolumn{1}{c}{ } &
      \multicolumn{1}{c}{ } &
      \multicolumn{1}{c}{ } &
      \multicolumn{4}{c}{-- -- -- -- -- -- -- -- -- -- -- 2001 -- -- -- -- -- -- -- -- -- -- --} &
      \multicolumn{5}{c}{-- -- -- -- -- -- -- -- -- -- -- -- -- -- -- -- 2013 -- -- -- -- -- -- -- -- -- -- -- -- -- -- -- --}\\
      \multicolumn{1}{l}{Element} &
      \multicolumn{1}{l}{Species} &
      \multicolumn{1}{l}{Arm} &
      \multicolumn{1}{c}{$\sigma\sb{\rm inst}$} &
      \multicolumn{1}{c}{$v\sb{\rm shift}$} &
      \multicolumn{1}{c}{$\sigma\sb{\rm obs}$} &
      \multicolumn{1}{c}{Resolved?} &
      \multicolumn{1}{c}{$\sigma\sb{\rm true}$} &
      \multicolumn{1}{c}{$v\sb{\rm shift}$ (fixed)} &
      \multicolumn{1}{c}{$v\sb{\rm shift}$} &
      \multicolumn{1}{c}{$\sigma\sb{\rm obs}$} &
      \multicolumn{1}{c}{Resolved?} &
      \multicolumn{1}{c}{$\sigma\sb{\rm true}$} &
      \multicolumn{1}{c}{Consistent } \\
      \multicolumn{1}{l}{ } &
      \multicolumn{1}{l}{ } &
      \multicolumn{1}{l}{ } &
      \multicolumn{1}{c}{(${\rm km\,s^{-1}}$)} &
      \multicolumn{1}{c}{(${\rm km\,s^{-1}}$)} &
      \multicolumn{1}{c}{(${\rm km\,s^{-1}}$)} &
      \multicolumn{1}{c}{ } &
      \multicolumn{1}{c}{(${\rm km\,s^{-1}}$)} &
      \multicolumn{1}{c}{(${\rm km\,s^{-1}}$)} &
      \multicolumn{1}{c}{(${\rm km\,s^{-1}}$)} &     
      \multicolumn{1}{c}{(${\rm km\,s^{-1}}$)} &
      \multicolumn{1}{c}{ } &
      \multicolumn{1}{c}{(${\rm km\,s^{-1}}$)} &
      \multicolumn{1}{c}{ $v\sb{\rm shift}$?} \\[1.7ex]
      \hline\\[-1.7ex]
S  & H, He   & HEG & 383 & $-653\pm65$ & $376\pm66$ & $\times$     & N/A & $-571\pm200$  & \ldots       & \ldots      & \ldots       & \ldots & $0.4\sigma$ \\
   & H, He   & MEG & 734 & $-966\pm73$ & $525\pm75$ & $\times$     & N/A & $-847\pm229$  & \ldots       & \ldots      & \ldots       & \ldots & $0.5\sigma$ 
 \\[1.7ex]
Si & H, He   & HEG & 309 & $-603\pm19$ & $412\pm20$ & $\checkmark$ & 273 & $-630\pm75$   & $-610\pm61$  & $311\pm63$  & $\times$     & N/A    & $0.3\sigma$ ($0.1\sigma$) \\
   & H, He   & MEG & 592 & $-718\pm19$ & $525\pm20$ & $\times$     & N/A & $-757\pm69$   & $-758\pm71$  & $530\pm74$  & $\times$     & N/A    & $0.5\sigma$ ($0.5\sigma$) \\
   & Low     & HEG & 225 & $-597\pm28$ & $224\pm28$ & $\times$     & N/A & $-697\pm104$  & $-711\pm87$  & $251\pm91$  & $\times$     & N/A    & $0.9\sigma$ ($1.2\sigma$) \\
   & Low     & MEG & 432 & $-661\pm40$ & $322\pm41$ & $\times$     & N/A & \ldots        & $-488\pm70$  & $250\pm70$  & $\times$     & N/A    & \;\;\;\;\;\;\;\;($2.1\sigma$)
 \\[1.7ex]      
Al & H, He   & HEG & 266 & $-536\pm22$ & $ 78\pm20$ & $\times$     & N/A & $-510\pm67$   & \ldots       & \ldots      & \ldots       & \ldots & $0.4\sigma$ \\
   & H, He   & MEG & 510 & $-544\pm67$ & $261\pm68$ & $\times$     & N/A & $-877\pm262$  & \ldots       & \ldots      & \ldots       & \ldots & $1.2\sigma$ 
 \\[1.7ex]     
Mg & H, He   & HEG & 215 & $-587\pm19$ & $297\pm19$ & $\checkmark$ & 205 & $-630\pm79$   & $-613\pm50$  & $179\pm52$  & $\times$     & N/A    & $0.5\sigma$ ($0.5\sigma$) \\
   & H, He   & MEG & 412 & $-596\pm21$ & $395\pm21$ & $\times$     & N/A & $-773\pm81$   & $-759\pm99$  & $352\pm101$ & $\times$     & N/A    & $2.1\sigma$ ($1.6\sigma$) 
 \\[1.7ex]
Ne & H, He   & HEG & 157 & $-727\pm25$ & $406\pm28$ & $\checkmark$ & 374 & \ldots        & \ldots       & \ldots      & \ldots       & \ldots & \ldots \\
   & H, He   & MEG & 301 & $-734\pm18$ & $380\pm19$ & $\checkmark$ & 231 & $-947\pm99$   & $-948\pm138$ & $379\pm153$ & $\times$     & N/A    & $2.1\sigma$ ($1.5\sigma$)
 \\[1.7ex]
\textit{All} & H, He & HEG & 383 & $-612\pm13$ & $345\pm13$ & $\times$ & N/A & $-603\pm47$  & $-582\pm34$  & $220\pm34$  & $\times$  & N/A  & $0.2\sigma$ ($0.8\sigma$) \\
             & H, He & MEG & 734 & $-686\pm13$ & $429\pm13$ & $\times$ & N/A & $-782\pm45$  & $-779\pm53$  & $413\pm54$  & $\times$  & N/A  & $2.0\sigma$ ($1.7\sigma$)
 \\[1.7ex]
     \hline\\
    \end{tabular}
    \label{table:vel_profiles}
     \textbf{Notes.} $\sigma\sb{\rm inst}$ is the instrumental
     resolution for the shortest wavelength line included in the
     stack.  Blank entries indicate no good Gaussian fit was obtained.
     In the case of Ne, this is because the 2013 HEG velocity profile
     contains only $\sim5$ counts per bin. The final column lists
     whether the velocity shifts measured for each profile are
     consistent between the two epochs.  Values quoted are the
     significance of differences between the fit to the 2001 data and
     the fixed profile fit to the 2013 data.  Where possible, a
     comparison with a free profile fit is listed in parentheses.
     `Low' indicates the best-fitting results to the Si profiles
     constructed from low-ionization absorption lines only, rather
     than those from H- and He-like ions.
\vspace*{-0.2cm}
\end{table*}

The velocity profiles shown in Figures~\ref{fig:vel}
and~\ref{fig:vel2} are modelled with a Gaussian plus a constant fixed
at 1.0 by definition, using a \chisq minimization technique.  We model
the 2001 data with both the centroid, $v\sb{\rm shift}$, and width,
$\sigma\sb{\rm obs}$, of the Gaussian as free parameters.  Where
possible, we also leave these parameters free to vary in the fit to
the 2013 data but we also fix the width of the Gaussian to that
determined in the 2001 data and allow only the centroid to vary.  For
a velocity profile to be resolved, it must be broader than
$\sigma\sb{\rm inst}$ i.e., $\sigma\sb{\rm obs}-\Delta\sigma\sb{\rm
  obs} > \sigma\sb{\rm inst}$ where $\Delta\sigma\sb{\rm obs}$ is the
error on $\sigma\sb{\rm obs}$.  In cases where this is true, the
intrinsic width of the profile is determined from $\sigma\sb{\rm
  true}^{2}=\sigma\sb{\rm obs}^{2}-\sigma\sb{\rm inst}^{2}$.  The
best-fitting Gaussian parameters from this modeling, along with the
instrumental resolutions for each profile, and an indication of
whether a particular profile is resolved or not, are listed in
Table~\ref{table:vel_profiles}.  The final column of this table lists
whether the velocity profiles have a consistent velocity shift between
the two epochs.  This is determined based on the values from the fixed
shape profile fits to the 2013 data; if a free fit was possible this
value is also quoted in parentheses.  These values are all $\le
2.1\sigma$ showing that in general, the X-ray outflow velocities are
consistent in each epoch and there are no significant changes in the
kinematic structure of the absorbing material.  Profiles showing the
highest significance changes in velocity shift include the MEG profile
of H- and He-like Mg which shows a slight shift toward a larger
outflow velocity in 2013, however, in this case (and many others) the
profile for neither epoch is instrumentally resolved and the velocity
shifts obtained using the HEG profiles are consistent.  The MEG
profile of Ne only includes $\sim 14$ counts per bin in the 2013
spectrum, resulting in a poorly defined profile.  The MEG velocity
profile produced from all 23 H- and He-like absorption lines of S, Si,
Al, Mg, and Ne shows a change in the velocity shift between epochs
with a significance of $\sim 2\sigma$, but again, the HEG profiles
have consistent velocities.  With 10 profiles being considered, we
might expect to detect a velocity shift with $>2\sigma$ significance
in at least one case.

The weighted average velocity shift of the profiles listed in
Table~\ref{table:vel_profiles} (H- and He-like only) is
$v=-647\pm22\;\kms$ in 2001 and $v=-672\pm66\;\kms$ in 2013, which are
consistent with each other.  The average velocity shift for 2013 from
the fixed-profile fits is also consistent, although it does have a
larger error; $-688\pm87\;\kms$.  These values also agree with the
average velocity shifts determined from the individual lines as listed
in Table~\ref{table:lines}; $v=-654\pm14\;\kms$ (2001),
$v=-693\pm58\;\kms$ (2013).  The average widths of the velocity
profiles are also consistent, $\sigma\sb{\rm obs}=352\pm22\;\kms$ in
2001 and $\sigma\sb{\rm obs}=309\pm69\;\kms$ in 2013, although none of
the profiles in the 2013 data are actually resolved.  The asymmetry of
the profiles observed by K02, possibly as a result of another
absorption system with a higher outflow velocity, is not apparent in
the less well defined 2013 profiles.  This justifies the use of a
single Gaussian in our fitting, which although perhaps not strictly
representative of the profile, allows us to make simple comparisons
between the two epochs.

In the case of Si we are able to construct velocity profiles using
lower ionization absorption lines e.g., those from Li-like ions etc
(rather than H- and He-like), which may originate from different
absorbing material.  The best-fitting Gaussian parameters for these
profiles are also listed in Table~\ref{table:vel_profiles} (as `Low'),
and are shown in Figure~\ref{fig:vel2} (bottom).  The MEG profile
shows a velocity shift from $v=-661\pm40\;\kms$ (2001) to
$v=-488\pm70\;\kms$ (2013) at $2.1\sigma$ significance, but neither
profile is resolved.\\

\subsubsection{UV kinematics}
\label{section:uv_kin}
\begin{figure*}
  \centering 
    \begin{tabular}{cc}
      \vspace*{0.3cm}
      \includegraphics[width=0.49\textwidth]{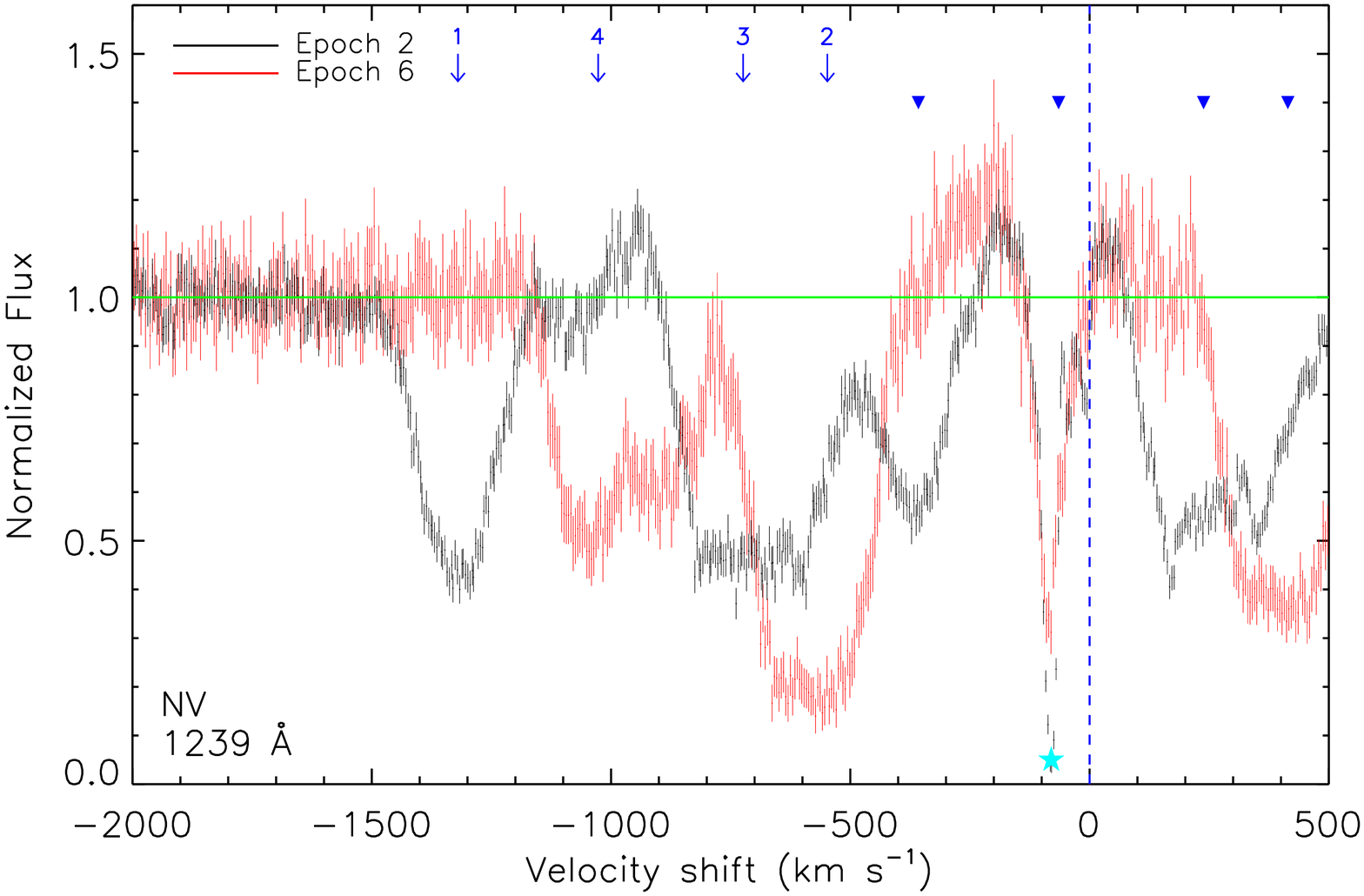} &
      \includegraphics[width=0.49\textwidth]{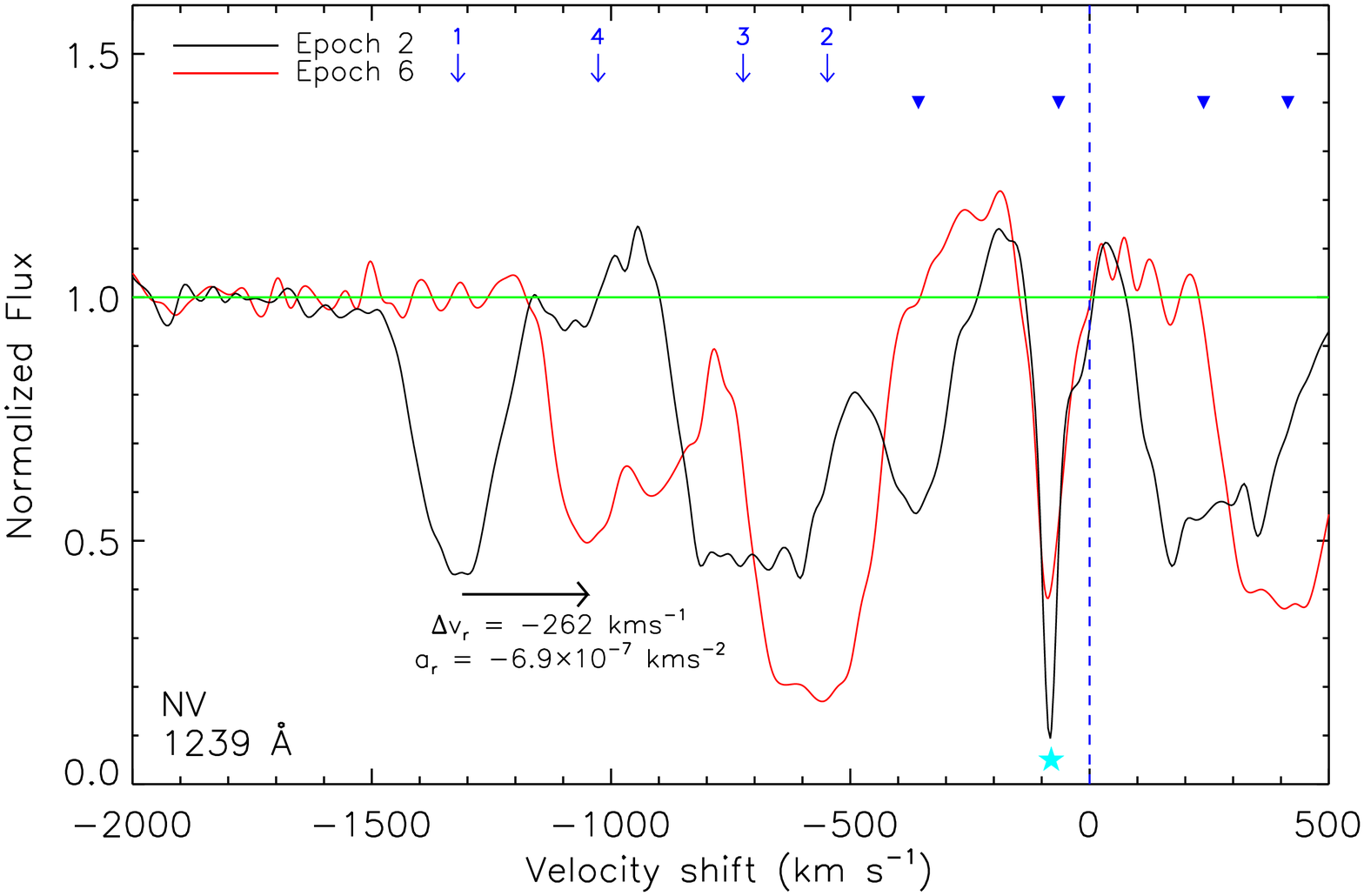} \\
      \vspace*{0.3cm}
      \includegraphics[width=0.49\textwidth]{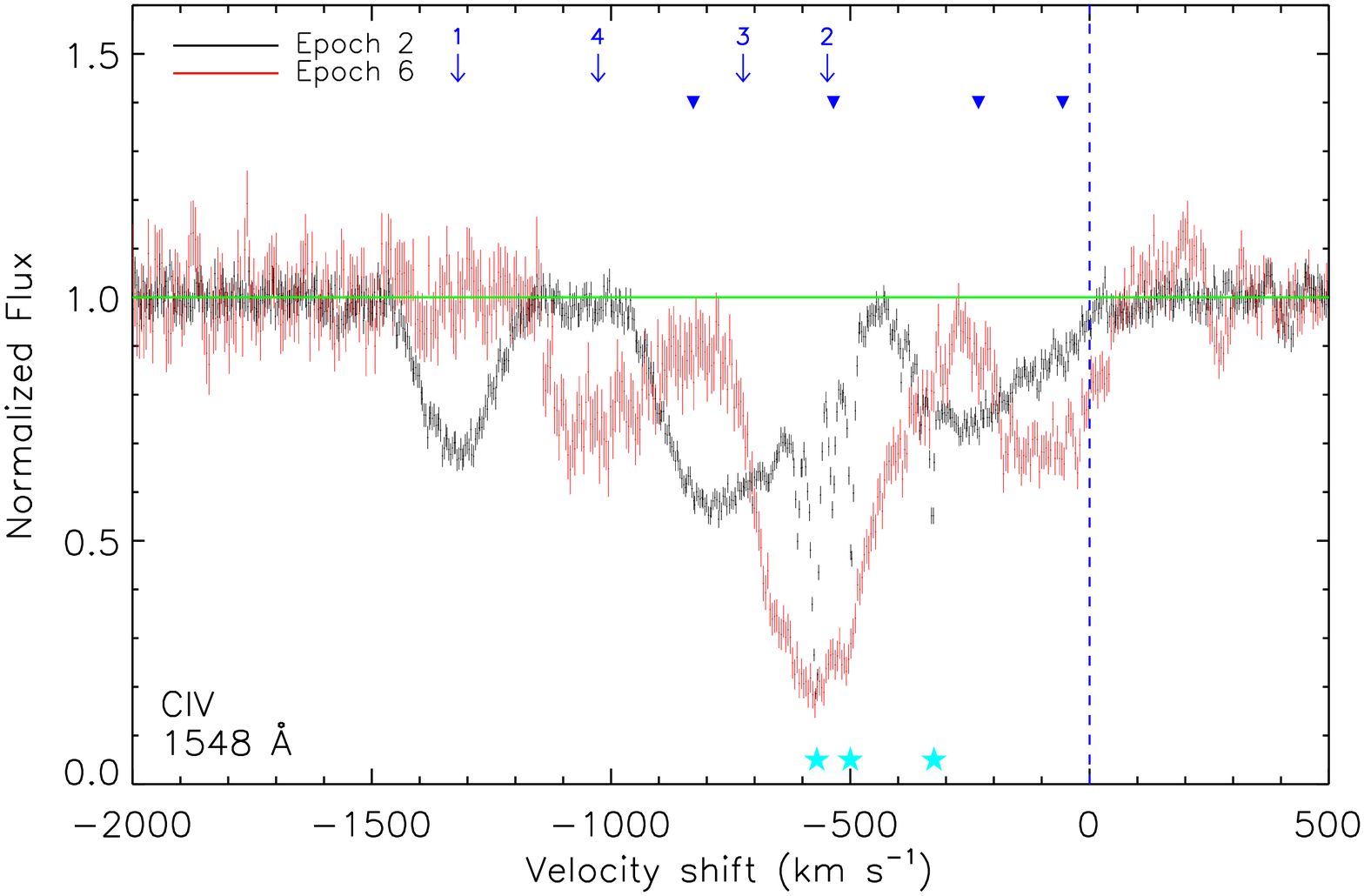} &
      \includegraphics[width=0.49\textwidth]{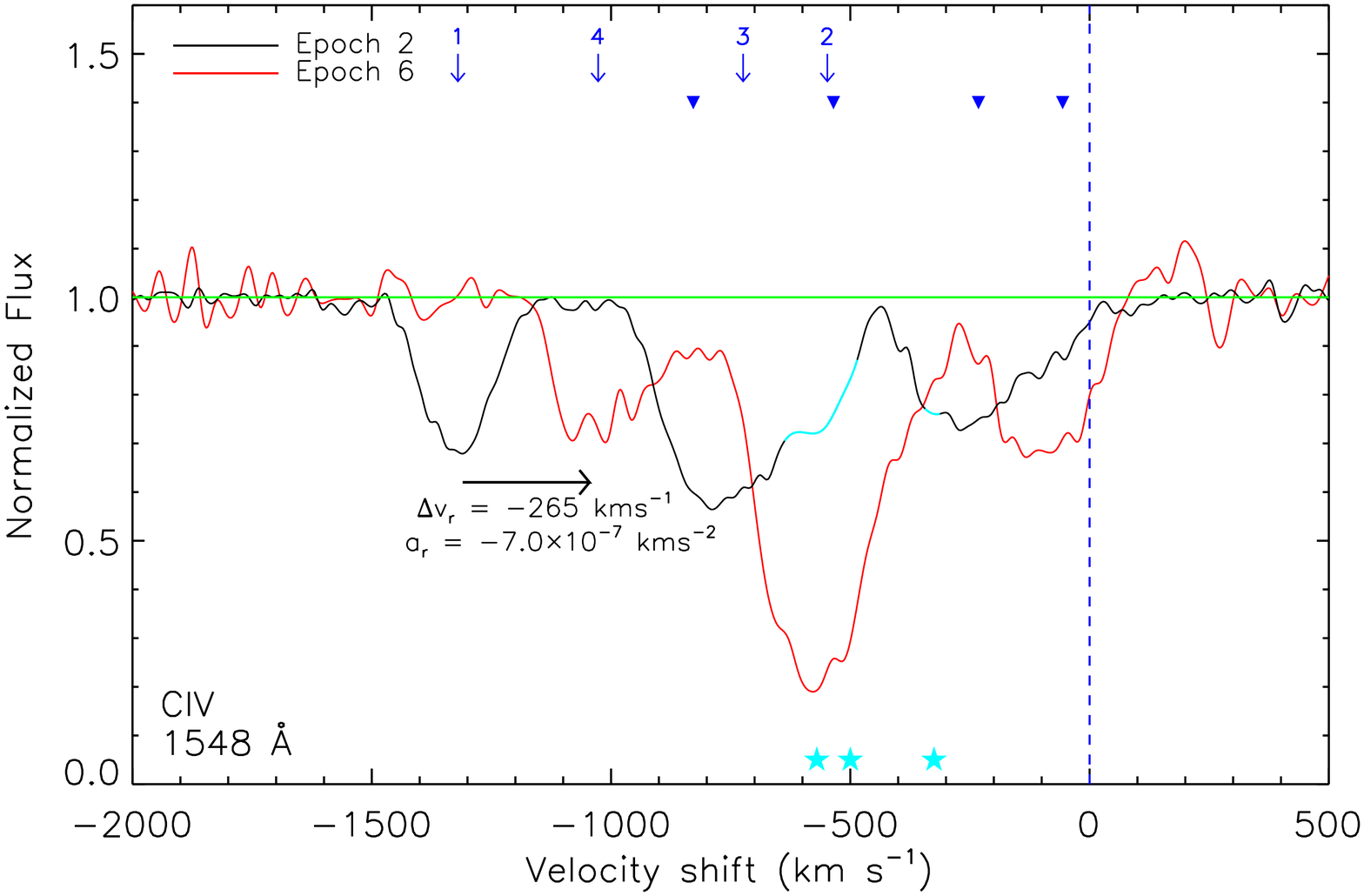} \\
      \includegraphics[width=0.49\textwidth]{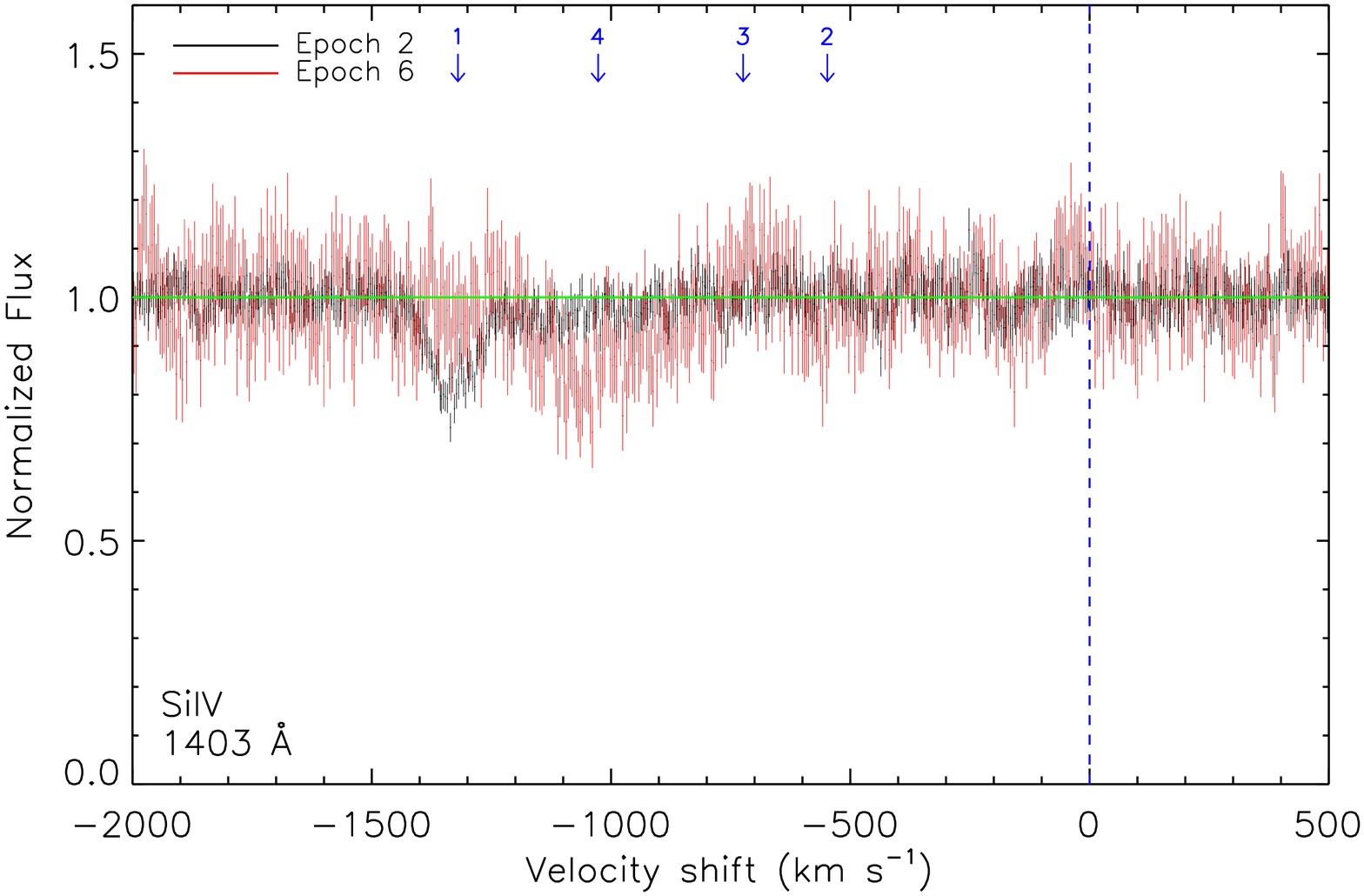} &
      \includegraphics[width=0.49\textwidth]{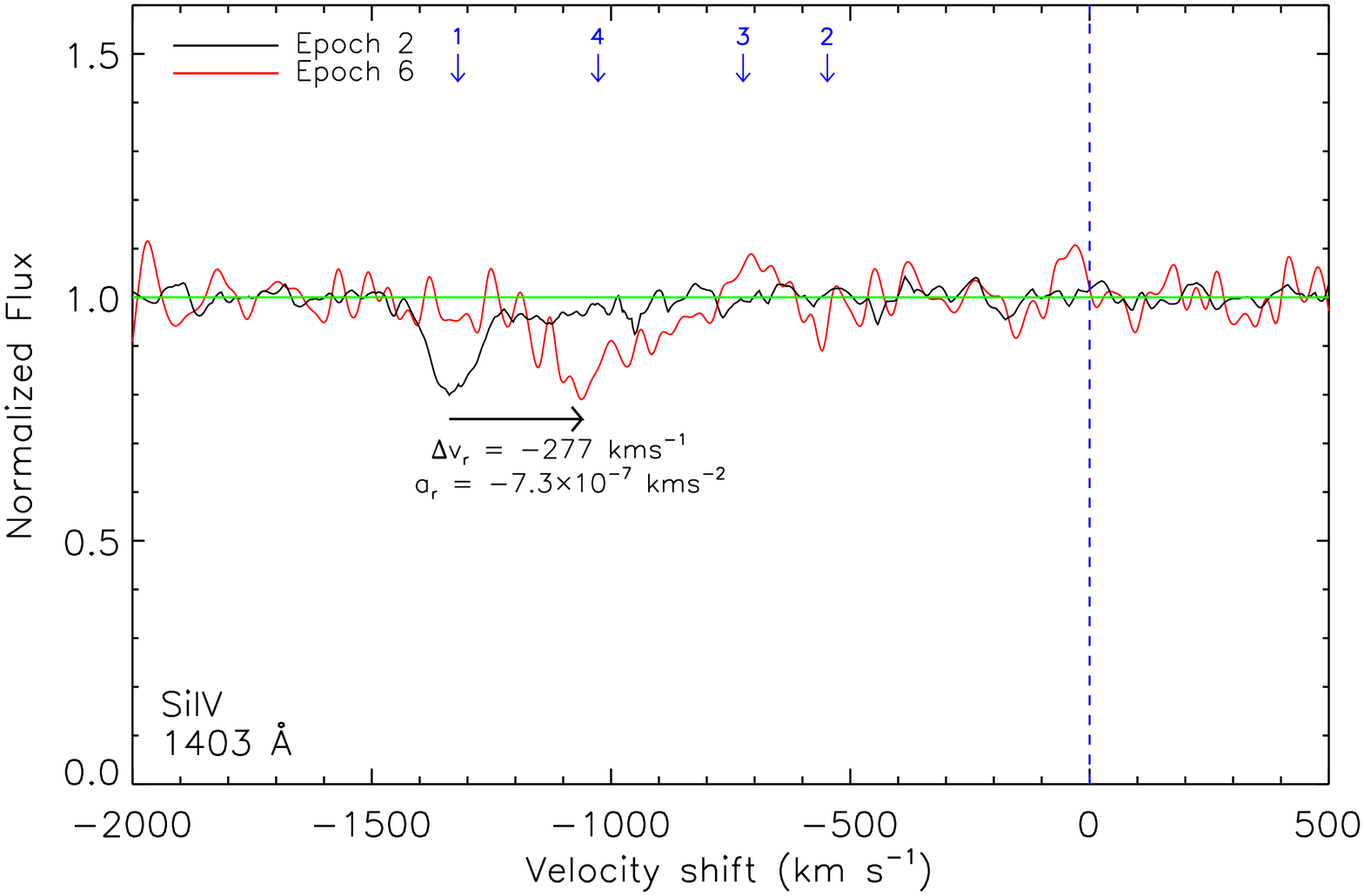} \\
    \end{tabular}
    \caption{Velocity profiles for the UV absorption lines \nv\,(top),
      \civ\,(middle), and \siiv\,(bottom).  They are created after
      first dividing by the continuum level, estimated with a cubic
      spline fit to unabsorbed regions either side of the lines,
      before converting the observed wavelength to a velocity
      blueshift with respect to the rest-frame wavelength of the
      lines.  Epoch~2 data (2001) are shown in black, Epoch~6 data
      (2013) are shown in red.  The velocity bin size is $4-5\;\kms$
      for the left-hand plots.  Dark blue arrows indicate the
      locations of the UV kinematic components using values from G03a.
      In the case of \nv\, and \civ, the profile includes features
      from both doublet members.  In these cases, the expected
      location of the components from the longer wavelength line are
      marked with the dark blue triangles.  Cyan stars indicate the
      locations of interstellar absorption lines.  Right hand panels
      show the same velocity profiles, but with a cubic spline fitted
      to the data with the binning increased to $5\times$ the
      original.  In the case of \civ, the ISM absorption lines in the
      Epoch~2 data have been excluded from the spline fit; these
      portions are highlighted in cyan.  The velocity shifts and
      decelerations quoted on each figure are determined from the
      minima of the absorption troughs assuming the physical
      absorption component originally at the velocity shift of
      component~1 has decelerated at a constant rate to the velocity
      shift of component~4.}
  \label{fig:hst_vel}
\end{figure*}

We also produce absorption profiles plotted in velocity space using
the UV spectra.  The original wavelength-binned spectra are first
normalized by dividing by the estimated continuum level.  This is
determined from a cubic-spline fit to unabsorbed regions of the
spectra on either side of absorption features, using the data binned
to $\sim 0.4$\,\ang\, ($\sim 0.8$\,\ang\, in the case of Si) as the
spline nodes.  The observed wavelengths are then converted to a
velocity shift with respect to the rest-frame wavelengths of the
absorption lines.  The resulting profiles for \nv\,(1239\,\ang),
\civ\,(1548\,\ang) and \siiv\,(1403\,\ang) are shown in
Figure~\ref{fig:hst_vel}.  Epoch~2 data are shown in black and Epoch~6
in red.  The dark blue numbered arrows indicate the expected locations
of the UV kinematic components based on the the values previously
reported by G03a; i.e., (1) $-1320\;\kms$, (2) $-548\;\kms$, (3)
$-724\;\kms$, and (4) $-1027\;\kms$.  In the cases of \nv\, and \civ,
the profiles created are a superposition of features due to each of
the lines in the doublet.  We therefore create the profile with
respect to the shortest wavelength doublet member and indicate the
expected locations of the kinematic components of the longer
wavelength doublet member with dark blue triangles.  For \siiv, the
profiles from the two doublet members do not overlap.  We show only
the profile from the line at 1403\,\ang\, as the profile from the
shorter wavelength line contains a strong Galactic absorption feature.
Blue cyan stars on each figure indicate the location of such
interstellar features.  The right hand plots in
Figure~\ref{fig:hst_vel} show a smoothed version of each velocity
profile, produced by a cubic spline fit to the data binned to
$5\times$ the original velocity bin size.

In all three of the UV absorption profiles, kinematic component~1,
which was very strong in Epoch~2 (2001) has completely disappeared by
Epoch~6 (2013).  Similarly, component~4, which was not present in
Epoch~2, appears in all three profiles in Epoch~6.  While this may
simply be due to the strengthening and weakening of physically
distinct absorption components, G03b reported a deceleration of
component~1.  Quoted on Figure~\ref{fig:hst_vel} are the velocity
shifts measured from the minima of the absorption troughs and the
inferred decelerations, assuming these remain constant over the full
12 yr period.  We find a velocity shift of $\Delta
v\sb{r}\sim-268\;\kms$ corresponding to a deceleration of $a_{r}\sim
7\times10^{-7}\;\kmss$.  This is lower than the average deceleration
reported by G03b, but they did find a change in the deceleration
within their observations ($a_{r}\sim 8\times10^{-7}\;\kmss$ between
Epochs $1 \rightarrow 2$ and $a_{r}\sim 18\times10^{-7}\;\kmss$
between $2 \rightarrow 3$).  The apparent depth of the absorption
trough now at component~4 is similar to that which was at component~1,
supporting the idea that it may be due to the same physical absorption
component.  However, the profile shape appears strongly asymmetric in
Epoch~6 and the \nv\, profile suggests this may be due to a
superposition of two separate components.  While this could be due to
the sub-components of component~1 having different decelerations, the
asymmetry is also clearly observed in the \siiv\, profile which is not
expected to be the case in such a scenario.  This will be discussed
further in J.~Gabel et al. (in preparation).

The \nv\, and \civ\, profiles both show a strong velocity component~3
in their Epoch~2 spectra, which is weak, or even no longer present in
Epoch~6.  Similarly, component~2 is weak in Epoch~2, but strong in
Epoch~6.  Component~2 is also weakly present in the \siiv\, profile in
Epoch~6.  In the case of \civ, component~4 from the longer-wavelength
member of the doublet is expected to appear at the same outflow
velocity as component~2, possibly distorting the profile, but the
\nv\, profile which shows similar general effects, suffers from no
such contamination.  These observations may again be due to the
independent strengthening and weakening of different absorption
components, particularly as the the two absorption profiles show
different shapes, the one in Epoch~6 being both narrower and deeper;
however the total area of each remains similar.  This may suggest that
component~3 has decelerated to the velocity shift of component~2.  No
deceleration of this component was found in the initial study by G03b,
but the deceleration implied would be
$a\sb{r}\sim3.3\times10^{-7}\;\kmss$, corresponding to a velocity
change of only $\Delta v\sb{r}=-19\;\kms$ between Epochs~1 and 3
investigated by G03b.  Such a shift would not have been detected due
to the wide absorption trough of this component ($\sim200\;\kms$).\\


\subsection{X-ray/UV absorber connection}
\label{section:joint}
The UV velocity profiles indicate significant changes in the 12 yr
period between the observations, possibly due to radial deceleration
of the absorbing material.  Such changes are not apparent in the X-ray
velocity profiles; the broad profile shapes remain similar between
epochs and no velocity shifts are detected significant at $>3\sigma$.
The possible velocity shifts in the UV are $\Delta
v\sb{r,\,1\rightarrow4}\sim-268\,\kms$ and $\Delta
v\sb{r,\,3\rightarrow 2}\sim-125\;\kms$.  Although the typical errors
in the X-ray velocity estimates are only $50-100\;\kms$, the widths of
the profiles are very broad, $\sigma\sb{\rm obs}>300\;\kms$ in each
epoch, and few are instrumentally resolved.  Therefore it is unlikely
that velocity shifts of these sizes would be detectable in the X-ray
velocity profiles.  Assuming typical errors on the central velocity of
$40\;\kms$ and $85\;\kms$ on the 2001 and 2013 MEG profiles,
respectively, only velocity shifts as large as $\Delta
v\simeq282\;\kms$ would be detected with $3\sigma$ significance in the
X-ray data.  The UV data indicate shifts smaller than this value,
which are therefore unlikely to be detected in the X-ray data whether
present or not.

\begin{figure}[h]
  \centering 
    \includegraphics[width=0.48\textwidth]{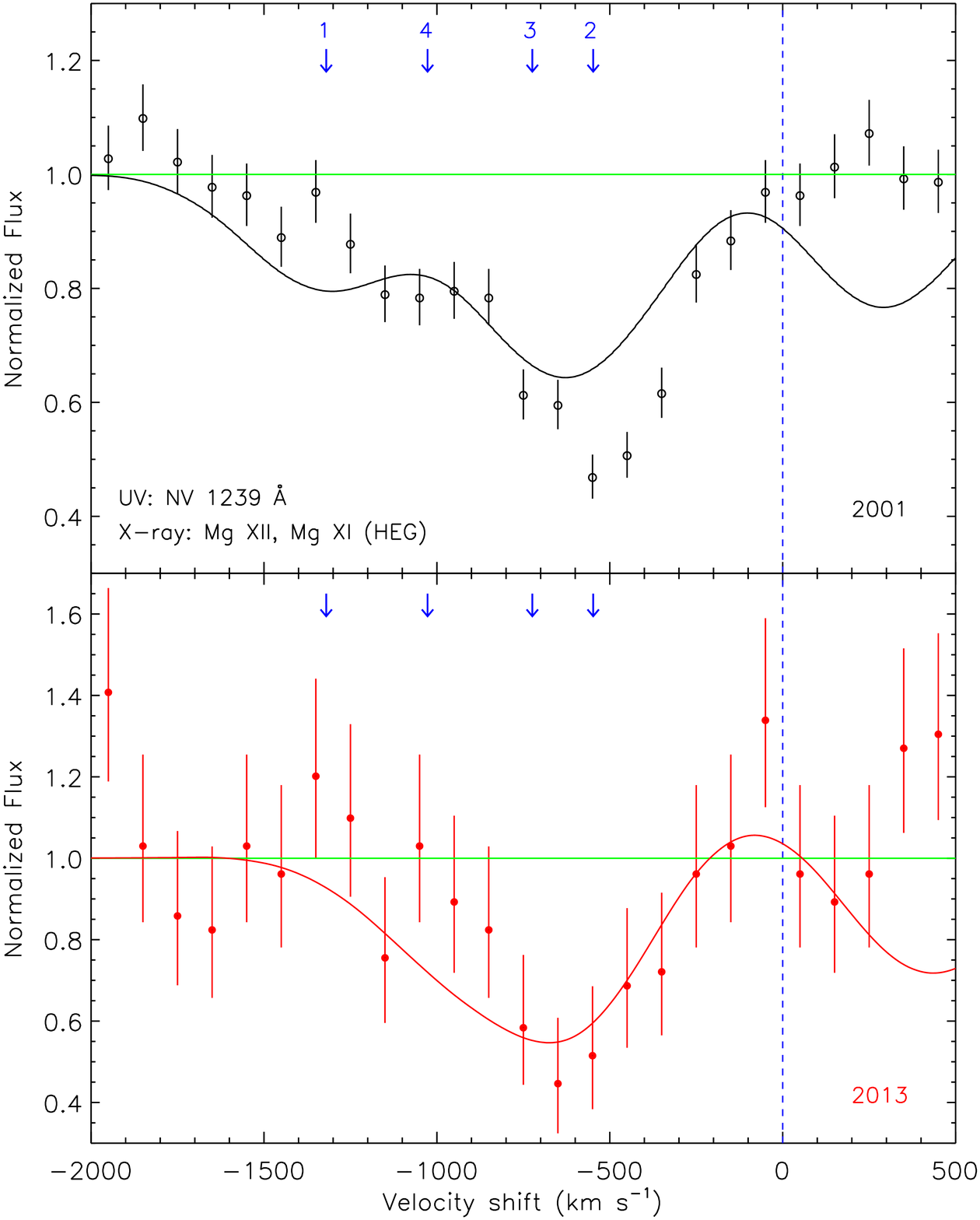} 
    \caption{A combined X-ray and UV velocity profile.  The top panel
      shows the data from 2001; black, open circles are the X-ray
      velocity profile constructed from absorption lines of H- and
      He-like Mg (Mg~\textsc{xii} and Mg~\textsc{xi}) detected in the
      HEG.  The smooth black curve represents the UV velocity profile
      from the absorption line N~\textsc{v} at 1239\,\ang\, which has
      been convolved with a Gaussian of width $\sigma=215\;\kms$ in
      order to downgrade the spectrum to match the velocity resolution
      of the HEG at these wavelengths ($\sim 7.1\,\ang$).  Note that
      the dip in the UV profile at $v>0\;\kms$ is due to the longer
      wavelength doublet member of N~\textsc{v} at 1243\,\ang.  The
      bottom panel considers the data from 2013, where the UV profile
      is indicated by the smooth red curve and the X-ray profile is
      shown by the red, filled circles and $1\sigma$ error bars.  The
      X-ray velocity profiles are binned using a bin width of
      $100\;\kms$, although we note that this over-samples the
      velocity resolution.  The blue arrows in each panel indicate the
      location of the four UV kinematic components using the values
      quoted in G03a.  Note that the two panels have different {\it
        y}-axis ranges.}
  \label{fig:vel_combined}
\end{figure}

In Figure~\ref{fig:vel_combined} we compare the X-ray velocity profile
produced from the absorption lines of H- and He-like Mg to the UV
velocity profile for \nv\,(1239\,\AA).  We use Mg as it has one of the
highest S/N X-ray profiles (along with Si), but a higher resolution
($215\;\kms$ compared with $309\;\kms$).  We use \nv\, as the overlap
of the features from the two doublet members is lower than for \civ.
While we show only one profile, the combined profiles from each of the
four combinations of Mg/Si and \civ/\nv, yield similar results.  The
top panel shows the 2001 data, with the X-ray data shown by the black
open circles and the UV data shown by the smooth black curve.  The UV
profile has been convolved with a Gaussian of width
$\sigma=215\;\kms$, to downgrade its velocity resolution to match that
of the HEG at these wavelengths (${\rm FWHM}=507\;\kms$).  The bottom
panel shows the 2013 data with red filled circles showing the X-ray
data, and the red curve showing the convolved UV velocity profile.
The X-ray profiles are binned using a bin size of $100\;\kms$ as in
Figures~\ref{fig:vel} and \ref{fig:vel2}, which over-samples the
instrumental resolution.  The blue arrows indicate the locations of
the four UV kinematic components identified in the 2001 data by G03a.

As seen in previous works (e.g.,~K02; G03a), the X-ray and UV velocity
profiles appear broadly similar, suggesting that it may simply be the
poorer resolution of the X-ray profile which limits our detection of
relatively small kinematic changes.  Figure~\ref{fig:vel_combined}
shows that the broad X-ray profile may be a superposition of all four
UV kinematic components, with the asymmetry in the 2001 \mbox{X-ray}
profiles being due to a strong UV component~1.  The small differences
observed between the profiles are likely due to them probing different
material.  The X-ray profiles are constructed primarily of absorption
lines from higher ionization state ions, whereas the UV profiles show
absorption due to lower ionization state ions which may reside in
different physical absorption components.  However, photoionization
modelling of both the 2001 X-ray and UV spectra indicates that the
lowest-ionization X-ray absorber is consistent with the higher
ionization UV absorbers (components 1b, 2, and 3), suggesting a
physical connection between at least some of the X-ray and UV
absorbing material \citep{netzer03,gabel05}.\\

\subsection{Iron K region}
\label{section:iron}
In this section, we investigate possible changes in the iron~K region
of the spectra between the two epochs.

A relativistically broadened iron \ka emission line at 6.4~keV has
been detected in observations of NGC~3783 by \asca \citep{nandra97},
\textit{BeppoSAX} \citep{derosa02}, \suzaku \citep{brenneman11}, and
\xmm \citep{blustin02}, although analysis of the \xmm spectrum by
\citet{reeves04} showed that when the warm absorbers were thoroughly
modelled in the spectra, the requirement for a relativistic iron line
was greatly reduced.  The \chandra data also did not require a broad
line in the modelling, showing only a narrow core likely produced in
distant material \citep{kaspi02}.  However, analysis by
\citet{yaqoob05} found that the excess of emission at the base of the
line could be well described by either a Compton-scattering `shoulder'
or with a disk-line model.  A weaker emission line due to a blend of
Fe \kb and Fe~\textsc{xxvi} emission has been detected in previous
observations \citep{blustin02,kaspi02,reeves04,yaqoob05,brenneman11},
although due to the low branching ratio (150:17 between the \ka and
\kb transitions), this feature is unlikely to be significantly
detected in the \chandra HEG data (2001 or 2013).  An absorption line
due to highly ionized iron has been detected in \chandra and \suzaku
observations \citep{kaspi02,yaqoob05,brenneman11} and \xmm
observations showed it to be variable on timescales of $\sim 10^{5}$ s
\citep{reeves04}.  The depth of the line varied from one orbit to the
next, appearing strongest when the continuum flux was higher.  It is
likely due to the highest ionization component of the warm absorber
($\textrm{log}\,\xi \sim 3$, $N\sb{\rm
  H}=5\times10^{22}\,\textrm{cm}^{-2}$) located within 0.1~pc of the
nucleus, passing across the central X-ray emitting source
\citep{reeves04}.

\begin{figure}[h]
  \centering
  \includegraphics[angle=270, width=0.48\textwidth]{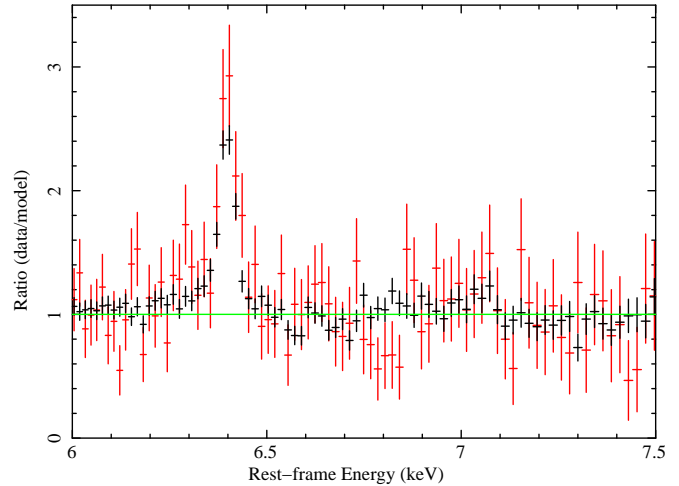}
  \caption{The iron \ka emission-line region of NGC~3783.  Shown is
    the ratio of the HEG data to a power-law model fit over the
    energies 5--6 and 6.5--7.5~keV.  This simple model is used here
    for illustrative purposes only.  The data from 2001 are shown in
    black and the 2013 data are shown in red.  The spectra from both
    epochs are modelled simultaneously, with \gmm tied to a common
    value, but the normalizations allowed to vary independently.  The
    data has been displayed with an increased binning for clarity.\\}
  \label{fig:iron}
\end{figure}

For our analysis, only data from the HEG were used, binned at
0.0025\,\ang\, to give the highest resolution spectrum in this region.
The spectral modelling was done in \xspec v12.7.1, using the
$C-$statistic \citep{cash79}.  Figure~\ref{fig:iron} shows the ratio
of the data to a simple power-law model fit over the energy range
5.0--7.5~keV ($1.6-2.5$\,\ang), excluding 6.0--6.5~keV.  Black
corresponds to the 2001 data, red shows the 2013 data, and the binning
in each data set has been increased in the figure for clarity.  The
power-law indices were initially left free to vary independently, but
were found to be consistent within their 90\% errors and were
subsequently tied to the same value.  The normalizations remain free
to vary independently.  Although the warm absorbers may introduce
subtle spectral curvature \citep{reeves04,patrick11} which is not
taken into account by our simple modelling of the continuum, this
should not significantly impact our primary goal of comparing the two
data sets obtained at different epochs.

Figure~\ref{fig:iron} clearly shows the iron \ka emission-line in both
epochs.  We therefore fit a power-law plus Gaussian emission line
model, \texttt{phabs*(po+zgauss)} in \textsc{xspec}, to data in the
energy range $5-7.5$ keV.  We continue to use the Cash statistic, and
errors on best-fitting spectral parameters are quoted at 90\%
significance.  The best-fitting power-law slopes over this range are
consistent in each epoch, as are the best-fitting parameters of the
iron \ka emission line, which are listed in Table~\ref{table:iron}.
The addition of the Gaussian emission-line to a simple power-law model
over this energy range results in an improvement in the Cash statistic
of $\Delta C=732$ in 2001 and $\Delta C=145$ in 2013, with $C=367$ in
2001 and $C=350$ in 2013, for 324 degrees of freedom ($\nu$).  The EW
of the line does not change significantly between epochs and is
consistent with literature values \citep{kaspi02,reeves04,yaqoob05}.
Although the underlying continuum flux in the \mbox{5--7.5}~keV energy
range drops significantly in the 2013 data, the line flux remains
consistent between epochs.  The width of the line is also consistent
for each epoch and corresponds to FWHM values that are slightly higher
than those in the literature \citep{kaspi02,yaqoob05}, but our
modelling does not take into account the Compton shoulder created at
$\sim 6.2$~keV due to the Compton scattering of the line photons.  The
width is much lower than that reported by \citet{reeves04}, although
this is likely a result of the lower energy resolution of the \xmm
EPIC-pn ($\Delta E \sim 120$ eV at 6~keV; \citealt{pn}).  The FWHM
resolution of the HETGS is $\Delta E \sim 35$ eV at 6~keV,
corresponding to $\sim 1750\;\kms$, and therefore the emission line is
resolved in both of our data sets.\\

\begin{table}[h]
\vspace*{-0.4cm}
\centering
\caption{Iron \ka line parameters}
\label{table:iron}
    \begin{tabular}{lllcc}
      \hline \hline \\[-2.0ex]
      \multicolumn{1}{l}{Parameter} &
      \multicolumn{1}{l}{2001} &
      \multicolumn{1}{l}{2013} \\[0.5ex]
      \hline\\[-2.0ex]
        $\Gamma$                                  & $1.62\sp{+0.09}\sb{-0.09}$     & $1.47\sp{+0.24}\sb{-0.26}$     \\[1.2ex]
        E (keV)                                   & $6.397\sp{+0.002}\sb{-0.003}$  & $6.400\sp{+0.011}\sb{-0.010}$  \\[1.2ex]
        $\sigma$ (eV)                             & $20\sp{+2}\sb{-4}$             & $20\sp{+6}\sb{-11}$            \\[1.2ex]
        FWHM ($\kms$)                             & $2209\sp{+212}\sb{-424}$       & $2208\sp{+636}\sb{-1413}$      \\[1.2ex]
        EW (eV)                                   & $86\sp{+9}\sb{-8}$             & $110\sp{+40}\sb{-20}$          \\[1.2ex]
        Continuum flux ($\times 10^{-11}\,\flux$)  & $1.87\sp{+0.03}\sb{-0.02}$     & $1.23\sp{+0.06}\sb{-0.02}$     \\[1.2ex]
        Line Flux ($\times 10^{-13}\,\flux$)       & $6.39\sp{+0.49}\sb{-0.51}$     & $5.55\sp{+1.03}\sb{-1.45}$     \\[1.2ex]
        C/$\nu$                                   & $367/324$                      & $350/324$                      \\[1.2ex]
        $\Delta C$                                & $732$                          & $145$                          \\[1.2ex] 
      \hline\\[1.2ex]
    \end{tabular}
    \textbf{Notes.} Each of these parameters are calculated from a fit
    to data in the $5-7.5$ keV energy range.  $\Delta C$ values are
    the improvement in the Cash statistic gained when the Gaussian is
    added to a simple power-law model fit to data in this energy only.
\end{table}

Using simple virial arguments, the distance to the material from which
the iron line is emitted is $31\sp{+7}\sb{-10}$ light days (2001) and
$31\sp{+14}\sb{-29}$ light days (2013), corresponding to $\sim0.026$
pc.  In this determination we have assumed that the material is moving
entirely perpendicular to our line-of-sight and hence assume the
minimum possible velocity and thus derive a maximum radius for the
location of the material.  Our estimate places the material beyond the
broad-line region which lies at a distance of $r=0.0015-0.0018$~pc
($\textrm{FWHM}\sb{H\beta}=2910-2650\;\kms$; \citealt{onken02}) but
within the inner edge of the torus at $r\sb{\rm in}=0.061$~pc, which
has been determined from near-infrared reverberation mapping
\citep{honig13}.  Given this distance we might expect to see a
response to changes in the continuum as the light travel time
($t\sb{\rm cross}=31$ days) is much shorter than the time between our
observations ($\sim 12$ yrs).  However, since we likely observe a
combination of emission from material at different locations around
the AGN, such variations may be `washed out' by the time they reach
us.  \rxte monitoring shows that the hard X-ray flux varies
considerably with the largest fluctuations being $\Delta F\sb{\rm
  2-10\, keV}\sim 5\times10\sp{-11}\;\flux$ on timescales of $\sim 20$
days.  Since no \rxte monitoring is available after January 2012, we
do not know the behavior of the continuum fluctuations immediately
preceding our \chandra observation.  We could therefore be seeing the
spectrum responding to a flare in the continuum emission, an unknown
time earlier.  This limits our ability to put a strong constraint on
variations of the iron line.

A second Gaussian emission line was included to model Fe K$\beta$
which is expected to appear in the spectrum at $E=7.06$~keV with 11\%
of the Fe \ka flux.  There is a hint of this line in both data sets,
but only upper limits on the EW were obtained; EW$\sb{\rm
  2001}<40$~eV, EW$\sb{\rm 2013}<80$~eV.

A Gaussian was also added to the model to look for the variable
absorption line previously reported at 6.7~keV \citep{reeves04}.
Leaving the central energy of the line as a free parameter does not
result in lines at this expected energy, particularly in the case of
the 2013 data due to the potential feature at 6.8~keV, the nature of
which is unclear (however we note that the bins in the 2013 spectrum
contain only $10-15$ counts at these energies).  Constraining both the
energy and width of the 6.7~keV absorption line to the parameters
given by \citet{yaqoob05} and allowing only the normalizations to
vary, we measure EW$\sb{\rm 2001}=14\sp{+7}\sb{-8}$~eV and EW$\sb{\rm
  2013}<30$~eV.  This limit is greater than the EW reported by
\citet{reeves04} for the \xmm orbit in which the line was strongest
($17\pm5$~eV); therefore while the line may have become stronger, we
cannot rule out it becoming weaker, in agreement with the observed
dependence on continuum flux seen by \citet{reeves04}.\\


\section{Summary}
\label{section:sum}
In this paper we have presented recent (2013) \chandra HETGS and \hst
COS observations of the nearby Seyfert galaxy NGC~3783 which is known
to show strong absorption features in its X-ray and UV spectra.  We
compare these to the high-resolution, high S/N archival data from
2001, allowing us to investigate variations in the absorption over a
12 yr period.  We summarize our main results below.

\begin{enumerate}

\item In Figure~\ref{fig:full_x_spec} we present the broadband X-ray
  spectra of NGC~3783 from both epochs.  There is a significant drop
  in flux from $F\sb{\rm 2-6}=(3.84\pm0.01)\times 10\sp{-11}\,\flux$
  in 2001, to $F\sb{\rm 2-6}=(2.25\pm0.02)\times 10\sp{-11}\,\flux$ in
  2013.  A similar ($40\pm4\%$) drop in the $1270-1520$\,\ang\, UV
  flux is shown in Figure~\ref{fig:uv_spec}.  The change in X-ray flux
  is consistent with previous variations observed by \textit{RXTE}
  (see the long-term lightcurve presented in Figure~\ref{fig:lc}).

\item Figure~\ref{fig:full_x_spec} also shows a significant flattening
  of the \mbox{X-ray} power-law slope from $\Gamma\sb{\rm
    2-6}=1.37\pm0.01$ in 2001, to $\Gamma\sb{\rm 2-6}=1.07\pm0.04$ in
  2013.  We show that this is unlikely to be the result of an
  increased reflection component, although our data do not extend to
  the higher energies required to model this in detail.  We also found
  no strong evidence for a large increase in the column density of the
  absorbing material, although smaller variations which we are unable
  to detect could contribute to a change in $\Gamma$.  The observed
  variation could also simply be due to intrinsic variations within
  the continuum source itself.

\item Of the 19 individual absorption lines that we consider in
  \S\ref{section:lines}, we find none with a significant variation in
  their line flux or EW between the 2001 and 2013 spectra.  We
  therefore suggest that there are no dramatic changes in the covering
  factor, column density or ionization state of the absorbing
  material.

\item We do not find any significant change in the X-ray kinematics
  between the two epochs.  The average outflow velocities inferred
  from individual absorption lines ($v\sb{\rm
    2001,\,lines}=-654\pm14\;\kms$, $v\sb{\rm
    2013,\,lines}=-693\pm58\;\kms$) and the average values determined
  from the velocity profiles (see \S\ref{section:kinematics};
  $v\sb{\rm 2001,\,profiles}=-647\pm22\;\kms$, $v\sb{\rm
    2013,\,profiles}=-672\pm66\;\kms$) are all consistent.  The widths
  of the velocity profiles are also consistent in each epoch;
  $\sigma\sb{\rm obs,\,2001}=352\pm22\;\kms$, $\sigma\sb{\rm
    obs,\,2013}=309\pm69\;\kms$.

\item We do see significant changes in the UV kinematics between the
  two epochs, with kinematic components~1 and 3 being much weaker in
  2013, while components~4 and 2 are much stronger.  This may be due
  to radial deceleration of the material with component~1 decelerating
  to the location of component~4 with $a\sb{1\rightarrow4}=7\times
  10\sp{-7}\;\kmss$ ($\Delta v\sb{1\rightarrow4}\sim-268\;\kms$) and
  component~3 decelerating to the location of component~2 with
  $a\sb{3\rightarrow2}=3.3\times 10\sp{-7}\;\kmss$ ($\Delta
  v\sb{3\rightarrow2}\sim-125\;\kms$).

\item Despite the different behavior apparently observed in the X-ray
  and UV kinematics, we show that when the UV data are downgraded to
  the velocity resolution of the X-ray data, very similar velocity
  profiles are obtained.  This suggests a connection between at least
  some of the X-ray and UV outflowing material, in agreement with
  previous works.

\item We detect a narrow Fe \ka emission line at 6.4~keV in both data
  sets.  We find no significant variation between the epochs, with the
  EW and flux estimates consistent within $1\sigma$; ${\rm
    EW\sb{2001}}=86\sp{+9}\sb{-8}$ eV, ${\rm
    EW\sb{2013}}=110\sp{+40}\sb{-20}$ eV and $F\sb{\rm
    line,\,2001}=6.39\sp{+0.49}\sb{-0.51}\times10^{-13}\,\flux$,
  $F\sb{\rm
    line,\,2013}=5.55\sp{+1.03}\sb{-1.45}\times10^{-13}\,\flux$.  The
  widths of the lines are also consistent (${\rm FWHM\sb{\rm
      2001}}=2209\sp{+212}\sb{-424}\,\kms$, ${\rm FWHM\sb{\rm
      2013}}=2208\sp{+636}\sb{-1413}\,\kms$) and using simple virial
  arguments, the radial distance of the material producing the iron
  line is found to be $\sim 31$ light days or 0.026~pc from the
  central black hole.  This places the material within the inner edge
  of the torus but beyond the broad-line region.

\end{enumerate}

In general, the physical properties of the X-ray WA in NGC~3783 appear
to have remained remarkably stable between the \chandra HETGS
observations in 2001 and 2013.  This lack of significant absorption
variation suggests that the absorbers lie far from the central SMBH,
consistent with previous lower limits i.e.,~$r\ge3.2,0.6,0.2$~pc
\citep{netzer03} and $r\ge2.8,0.5$~pc \citep{behar03}.  This places
the absorbers at distances much greater than the BLR
($r=0.0015-0.0018$~pc; \citealt{onken02}) and the inner edge of the
torus ($r\sb{\rm in}=0.061$~pc; \citealt{honig13}) and hence, taken at
face value, they do not appear to be commensurate with either a
disk-wind origin \citep{elvis00} or photoionization off of the inner
torus edge \citep{krolik01}.  However, the torus does extend to larger
distances, with half-light radii of $r\sb{1/2}=4.23\pm0.63$~pc (major
axis) and $r\sb{1/2}=1.42\pm0.21$~pc (minor axis) estimated by
mid-infrared interferometry \citep{honig13}, suggesting that the WA
may still be related to this structure.  The WA we currently observe
could also have originated as a disk-wind, but travelled for many
years to its current, more distant location further from the AGN.
This is a scenario more consistent with that suggested by
\citet{king14} in which the wind collides with a shell of gas
surrounding the SMBH, which rapidly slows and cools, producing an
observable WA.  For NGC~3783 this gas is expected to lie at
$r\le14$~pc (determined from equation 8 of \citealt{king14}, and using
a velocity dispersion of $\sigma=95\pm10\;\kms$ from
\citealt{onken04}), which may be too distant to explain the WA
observed here, but since we only have lower limits on its distance,
this cannot be stated with certainty. \citet{netzer03} also gave an
upper limit of $r\le25$~pc, consistent with both this scenario, and
with the expected location of the UV absorbers \citep{gabel05}.

While we find no statistically significant variations in the physical
properties of the X-ray WA between our observations, we note that
\citet{george98} did report a decrease in the opacity by a factor of
$\approx 2$ over an 18 month period.  This is not necessarily in
conflict with our results as the observations occurred over a
different time interval (1993--1996) to the data we consider here.  It
is also not unprecedented for AGN to show periods of stability and
variability with regards to their absorption properties; e.g.~NGC~5548
\citep{steenbrugge05,detmers08,krongold10,kaastra14}.  This suggests
that AGN can host absorbing material with a range of physical
properties and distances from the central SMBH, producing variations
on a wide range of timescales.

While the X-ray absorption properties lack significant variation, the
UV absorbers show clear kinematic changes.  However, this does not
necessarily indicate that the X-ray and UV absorbing material is not
physically related as we show that such changes would simply not be
detected in the lower velocity resolution X-ray data.  The dramatic UV
variability observed on $>10$~yr timescales can be reasonably expected
to continue, suggesting that further monitoring of NGC~3783 in the UV
is likely to yield stronger constraints on the outflow physics.  The
kinematic change in UV component 1 was first observed in data from
2001 but the potential deceleration of UV component 3 has required a
longer baseline for detection due to its smaller magnitude.
Constraining changes in this deceleration, as seen in component 1,
would require an even longer baseline.  Further observations would
also allow more detailed discrimination between the various scenarios
suggested for the deceleration; e.g.~a bulk deceleration of the
material or a directional shift in the motion of the absorber with
respect to our line of sight.

The lack of variation seen in the X-ray WA suggests that constant
monitoring will be required over many years to constrain fully its
nature.  Such a campaign is impractical with \textit{Chandra} owing to
the long required exposures; a better approach would be to trigger
grating observations in response to continuous monitoring at lower
spectral resolution with, e.g.,~\textit{Swift}.  Regular monitoring at
high resolution would be an excellent and feasible program for the
forthcoming \textit{Athena} X-ray observatory.


\acknowledgments
We acknowledge the support of \chandra X-ray Center grant GO3-14114X,
Space Telescope Science Institute grant HST-GO-13115.01-A, and NASA
ADP grant NNX11AJ59G (AES, WNB).  The Technion group is supported by
the I-CORE program of the Planning and Budgeting Committee and the
Israel Science Foundation (grant numbers 1937/12 and 1163/10), and by
a grant from Israel's Ministry of Science and Technology.  This work
has made used of lightcurves provided by the University of California,
San Diego Center for Astrophysics and Space Sciences, X-ray Group
(R.~E.~Rothschild, A.~G.~Markowitz, E.~S.~Rivers, and B.~A.~McKim),
obtained at http://cass.ucsd.edu/\textasciitilde{}rxteagn/.  We also
thank Ian McHardy and Phil Uttley for providing data in advance of
publication, and the anonymous referee for helpful comments on the 
original manuscript.\\

{\it Facilities:} \facility{CXO (ACIS)}, \facility{HST (COS)}.\\




\begin{thebibliography}{99}
\expandafter\ifx\csname natexlab\endcsname\relax\def\natexlab#1{#1}\fi

\bibitem[{{Alloin} {et~al.}(1995){Alloin}, {Santos-Lleo}, {Peterson},
  {Wamsteker}, {Altieri}, {Brinkmann}, {Clavel}, {Crenshaw}, {George}, {Glass},
  {Johnson}, {Kriss}, {Malkan}, {Polidan}, {Reichert}, {Rodriguez-Pascual},
  {Romanishin}, {Starr}, {Stirpe}, {Taylor}, {Turner}, {Vega}, {Winge}, \&
  {Wood}}]{alloin95}
{Alloin}, D., {Santos-Lleo}, M., {Peterson}, B.~M., {et~al.} 1995, \aap, 293,
  293

\bibitem[{{Behar}(2009)}]{behar09}
{Behar}, E. 2009, \apj, 703, 1346

\bibitem[{{Behar} {et~al.}(2003){Behar}, {Rasmussen}, {Blustin}, {Sako},
  {Kahn}, {Kaastra}, {Branduardi-Raymont}, \& {Steenbrugge}}]{behar03}
{Behar}, E., {Rasmussen}, A.~P., {Blustin}, A.~J., {et~al.} 2003, \apj, 598,
  232

\bibitem[{{Blustin} {et~al.}(2002){Blustin}, {Branduardi-Raymont}, {Behar},
  {Kaastra}, {Kahn}, {Page}, {Sako}, \& {Steenbrugge}}]{blustin02}
{Blustin}, A.~J., {Branduardi-Raymont}, G., {Behar}, E., {et~al.} 2002, \aap,
  392, 453

\bibitem[{{Brenneman} {et~al.}(2011){Brenneman}, {Reynolds}, {Nowak}, {Reis},
  {Trippe}, {Fabian}, {Iwasawa}, {Lee}, {Miller}, {Mushotzky}, {Nandra}, \&
  {Volonteri}}]{brenneman11}
{Brenneman}, L.~W., {Reynolds}, C.~S., {Nowak}, M.~A., {et~al.} 2011, \apj,
  736, 103

\bibitem[{{Cash}(1979)}]{cash79}
{Cash}, W. 1979, \apj, 228, 939

\bibitem[{{Crenshaw} \& {Kraemer}(2012)}]{crenshaw12}
{Crenshaw}, D.~M., \& {Kraemer}, S.~B. 2012, \apj, 753, 75

\bibitem[{{Crenshaw} {et~al.}(1999){Crenshaw}, {Kraemer}, {Boggess}, {Maran},
  {Mushotzky}, \& {Wu}}]{crenshaw99}
{Crenshaw}, D.~M., {Kraemer}, S.~B., {Boggess}, A., {et~al.} 1999, \apj, 516,
  750

\bibitem[{{De Rosa} {et~al.}(2002){De Rosa}, {Piro}, {Fiore}, {Grandi},
  {Maraschi}, {Matt}, {Nicastro}, \& {Petrucci}}]{derosa02}
{De Rosa}, A., {Piro}, L., {Fiore}, F., {et~al.} 2002, \aap, 387, 838

\bibitem[{{de Vaucouleurs} {et~al.}(1991){de Vaucouleurs}, {de Vaucouleurs},
  {Corwin}, {Buta}, {Paturel}, \& {Fouqu{\'e}}}]{RC3}
{de Vaucouleurs}, G., {de Vaucouleurs}, A., {Corwin}, Jr., H.~G., {et~al.}
  1991, {Third Reference Catalogue of Bright Galaxies. Volume I: Explanations
  and references. Volume II: Data for galaxies between 0$^{h}$ and 12$^{h}$.
  Volume III: Data for galaxies between 12$^{h}$ and 24$^{h}$.}

\bibitem[{{den Herder} {et~al.}(2001){den Herder}, {Brinkman}, {Kahn},
  {Branduardi-Raymont}, {Thomsen}, {Aarts}, {Audard}, {Bixler}, {den Boggende},
  {Cottam}, {Decker}, {Dubbeldam}, {Erd}, {Goulooze}, {G{\"u}del}, {Guttridge},
  {Hailey}, {Janabi}, {Kaastra}, {de Korte}, {van Leeuwen}, {Mauche},
  {McCalden}, {Mewe}, {Naber}, {Paerels}, {Peterson}, {Rasmussen}, {Rees},
  {Sakelliou}, {Sako}, {Spodek}, {Stern}, {Tamura}, {Tandy}, {de Vries},
  {Welch}, \& {Zehnder}}]{RGS}
{den Herder}, J.~W., {Brinkman}, A.~C., {Kahn}, S.~M., {et~al.} 2001, \aap,
  365, L7

\bibitem[{{Detmers} {et~al.}(2008){Detmers}, {Kaastra}, {Costantini},
  {McHardy}, \& {Verbunt}}]{detmers08}
{Detmers}, R.~G., {Kaastra}, J.~S., {Costantini}, E., {McHardy}, I.~M., \&
  {Verbunt}, F. 2008, \aap, 488, 67

\bibitem[{{Elvis}(2000)}]{elvis00}
{Elvis}, M. 2000, \apj, 545, 63

\bibitem[{{Fabian} {et~al.}(1994){Fabian}, {Kunieda}, {Inoue}, {Matsuoka},
  {Mihara}, {Miyamoto}, {Otani}, {Ricker}, {Tanaka}, {Yamauchi}, \&
  {Yaqoob}}]{fabian94}
{Fabian}, A.~C., {Kunieda}, H., {Inoue}, S., {et~al.} 1994, \pasj, 46, L59

\bibitem[{{Ferrarese} \& {Merritt}(2000)}]{ferrarese00}
{Ferrarese}, L., \& {Merritt}, D. 2000, \apjl, 539, L9

\bibitem[{{Filiz Ak} {et~al.}(2013){Filiz Ak}, {Brandt}, {Hall}, {Schneider},
  {Anderson}, {Hamann}, {Lundgren}, {Myers}, {P{\^a}ris}, {Petitjean}, {Ross},
  {Shen}, \& {York}}]{filizak13}
{Filiz Ak}, N., {Brandt}, W.~N., {Hall}, P.~B., {et~al.} 2013, \apj, 777, 168

\bibitem[{{Fruscione} {et~al.}(2006){Fruscione}, {McDowell}, {Allen},
  {Brickhouse}, {Burke}, {Davis}, {Durham}, {Elvis}, {Galle}, {Harris},
  {Huenemoerder}, {Houck}, {Ishibashi}, {Karovska}, {Nicastro}, {Noble},
  {Nowak}, {Primini}, {Siemiginowska}, {Smith}, \& {Wise}}]{ciao}
{Fruscione}, A., {McDowell}, J.~C., {Allen}, G.~E., {et~al.} 2006, in Society
  of Photo-Optical Instrumentation Engineers (SPIE) Conference Series, Vol.
  6270, Society of Photo-Optical Instrumentation Engineers (SPIE) Conference
  Series

\bibitem[{{Gabel} {et~al.}(2002){Gabel}, {Kraemer}, \& {Crenshaw}}]{gabel02}
{Gabel}, J.~R., {Kraemer}, S.~B., \& {Crenshaw}, D.~M. 2002, in Astronomical
  Society of the Pacific Conference Series, Vol. 255, Mass Outflow in Active
  Galactic Nuclei: New Perspectives, ed. D.~M. {Crenshaw}, S.~B. {Kraemer}, \&
  I.~M. {George}, 81

\bibitem[{{Gabel} {et~al.}(2003{\natexlab{a}}){Gabel}, {Crenshaw}, {Kraemer},
  {Brandt}, {George}, {Hamann}, {Kaiser}, {Kaspi}, {Kriss}, {Mathur},
  {Mushotzky}, {Nandra}, {Netzer}, {Peterson}, {Shields}, {Turner}, \&
  {Zheng}}]{gabel03a}
{Gabel}, J.~R., {Crenshaw}, D.~M., {Kraemer}, S.~B., {et~al.}
  2003{\natexlab{a}}, \apj, 583, 178, (G03a)

\bibitem[{{Gabel} {et~al.}(2003{\natexlab{b}}){Gabel}, {Crenshaw}, {Kraemer},
  {Brandt}, {George}, {Hamann}, {Kaiser}, {Kaspi}, {Kriss}, {Mathur},
  {Mushotzky}, {Nandra}, {Netzer}, {Peterson}, {Shields}, {Turner}, \&
  {Zheng}}]{gabel03b}
---. 2003{\natexlab{b}}, \apj, 595, 120, (G03b)

\bibitem[{{Gabel} {et~al.}(2005){Gabel}, {Kraemer}, {Crenshaw}, {George},
  {Brandt}, {Hamann}, {Kaiser}, {Kaspi}, {Kriss}, {Mathur}, {Nandra}, {Netzer},
  {Peterson}, {Shields}, {Turner}, \& {Zheng}}]{gabel05}
{Gabel}, J.~R., {Kraemer}, S.~B., {Crenshaw}, D.~M., {et~al.} 2005, \apj, 631,
  741

\bibitem[{{Garmire} {et~al.}(2003){Garmire}, {Bautz}, {Ford}, {Nousek}, \&
  {Ricker}}]{acis}
{Garmire}, G.~P., {Bautz}, M.~W., {Ford}, P.~G., {Nousek}, J.~A., \& {Ricker},
  Jr., G.~R. 2003, in Society of Photo-Optical Instrumentation Engineers (SPIE)
  Conference Series, Vol. 4851, Society of Photo-Optical Instrumentation
  Engineers (SPIE) Conference Series, ed. J.~E. {Truemper} \& H.~D.
  {Tananbaum}, 28--44

\bibitem[{{Gebhardt} {et~al.}(2000){Gebhardt}, {Bender}, {Bower}, {Dressler},
  {Faber}, {Filippenko}, {Green}, {Grillmair}, {Ho}, {Kormendy}, {Lauer},
  {Magorrian}, {Pinkney}, {Richstone}, \& {Tremaine}}]{gebhardt00}
{Gebhardt}, K., {Bender}, R., {Bower}, G., {et~al.} 2000, \apjl, 539, L13

\bibitem[{{Gehrels}(1986)}]{gehrels86}
{Gehrels}, N. 1986, \apj, 303, 336

\bibitem[{{George} {et~al.}(1998){George}, {Turner}, {Mushotzky}, {Nandra}, \&
  {Netzer}}]{george98}
{George}, I.~M., {Turner}, T.~J., {Mushotzky}, R., {Nandra}, K., \& {Netzer},
  H. 1998, \apj, 503, 174

\bibitem[{{Gibson} \& {Brandt}(2012)}]{gibson12}
{Gibson}, R.~R., \& {Brandt}, W.~N. 2012, \apj, 746, 54

\bibitem[{{Gibson} {et~al.}(2007){Gibson}, {Canizares}, {Marshall}, {Young}, \&
  {Lee}}]{gibson07}
{Gibson}, R.~R., {Canizares}, C.~R., {Marshall}, H.~L., {Young}, A.~J., \&
  {Lee}, J.~C. 2007, \apj, 655, 749

\bibitem[{{Green} {et~al.}(2012){Green}, {Froning}, {Osterman}, {Ebbets},
  {Heap}, {Leitherer}, {Linsky}, {Savage}, {Sembach}, {Shull}, {Siegmund},
  {Snow}, {Spencer}, {Stern}, {Stocke}, {Welsh}, {B{\'e}land}, {Burgh},
  {Danforth}, {France}, {Keeney}, {McPhate}, {Penton}, {Andrews},
  {Brownsberger}, {Morse}, \& {Wilkinson}}]{COS}
{Green}, J.~C., {Froning}, C.~S., {Osterman}, S., {et~al.} 2012, \apj, 744, 60

\bibitem[{{G{\"u}ltekin} {et~al.}(2009){G{\"u}ltekin}, {Richstone}, {Gebhardt},
  {Lauer}, {Tremaine}, {Aller}, {Bender}, {Dressler}, {Faber}, {Filippenko},
  {Green}, {Ho}, {Kormendy}, {Magorrian}, {Pinkney}, \& {Siopis}}]{gultekin09}
{G{\"u}ltekin}, K., {Richstone}, D.~O., {Gebhardt}, K., {et~al.} 2009, \apj,
  698, 198

\bibitem[{{Halpern}(1984)}]{halpern84}
{Halpern}, J.~P. 1984, \apj, 281, 90

\bibitem[{{Holczer} \& {Behar}(2012)}]{holczer12}
{Holczer}, T., \& {Behar}, E. 2012, \apj, 747, 71

\bibitem[{{Holczer} {et~al.}(2007){Holczer}, {Behar}, \& {Kaspi}}]{holczer07}
{Holczer}, T., {Behar}, E., \& {Kaspi}, S. 2007, \apj, 663, 799

\bibitem[{{H{\"o}nig} {et~al.}(2013){H{\"o}nig}, {Kishimoto}, {Tristram},
  {Prieto}, {Gandhi}, {Asmus}, {Antonucci}, {Burtscher}, {Duschl}, \&
  {Weigelt}}]{honig13}
{H{\"o}nig}, S.~F., {Kishimoto}, M., {Tristram}, K.~R.~W., {et~al.} 2013, \apj,
  771, 87

\bibitem[{{Kaastra} {et~al.}(2004){Kaastra}, {Raassen}, {Mewe}, {Arav},
  {Behar}, {Costantini}, {Gabel}, {Kriss}, {Proga}, {Sako}, \&
  {Steenbrugge}}]{kaastra04}
{Kaastra}, J.~S., {Raassen}, A.~J.~J., {Mewe}, R., {et~al.} 2004, \aap, 428, 57

\bibitem[{{Kaastra} {et~al.}(2012){Kaastra}, {Detmers}, {Mehdipour}, {Arav},
  {Behar}, {Bianchi}, {Branduardi-Raymont}, {Cappi}, {Costantini}, {Ebrero},
  {Kriss}, {Paltani}, {Petrucci}, {Pinto}, {Ponti}, {Steenbrugge}, \& {de
  Vries}}]{kaastra12}
{Kaastra}, J.~S., {Detmers}, R.~G., {Mehdipour}, M., {et~al.} 2012, \aap, 539,
  A117

\bibitem[{{Kaastra} {et~al.}(2014){Kaastra}, {Kriss}, {Cappi}, {Mehdipour},
  {Petrucci}, {Steenbrugge}, {Arav}, {Behar}, {Bianchi}, {Boissay},
  {Branduardi-Raymont}, {Chamberlain}, {Costantini}, {Ely}, {Ebrero}, {Di
  Gesu}, {Harrison}, {Kaspi}, {Malzac}, {De Marco}, {Matt}, {Nandra},
  {Paltani}, {Person}, {Peterson}, {Pinto}, {Ponti}, {Pozo Nu{\~n}ez}, {De
  Rosa}, {Seta}, {Ursini}, {de Vries}, {Walton}, \& {Whewell}}]{kaastra14}
{Kaastra}, J.~S., {Kriss}, G.~A., {Cappi}, M., {et~al.} 2014, Science, in press

\bibitem[{{Kaiser} {et~al.}(2002){Kaiser}, {Kriss}, \& {Sembach}}]{kaiser02}
{Kaiser}, M.~E., {Kriss}, G.~A., \& {Sembach}, K.~R. 2002, in Astronomical
  Society of the Pacific Conference Series, Vol. 255, Mass Outflow in Active
  Galactic Nuclei: New Perspectives, ed. D.~M. {Crenshaw}, S.~B. {Kraemer}, \&
  I.~M. {George}, 75

\bibitem[{{Kalberla} {et~al.}(2005){Kalberla}, {Burton}, {Hartmann}, {Arnal},
  {Bajaja}, {Morras}, \& {P{\"o}ppel}}]{kalberla05}
{Kalberla}, P.~M.~W., {Burton}, W.~B., {Hartmann}, D., {et~al.} 2005, \aap,
  440, 775

\bibitem[{{Kaspi} {et~al.}(2000){Kaspi}, {Brandt}, {Netzer}, {Sambruna},
  {Chartas}, {Garmire}, \& {Nousek}}]{kaspi00}
{Kaspi}, S., {Brandt}, W.~N., {Netzer}, H., {et~al.} 2000, \apjl, 535, L17

\bibitem[{{Kaspi} {et~al.}(2004){Kaspi}, {Netzer}, {Chelouche}, {George},
  {Nandra}, \& {Turner}}]{kaspi04}
{Kaspi}, S., {Netzer}, H., {Chelouche}, D., {et~al.} 2004, \apj, 611, 68

\bibitem[{{Kaspi} {et~al.}(2001){Kaspi}, {Brandt}, {Netzer}, {George},
  {Chartas}, {Behar}, {Sambruna}, {Garmire}, \& {Nousek}}]{kaspi01}
{Kaspi}, S., {Brandt}, W.~N., {Netzer}, H., {et~al.} 2001, \apj, 554, 216

\bibitem[{{Kaspi} {et~al.}(2002){Kaspi}, {Brandt}, {George}, {Netzer},
  {Crenshaw}, {Gabel}, {Hamann}, {Kaiser}, {Koratkar}, {Kraemer}, {Kriss},
  {Mathur}, {Mushotzky}, {Nandra}, {Peterson}, {Shields}, {Turner}, \&
  {Zheng}}]{kaspi02}
{Kaspi}, S., {Brandt}, W.~N., {George}, I.~M., {et~al.} 2002, \apj, 574, 643,
  (K02)

\bibitem[{{King}(2003)}]{king03}
{King}, A. 2003, \apjl, 596, L27

\bibitem[{{King}(2005)}]{king05}
---. 2005, \apjl, 635, L121

\bibitem[{{King} \& {Pounds}(2014)}]{king14}
{King}, A.~R., \& {Pounds}, K.~A. 2014, \mnras, 437, L81

\bibitem[{{Kormendy} \& {Ho}(2013)}]{kormendy13}
{Kormendy}, J., \& {Ho}, L.~C. 2013, \araa, 51, 511

\bibitem[{{Kraemer} {et~al.}(2001){Kraemer}, {Crenshaw}, \&
  {Gabel}}]{kraemer01}
{Kraemer}, S.~B., {Crenshaw}, D.~M., \& {Gabel}, J.~R. 2001, \apj, 557, 30

\bibitem[{{Kraemer} {et~al.}(2002){Kraemer}, {Crenshaw}, {George}, {Netzer},
  {Turner}, \& {Gabel}}]{kraemer02}
{Kraemer}, S.~B., {Crenshaw}, D.~M., {George}, I.~M., {et~al.} 2002, \apj, 577,
  98

\bibitem[{{Krolik} \& {Kriss}(2001)}]{krolik01}
{Krolik}, J.~H., \& {Kriss}, G.~A. 2001, \apj, 561, 684

\bibitem[{{Krongold} {et~al.}(2003){Krongold}, {Nicastro}, {Brickhouse},
  {Elvis}, {Liedahl}, \& {Mathur}}]{krongold03}
{Krongold}, Y., {Nicastro}, F., {Brickhouse}, N.~S., {et~al.} 2003, \apj, 597,
  832

\bibitem[{{Krongold} {et~al.}(2010){Krongold}, {Elvis}, {Andrade-Velazquez},
  {Nicastro}, {Mathur}, {Reeves}, {Brickhouse}, {Binette}, {Jimenez-Bailon},
  {Grupe}, {Liu}, {McHardy}, {Minezaki}, {Yoshii}, \& {Wilkes}}]{krongold10}
{Krongold}, Y., {Elvis}, M., {Andrade-Velazquez}, M., {et~al.} 2010, \apj, 710,
  360

\bibitem[{{Lamer} {et~al.}(2003){Lamer}, {McHardy}, {Uttley}, \&
  {Jahoda}}]{lamer03}
{Lamer}, G., {McHardy}, I.~M., {Uttley}, P., \& {Jahoda}, K. 2003, \mnras, 338,
  323

\bibitem[{{Longinotti} {et~al.}(2013){Longinotti}, {Krongold}, {Kriss}, {Ely},
  {Gallo}, {Grupe}, {Komossa}, {Mathur}, \& {Pradhan}}]{longinotti13}
{Longinotti}, A.~L., {Krongold}, Y., {Kriss}, G.~A., {et~al.} 2013, \apj, 766,
  104

\bibitem[{{Maran} {et~al.}(1996){Maran}, {Crenshaw}, {Mushotzky}, {Reichert},
  {Carpenter}, {Smith}, {Hutchings}, \& {Weymann}}]{maran96}
{Maran}, S.~P., {Crenshaw}, D.~M., {Mushotzky}, R.~F., {et~al.} 1996, \apj,
  465, 733

\bibitem[{{Marchese} {et~al.}(2014){Marchese}, {Braito}, {Reeves}, {Ceca},
  {Caccianiga}, {Markowitz}, {Risaliti}, {Severgnini}, \&
  {Turner}}]{marchese14}
{Marchese}, E., {Braito}, V., {Reeves}, J.~N., {et~al.} 2014, \mnras, 437, 2806

\bibitem[{{Marinucci} {et~al.}(2014){Marinucci}, {Matt}, {Miniutti},
  {Guainazzi}, {Parker}, {Brenneman}, {Fabian}, {Kara}, {Arevalo},
  {Ballantyne}, {Boggs}, {Cappi}, {Christensen}, {Craig}, {Elvis}, {Hailey},
  {Harrison}, {Reynolds}, {Risaliti}, {Stern}, {Walton}, \&
  {Zhang}}]{marinucci14}
{Marinucci}, A., {Matt}, G., {Miniutti}, G., {et~al.} 2014, \apj, 787, 83

\bibitem[{{Markert} {et~al.}(1994){Markert}, {Canizares}, {Dewey}, {McGuirk},
  {Pak}, \& {Schattenburg}}]{hetgs}
{Markert}, T.~H., {Canizares}, C.~R., {Dewey}, D., {et~al.} 1994, in Society of
  Photo-Optical Instrumentation Engineers (SPIE) Conference Series, Vol. 2280,
  Society of Photo-Optical Instrumentation Engineers (SPIE) Conference Series,
  ed. O.~H. {Siegmund} \& J.~V. {Vallerga}, 168--180

\bibitem[{{Markowitz} \& {Edelson}(2004)}]{markowitz04}
{Markowitz}, A., \& {Edelson}, R. 2004, \apj, 617, 939

\bibitem[{{Markowitz} {et~al.}(2014){Markowitz}, {Krumpe}, \&
  {Nikutta}}]{markowitz14}
{Markowitz}, A.~G., {Krumpe}, M., \& {Nikutta}, R. 2014, \mnras, 439, 1403

\bibitem[{{Mathur} {et~al.}(1994){Mathur}, {Wilkes}, {Elvis}, \&
  {Fiore}}]{mathur94}
{Mathur}, S., {Wilkes}, B., {Elvis}, M., \& {Fiore}, F. 1994, \apj, 434, 493

\bibitem[{{Matt} {et~al.}(2011){Matt}, {Bianchi}, {Guainazzi}, {Longinotti},
  {Dadina}, {Karas}, {Malaguti}, {Miniutti}, {Petrucci}, {Piconcelli}, \&
  {Ponti}}]{matt11}
{Matt}, G., {Bianchi}, S., {Guainazzi}, M., {et~al.} 2011, \aap, 533, A1

\bibitem[{{McConnell} \& {Ma}(2013)}]{mcconnell13}
{McConnell}, N.~J., \& {Ma}, C.-P. 2013, \apj, 764, 184

\bibitem[{{McKernan} \& {Yaqoob}(1998)}]{mckernan98}
{McKernan}, B., \& {Yaqoob}, T. 1998, \apjl, 501, L29

\bibitem[{{McQuillin} \& {McLaughlin}(2013)}]{mcquillin13}
{McQuillin}, R.~C., \& {McLaughlin}, D.~E. 2013, \mnras, 434, 1332

\bibitem[{{Miller} {et~al.}(2007){Miller}, {Turner}, {Reeves}, {George},
  {Kraemer}, \& {Wingert}}]{miller07}
{Miller}, L., {Turner}, T.~J., {Reeves}, J.~N., {et~al.} 2007, \aap, 463, 131

\bibitem[{{Miyazawa} {et~al.}(2009){Miyazawa}, {Haba}, \&
  {Kunieda}}]{miyazawa09}
{Miyazawa}, T., {Haba}, Y., \& {Kunieda}, H. 2009, \pasj, 61, 1331

\bibitem[{{Mushotzky} {et~al.}(1980){Mushotzky}, {Marshall}, {Boldt}, {Holt},
  \& {Serlemitsos}}]{mushotzky80}
{Mushotzky}, R.~F., {Marshall}, F.~E., {Boldt}, E.~A., {Holt}, S.~S., \&
  {Serlemitsos}, P.~J. 1980, \apj, 235, 377

\bibitem[{{Nandra} {et~al.}(1997){Nandra}, {George}, {Mushotzky}, {Turner}, \&
  {Yaqoob}}]{nandra97}
{Nandra}, K., {George}, I.~M., {Mushotzky}, R.~F., {Turner}, T.~J., \&
  {Yaqoob}, T. 1997, \apj, 477, 602

\bibitem[{{Nandra} \& {Pounds}(1994)}]{nandra94}
{Nandra}, K., \& {Pounds}, K.~A. 1994, \mnras, 268, 405

\bibitem[{{Netzer} {et~al.}(2003){Netzer}, {Kaspi}, {Behar}, {Brandt},
  {Chelouche}, {George}, {Crenshaw}, {Gabel}, {Hamann}, {Kraemer}, {Kriss},
  {Nandra}, {Peterson}, {Shields}, \& {Turner}}]{netzer03}
{Netzer}, H., {Kaspi}, S., {Behar}, E., {et~al.} 2003, \apj, 599, 933

\bibitem[{{Onken} {et~al.}(2004){Onken}, {Ferrarese}, {Merritt}, {Peterson},
  {Pogge}, {Vestergaard}, \& {Wandel}}]{onken04}
{Onken}, C.~A., {Ferrarese}, L., {Merritt}, D., {et~al.} 2004, \apj, 615, 645

\bibitem[{{Onken} \& {Peterson}(2002)}]{onken02}
{Onken}, C.~A., \& {Peterson}, B.~M. 2002, \apj, 572, 746

\bibitem[{{Pan} {et~al.}(1990){Pan}, {Stewart}, \& {Pounds}}]{pan90}
{Pan}, H.~C., {Stewart}, G.~C., \& {Pounds}, K.~A. 1990, \mnras, 242, 177

\bibitem[{{Patrick} {et~al.}(2011){Patrick}, {Reeves}, {Lobban}, {Porquet}, \&
  {Markowitz}}]{patrick11}
{Patrick}, A.~R., {Reeves}, J.~N., {Lobban}, A.~P., {Porquet}, D., \&
  {Markowitz}, A.~G. 2011, \mnras, 416, 2725

\bibitem[{{Peterson} {et~al.}(2004){Peterson}, {Ferrarese}, {Gilbert}, {Kaspi},
  {Malkan}, {Maoz}, {Merritt}, {Netzer}, {Onken}, {Pogge}, {Vestergaard}, \&
  {Wandel}}]{peterson04}
{Peterson}, B.~M., {Ferrarese}, L., {Gilbert}, K.~M., {et~al.} 2004, \apj, 613,
  682

\bibitem[{{Pounds} {et~al.}(2004){Pounds}, {Reeves}, {Page}, \&
  {O'Brien}}]{pounds04}
{Pounds}, K.~A., {Reeves}, J.~N., {Page}, K.~L., \& {O'Brien}, P.~T. 2004,
  \apj, 605, 670

\bibitem[{{Reeves} {et~al.}(2004){Reeves}, {Nandra}, {George}, {Pounds},
  {Turner}, \& {Yaqoob}}]{reeves04}
{Reeves}, J.~N., {Nandra}, K., {George}, I.~M., {et~al.} 2004, \apj, 602, 648

\bibitem[{{Reeves} {et~al.}(2013){Reeves}, {Porquet}, {Braito}, {Gofford},
  {Nardini}, {Turner}, {Crenshaw}, \& {Kraemer}}]{reeves13}
{Reeves}, J.~N., {Porquet}, D., {Braito}, V., {et~al.} 2013, \apj, 776, 99

\bibitem[{{Reichert} {et~al.}(1994){Reichert}, {Rodriguez-Pascual}, {Alloin},
  {Clavel}, {Crenshaw}, {Kriss}, {Krolik}, {Malkan}, {Netzer}, {Peterson},
  {Wamsteker}, {Altamore}, {Altieri}, {Anderson}, {Blackwell}, {Boisson},
  {Brosch}, {Carone}, {Dietrich}, {England}, {Evans}, {Filippenko}, {Gaskell},
  {Goad}, {Gondhalekar}, {Horne}, {Kazanas}, {Kollatschny}, {Koratkar},
  {Korista}, {MacAlpine}, {Maoz}, {Mazeh}, {McCollum}, {Miller}, {Mendes de
  Oliveira}, {O'Brien}, {Pastoriza}, {Pelat}, {Perez}, {Perola}, {Pogge},
  {Ptak}, {Recondo-Gonzalez}, {Rodriguez-Espinosa}, {Rosenblatt}, {Sadun},
  {Santos-Lleo}, {Shields}, {Shrader}, {Shull}, {Simkin}, {Sitko}, {Snijders},
  {Sparke}, {Stirpe}, {Stoner}, {Storchi-Bergmann}, {Sun}, {Wang}, {Welsh},
  {White}, {Winge}, \& {Zheng}}]{reichert94}
{Reichert}, G.~A., {Rodriguez-Pascual}, P.~M., {Alloin}, D., {et~al.} 1994,
  \apj, 425, 582

\bibitem[{{Reis} {et~al.}(2012){Reis}, {Fabian}, {Reynolds}, {Brenneman},
  {Walton}, {Trippe}, {Miller}, {Mushotzky}, \& {Nowak}}]{reis12}
{Reis}, R.~C., {Fabian}, A.~C., {Reynolds}, C.~S., {et~al.} 2012, \apj, 745, 93

\bibitem[{{Reynolds}(1997)}]{reynolds97}
{Reynolds}, C.~S. 1997, \mnras, 286, 513

\bibitem[{{Reynolds} {et~al.}(2012){Reynolds}, {Brenneman}, {Lohfink},
  {Trippe}, {Miller}, {Fabian}, \& {Nowak}}]{reynolds12}
{Reynolds}, C.~S., {Brenneman}, L.~W., {Lohfink}, A.~M., {et~al.} 2012, \apj,
  755, 88

\bibitem[{{Rivers} {et~al.}(2013){Rivers}, {Markowitz}, \&
  {Rothschild}}]{rivers13}
{Rivers}, E., {Markowitz}, A., \& {Rothschild}, R. 2013, \apj, 772, 114

\bibitem[{{Shields} \& {Hamann}(1997)}]{shields97}
{Shields}, J.~C., \& {Hamann}, F. 1997, \apj, 481, 752

\bibitem[{{Silk} \& {Rees}(1998)}]{silk98}
{Silk}, J., \& {Rees}, M.~J. 1998, \aap, 331, L1

\bibitem[{{Steenbrugge} {et~al.}(2005){Steenbrugge}, {Kaastra}, {Crenshaw},
  {Kraemer}, {Arav}, {George}, {Liedahl}, {van der Meer}, {Paerels}, {Turner},
  \& {Yaqoob}}]{steenbrugge05}
{Steenbrugge}, K.~C., {Kaastra}, J.~S., {Crenshaw}, D.~M., {et~al.} 2005, \aap,
  434, 569

\bibitem[{{Str{\"u}der} {et~al.}(2001){Str{\"u}der}, {Briel}, {Dennerl},
  {Hartmann}, {Kendziorra}, {Meidinger}, {Pfeffermann}, {Reppin}, {Aschenbach},
  {Bornemann}, {Br{\"a}uninger}, {Burkert}, {Elender}, {Freyberg}, {Haberl},
  {Hartner}, {Heuschmann}, {Hippmann}, {Kastelic}, {Kemmer}, {Kettenring},
  {Kink}, {Krause}, {M{\"u}ller}, {Oppitz}, {Pietsch}, {Popp}, {Predehl},
  {Read}, {Stephan}, {St{\"o}tter}, {Tr{\"u}mper}, {Holl}, {Kemmer}, {Soltau},
  {St{\"o}tter}, {Weber}, {Weichert}, {von Zanthier}, {Carathanassis}, {Lutz},
  {Richter}, {Solc}, {B{\"o}ttcher}, {Kuster}, {Staubert}, {Abbey}, {Holland},
  {Turner}, {Balasini}, {Bignami}, {La Palombara}, {Villa}, {Buttler},
  {Gianini}, {Lain{\'e}}, {Lumb}, \& {Dhez}}]{pn}
{Str{\"u}der}, L., {Briel}, U., {Dennerl}, K., {et~al.} 2001, \aap, 365, L18

\bibitem[{{Taylor} {et~al.}(2003){Taylor}, {Uttley}, \& {McHardy}}]{taylor03}
{Taylor}, R.~D., {Uttley}, P., \& {McHardy}, I.~M. 2003, \mnras, 342, L31

\bibitem[{{Tombesi} {et~al.}(2013){Tombesi}, {Cappi}, {Reeves}, {Nemmen},
  {Braito}, {Gaspari}, \& {Reynolds}}]{tombesi13}
{Tombesi}, F., {Cappi}, M., {Reeves}, J.~N., {et~al.} 2013, \mnras, 430, 1102

\bibitem[{{Turner} {et~al.}(2001){Turner}, {Abbey}, {Arnaud}, {Balasini},
  {Barbera}, {Belsole}, {Bennie}, {Bernard}, {Bignami}, {Boer}, {Briel},
  {Butler}, {Cara}, {Chabaud}, {Cole}, {Collura}, {Conte}, {Cros}, {Denby},
  {Dhez}, {Di Coco}, {Dowson}, {Ferrando}, {Ghizzardi}, {Gianotti}, {Goodall},
  {Gretton}, {Griffiths}, {Hainaut}, {Hochedez}, {Holland}, {Jourdain},
  {Kendziorra}, {Lagostina}, {Laine}, {La Palombara}, {Lortholary}, {Lumb},
  {Marty}, {Molendi}, {Pigot}, {Poindron}, {Pounds}, {Reeves}, {Reppin},
  {Rothenflug}, {Salvetat}, {Sauvageot}, {Schmitt}, {Sembay}, {Short},
  {Spragg}, {Stephen}, {Str{\"u}der}, {Tiengo}, {Trifoglio}, {Tr{\"u}mper},
  {Vercellone}, {Vigroux}, {Villa}, {Ward}, {Whitehead}, \& {Zonca}}]{MOS}
{Turner}, M.~J.~L., {Abbey}, A., {Arnaud}, M., {et~al.} 2001, \aap, 365, L27

\bibitem[{{Turner} {et~al.}(2005){Turner}, {Kraemer}, {George}, {Reeves}, \&
  {Bottorff}}]{turner05}
{Turner}, T.~J., {Kraemer}, S.~B., {George}, I.~M., {Reeves}, J.~N., \&
  {Bottorff}, M.~C. 2005, \apj, 618, 155

\bibitem[{{Turner} {et~al.}(1993){Turner}, {Nandra}, {George}, {Fabian}, \&
  {Pounds}}]{turner93}
{Turner}, T.~J., {Nandra}, K., {George}, I.~M., {Fabian}, A.~C., \& {Pounds},
  K.~A. 1993, \apj, 419, 127

\bibitem[{{Turner} \& {Pounds}(1989)}]{turner89}
{Turner}, T.~J., \& {Pounds}, K.~A. 1989, \mnras, 240, 833

\bibitem[{{Turner} {et~al.}(2008){Turner}, {Reeves}, {Kraemer}, \&
  {Miller}}]{turner08}
{Turner}, T.~J., {Reeves}, J.~N., {Kraemer}, S.~B., \& {Miller}, L. 2008, \aap,
  483, 161

\bibitem[{{Vaughan} \& {Fabian}(2004)}]{vaughan04}
{Vaughan}, S., \& {Fabian}, A.~C. 2004, \mnras, 348, 1415

\bibitem[{{Weisskopf} {et~al.}(2000){Weisskopf}, {Tananbaum}, {Van Speybroeck},
  \& {O'Dell}}]{chandra}
{Weisskopf}, M.~C., {Tananbaum}, H.~D., {Van Speybroeck}, L.~P., \& {O'Dell},
  S.~L. 2000, in Society of Photo-Optical Instrumentation Engineers (SPIE)
  Conference Series, Vol. 4012, Society of Photo-Optical Instrumentation
  Engineers (SPIE) Conference Series, ed. J.~E. {Truemper} \& B.~{Aschenbach},
  2--16

\bibitem[{{Willingale} {et~al.}(2013){Willingale}, {Starling}, {Beardmore},
  {Tanvir}, \& {O'Brien}}]{willingale13}
{Willingale}, R., {Starling}, R.~L.~C., {Beardmore}, A.~P., {Tanvir}, N.~R., \&
  {O'Brien}, P.~T. 2013, \mnras, 431, 394

\bibitem[{{Woodgate} {et~al.}(1998){Woodgate}, {Kimble}, {Bowers}, {Kraemer},
  {Kaiser}, {Danks}, {Grady}, {Loiacono}, {Brumfield}, {Feinberg}, {Gull},
  {Heap}, {Maran}, {Lindler}, {Hood}, {Meyer}, {Vanhouten}, {Argabright},
  {Franka}, {Bybee}, {Dorn}, {Bottema}, {Woodruff}, {Michika}, {Sullivan},
  {Hetlinger}, {Ludtke}, {Stocker}, {Delamere}, {Rose}, {Becker}, {Garner},
  {Timothy}, {Blouke}, {Joseph}, {Hartig}, {Green}, {Jenkins}, {Linsky},
  {Hutchings}, {Moos}, {Boggess}, {Roesler}, \& {Weistrop}}]{STIS}
{Woodgate}, B.~E., {Kimble}, R.~A., {Bowers}, C.~W., {et~al.} 1998, \pasp, 110,
  1183

\bibitem[{{Yaqoob} {et~al.}(2005){Yaqoob}, {Reeves}, {Markowitz},
  {Serlemitsos}, \& {Padmanabhan}}]{yaqoob05}
{Yaqoob}, T., {Reeves}, J.~N., {Markowitz}, A., {Serlemitsos}, P.~J., \&
  {Padmanabhan}, U. 2005, \apj, 627, 156

\end{thebibliography}
\end{document}